\documentclass[%
reprint,
amsmath,amssymb,
aps,
]{revtex4-1}

\usepackage{graphicx}
\usepackage{dcolumn}
\usepackage{bm}
\usepackage[colorlinks,citecolor=blue,urlcolor=blue,linkcolor=blue]{hyperref}

\usepackage[normalem]{ulem}
\usepackage{gensymb}
\usepackage{xcolor}

\begin{document}
	
	\preprint{APS/123-QED}
	
	\title{Observation of Quantum metric and non-Hermitian Berry curvature in a plasmonic lattice}

	\author{Javier Cuerda}
	
	\email{javier.cuerda@aalto.fi}
	\author{Jani M. Taskinen}
	\author{Nicki K\"allman}
	\author{Leo Grabitz}
	\author{P\"aivi T\"orm\"a}%
	\email{paivi.torma@aalto.fi}
	\affiliation{%
		Department of Applied Physics, Aalto University School of Science, Aalto FI-00076, Finland	
	}%

	\date{\today}%
	
	\setlength{\parskip}{0pt} %
	
	\begin{abstract}
		We experimentally observe the quantum geometric tensor, namely the quantum metric and the Berry curvature, for a square lattice of radiatively coupled plasmonic nanoparticles. We observe a non-zero Berry curvature and show that it arises solely from non-Hermitian effects. The quantum metric is found to originate from a pseudospin-orbit coupling. The long-range nature of the radiative interaction renders the behavior distinct from tight-binding systems: Berry curvature and quantum metric are centered around high-symmetry lines of the Brillouin zone instead of high-symmetry points. Our results inspire new pathways in the design of topological systems by tailoring losses or gain.

	\end{abstract}

	\maketitle

	Classification of matter has traditionally relied on the energy dispersion relation. Increasing attention has been paid to the structure of the eigenfunctions (Bloch functions in a crystal), characterized by the Berry curvature and the Chern number, both in electronic~\cite{KlitzingPRL1980,ThoulessPRL1982,BernevigSuperconductors,HaldanePRL1988,Kane1PRL2005} and photonic~\cite{PriceJPhysPhot2022,OtaNanophotonics2020,OzawaRevModPhys2019,KhanikaevNatPhot2017,SunProgQuantElec2017,LuNatPhot2014,XuPRL2017,SmirnovaAPR2020} systems.
	Consequently, the concept of the quantum geometric tensor (QGT)~\cite{Provost1980} has gained importance. The real part of the QGT is the quantum (Fubini-Study) metric and the imaginary part gives the Berry curvature. While the Berry curvature and the Chern number have been broadly utilized, the importance of the quantum metric has only emerged recently. It has been predicted to be crucial in flat band superconductivity~\cite{PeottaNatComm2015,LiangPRB2017,HuhtinenPRB2022,TormaNatRevPhys2022} and other phenomena~\cite{PiechonPRB2016,BraunRevModPhys2018}. The first measurements of the full QGT have been achieved recently with superconducting circuits \cite{ZhengChinPhysLett2022}, coupled qubits in diamond \cite{YuNatSciRev2019}, an optical Raman lattice \cite{YiArxiv2023}, and exciton-polariton modes in a microcavity \cite{GianfrateNature2020,RenNatComm2021}.
	
	Studies of topological phenomena have extended to systems with gain or loss~\cite{ElGanainyNatPhys2018,ZhaoScience2019,Moiseyev2011,BergholtzRevModPhys2021,DingNatRevPhys2022,OkumaAnnuRevCondMat2023,NasariOptMatExp2023,LiuNanophotonics2023,WangJOSAB2023}. Non-hermiticity expands the classification of topological systems \cite{GongPRX2018,KawabataPRX2019} and brings new concepts and applications into play~\cite{YaoPRL2018,DopplerNature2016,ChenNature2017,WiersigPRL2014,MiriScience2019,OzdemirNatMat2019,DennerNatComm2021,SolnyshkovPRB2021,ShenPRL2018,LeykamPRL2017,SuSciAdv2021}. Thus far, the QGT has been measured mostly in Hermitian systems, and only very recently the quantum metric of a non-Hermitian system has been observed \cite{LiaoPRL2021}.  
	To our knowledge, non-Hermitian Berry curvature has not yet been experimentally measured. In this work, we experimentally measure the full quantum geometric tensor in a lattice of plasmonic nanoparticles and show that the quantum metric arises from (pseudo)spin-orbit coupling while the Berry curvature has a purely non-Hermitian origin. 
	
\begin{figure}[htp!]
		\includegraphics[width=0.90\columnwidth]{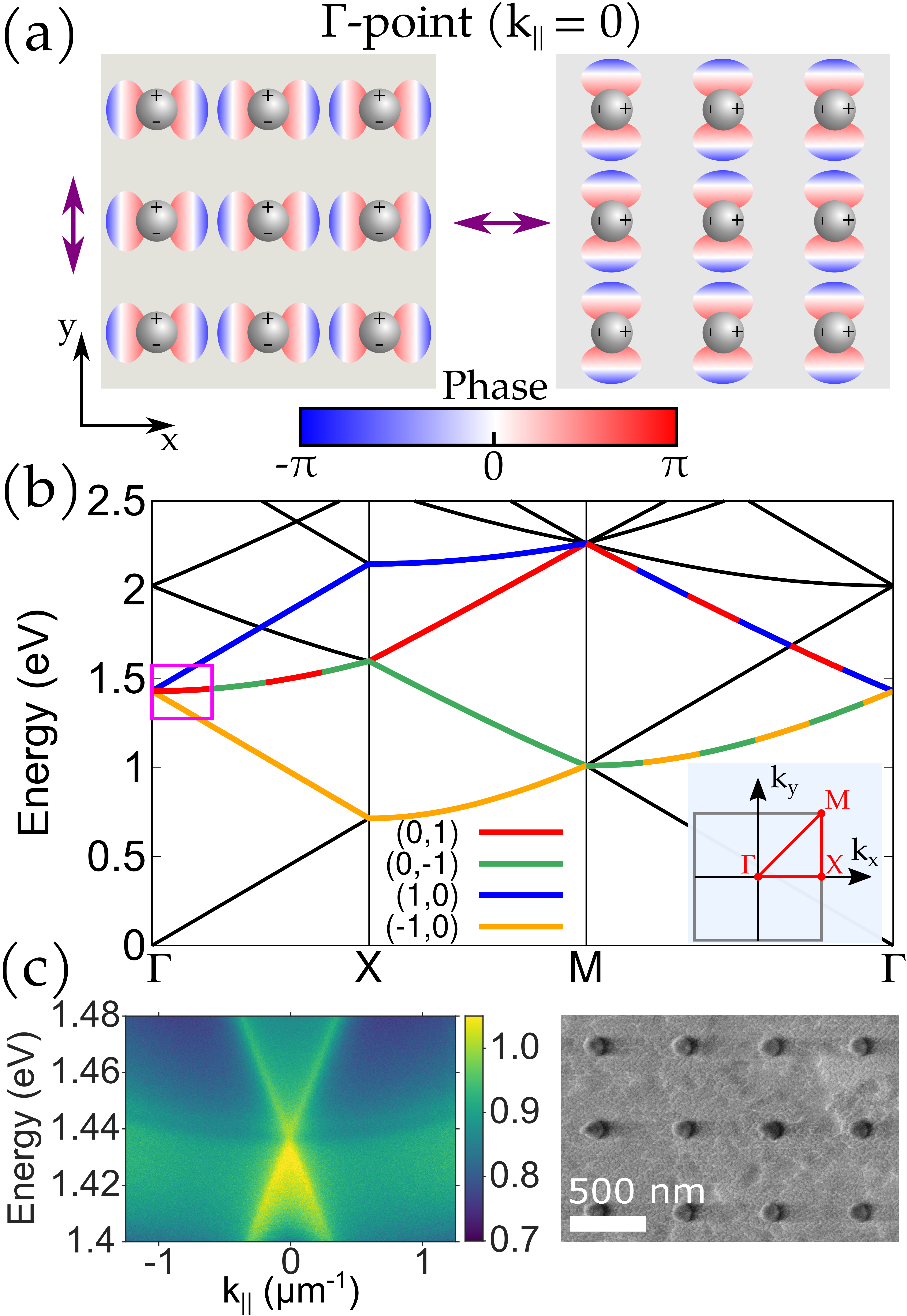}%
		\caption{\label{fig:soc_ela_transm} (a)  
			Schematic illustration of the SLR modes. Long-range radiative interactions (diffraction) couple the nanoparticles and create a collective mode, thus their radiation patterns acquire a phase that varies in real space depending on the lattice momentum $\mathbf{k}_{||}$ (the phase profile is only schematic, see Supplemental Material for realistic simulations). On the other hand, the individual nanoparticles radiate in a highly directional and polarization-dependent manner, thus the (typically dipolar) radiation patterns are very different for TE or TM (as defined with respect to $\mathbf{k}_{||}$) incoming light polarization. At normal incidence ($\mathbf{k}_{||}=0$), the phase difference of the in-plane electric field between different unit cells is zero and the dipolar moments oscillate all with the same phase (here denoted by having the same color at each nanoparticle). This yields two degenerate modes, identical by a $\pi/2$ rotation. (b) Band structure in the empty lattice approximation. Only the diffraction orders $(\pm 1,0)$ and $(0,\pm 1)$ are relevant in this work, and the rest are represented with black lines. The inset shows the trajectory $\Gamma-X-M-\Gamma$ in reciprocal space, where the Brillouin zone is delimited with grey lines. (c) Transmission measurement at the range of energies shown in the magenta square in (b) with $k_{||}$ pointing along the line $\Gamma-X$. A scanning electron micrograph of the experimental sample is provided.} 
		\vspace*{-4mm}
	\end{figure}

	Plasmonic metal and dielectric nanoparticles coupled radiatively in two-dimensional lattices and combined with active emitters have enabled strong light-matter coupling, lasing action and Bose-Einstein condensation~\cite{ZhouNatNano2013,SchokkerOptica2016,DeGiorgiACSPhot2018,WangMatToday2018,WuNanoLett2020,RamezaniOptica2017,HakalaNatPhys2018,VakevainenNatComm2020,KoshelevACSPhot2021,GuanChemRev2022,CastellanosAdvOptMat2023}, and represent intriguing potential for topological photonics~\cite{RiderACSPhot2022,TaskinenNanoLett2021,HaNatNano2018,MuraiACSPhot2020,HeilmannACSPhot2022,LiuNanophotonics2021,SalernoPRL2022}. These so-called plasmonic lattices sustain unique electromagnetic modes called surface lattice resonances (SLRs) \cite{KravetsChemRev2018,AuguiePRL2008,RodriguezPRX2011} that emerge from the long-range radiative interactions between the localized surface plasmon resonances (LSPRs) of individual nanoparticles. 
	SLRs have highly dispersive bands with polarization-dependent properties \cite{GuoPRB2017,KnudsonACSNano2019} since nanoparticles act like small (dipole or multipole) antenna. The strength of the coupling to different directions depends on the orientation of the dipole, i.e.~the polarization of the mode, see Fig.~\ref{fig:soc_ela_transm}(a). The long-range coupling between nanoparticles makes tight binding approximations non-applicable~\cite{WeickPRL2013}. The high dissipative losses of plasmonic systems, typically considered a caveat~\cite{KuznetsovScience2016,KrasnokIEEE2020}, here turn out to be the origin of an interesting feature, namely a non-zero Berry curvature
	despite a trivial lattice geometry and absence of magnetic field.

	We may understand the band structure of SLR modes in the two-dimensional $k-$space with an empty lattice approximation: the diffraction orders $(m,n)$ folded to the first Brillouin zone give the energy band dispersions of the system (see Supplemental Material~\cite{SupplementalMat}).
	In the square array of Fig.~\ref{fig:soc_ela_transm}(a), the energy of the empty lattice bands is \cite{GuoPRB2017,QuantumPlasmonics}
	\begin{equation}\label{energy_ela}
		E_{mn}=\dfrac{\hbar c}{n_{h}}\sqrt{(k_{x}+mG_{x})^{2}+(k_{y}+nG_{y})^{2}},
	\end{equation}
	where $m,n=0,\pm1,\pm2\ldots$, $G_{x,y}=2\pi/p$, $k_{x,y}$ define the direction of the in-plane $k-$vector, and $p=570$ nm is the lattice period. 
	Fig.~\ref{fig:soc_ela_transm}(b) shows the empty lattice bands along the path $\Gamma-X-M-\Gamma$ in reciprocal space. Along the trajectory $\Gamma-M$ (the diagonal of the first Brillouin zone), the empty lattice mode $(0,-1)$ becomes degenerate with $(-1,0)$; likewise $(0,1)$ is degenerate with $(1,0)$. For definiteness, we focus on the two transverse magnetic (TM) and transverse electric (TE) SLR bands related to empty lattice modes (0,-1) and (-1,0).
	
	The QGT is defined as
	\begin{equation}\label{def:qgt}
		T_{ij}^{n}=\langle\partial_{i}u_{n,\mathbf{k}}|\partial_{j}u_{n,\mathbf{k}}\rangle-\langle\partial_{i}u_{n,\mathbf{k}}|u_{n,\mathbf{k}}\rangle\langle u_{n,\mathbf{k}}|\partial_{j}u_{n,\mathbf{k}}\rangle,
	\end{equation}
	where $\partial_{i}\equiv\partial_{k_{i}}$ with $i=x,y$. The quantum metric is $g_{ij}^{n}=\Re T_{ij}^{n}$ and the Berry curvature is $\mathfrak{B}_{ij}^{n}=-2\Im T_{ij}^{n}$. 
	In our case, $|u_{n,\mathbf{k}}\rangle$ is the periodic part of the total Bloch function $e^{i\mathbf{k}\cdot\mathbf{r}}|u_{n,\mathbf{k}}\rangle$ of the band $n$. The QGT is non-trivial only in systems with multiple bands, i.e., multiple degrees of freedom (``orbitals'') associated with each unit cell. In our case, these degrees of freedom are the two polarization directions of light (giving for example the TE and TM modes of Fig.~\ref{fig:soc_ela_transm}(a)) and $|u_{n,\mathbf{k}}\rangle$ is a vector in the polarization basis. The QGT thus tells about how the polarization properties of a mode change across the Brillouin zone.
	
	We extract the QGT from experimentally obtained lattice dispersions by adapting the methods in Refs.~\cite{BleuPRB2018,GianfrateNature2020} to our system (see Supplemental Material~\cite{SupplementalMat}). Polarization-resolved transmission measurements are used to track the intensity maxima of the SLR bands (see Fig.~\ref{fig:soc_ela_transm}(c)), from these  the Stokes vector and the QGT are calculated. The samples consist of cylindrical gold nanoparticles (75~nm radius and 50 nm~height) arranged in a square lattice with a periodicity of 570~nm. The QGT is related to the single-particle band structure, thus all our theoretical results and the experimental extraction scheme apply for both quantum and classical light fields. Our experiments are in the classical (many-photon) regime. We note that transmission in our case, unlike in other non-Hermitian systems \cite{ZhenNature2015}, accurately describes the band structure and the polarization-dependent properties of the eigenmodes (see Supplemental Material~\cite{SupplementalMat}).
	
	\begin{figure*}[ht!]
		\includegraphics[width=0.99\textwidth]{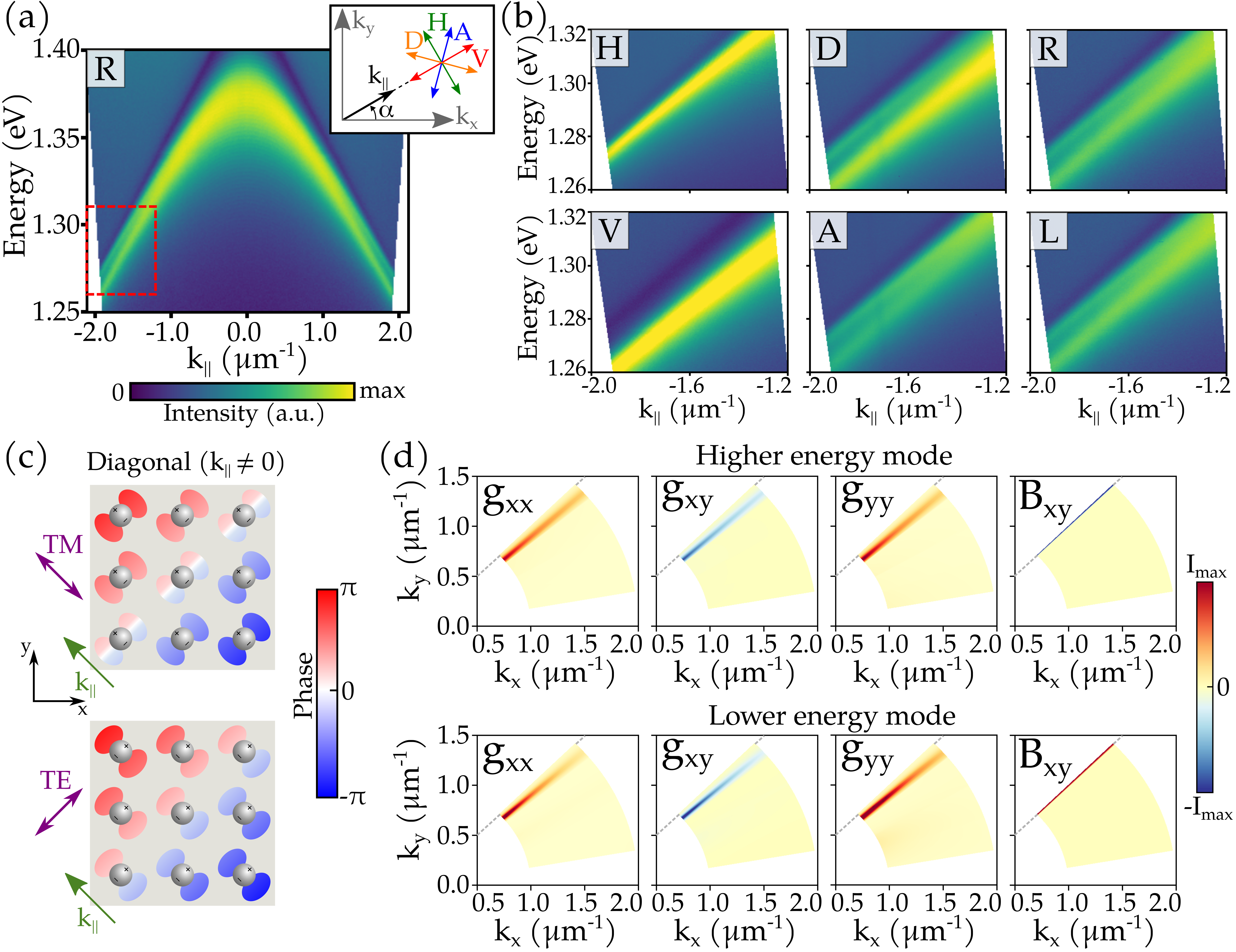}
		\caption{\label{fig:transm_exp_QGT} (a) Transmission measurements at the diagonal of the Brillouin zone ($\alpha=45\degree$) for right (R) circular polarization. (b) Measurements in the red dashed square of (a) with horizontal (H), vertical (V), diagonal (D), antidiagonal (A), right, and left (L) circular polarization filtering. (c) Explanation of TE-TM band splitting along one diagonal of the first Brillouin zone (with $\mathbf{k}_{||}\neq0$). The same phase distribution is created in real space for both TE and TM polarizations, but unlike in Fig.~\ref{fig:soc_ela_transm}(a), the radiation patterns are different at the individual nanoparticle level, giving rise to two non-degenerate bands. The modes that participate in the band splitting are only approximately TE or TM due to the non-Hermiticity, but we name them TE and TM for practical reasons. (d) Quantum metric and Berry curvature for the higher and lower energy modes in (a) and (b), for $10\degree<\alpha<45\degree$. The magnitude $I_{max}$ of the Berry curvature is $10^{-3}$ times smaller than the quantum metric components. The grey dashed line marks the diagonal of the Brillouin zone.
		}
	\vspace{-0.3cm}
	\end{figure*}

	Our experiments reveal a previously unnoticed band splitting at the diagonal of the Brillouin zone, which is not predicted by the empty lattice approximation in Fig.~\ref{fig:soc_ela_transm}(b). Fig.~\ref{fig:transm_exp_QGT}(a) shows the results for $k_{||}$ pointing along the trajectory $\Gamma-M$ (the diagonal), with filtering of right circularly polarized light. Interestingly, the higher (lower) energy band is TE (TM) polarized along the diagonal of the Brillouin zone, see  Fig.~\ref{fig:transm_exp_QGT}(b). The similar response to left and right circular polarization filtering indicates that the chiral symmetry is not lifted in our highly symmetric square lattice. The band splitting is also observed for diagonal and antidiagonal polarizations. These polarization properties are relevant for the quantum geometric phenomena analyzed below. 
	
	The origin of the band splitting of Fig.~\ref{fig:transm_exp_QGT}(b) is explained in Figs.~\ref{fig:soc_ela_transm}(a) and \ref{fig:transm_exp_QGT}(c). The field profile of the lattice modes is dominated by Bloch waves of the form $\Phi(\mathbf{r})\sim e^{i\varphi(\mathbf{r})}$ with a phase $\varphi(\mathbf{r})=\mathbf{k}\cdot\mathbf{r}$ that has contributions from both the diffraction orders and the in-plane momentum, i.e. $\mathbf{k}=\mathbf{k}_{||}+\mathbf{G}$, where $\mathbf{G}$ is the lattice vector~\cite{GuoPRB2017,QuantumPlasmonics}. At the $\Gamma$-point, the overall phase is zero, hence two degenerate modes with an in-phase radiation pattern, related by a $\pi/2$ rotation, appear (see Fig.~\ref{fig:soc_ela_transm}(a)) and TE-TM band splitting does not take place. Along the diagonal of the Brillouin zone, the phase is non-zero giving rise to a fixed phase distribution in real space. The mode polarization is given by the dipole orientation at the single particle level (see Fig.~\ref{fig:transm_exp_QGT}(c)), hence a given phase distribution $\varphi(\mathbf{r})$ in combination with two different polarizations (TE or TM with respect to $\mathbf{k}_{||}$) creates two fundamentally different modes and a TE-TM splitting. 
	A thorough analysis of the band splittings using a T-matrix approach is presented in Ref.~\cite{CuerdaPRB2023}. Below we provide a simplified model that relates this phenomenon with (pseudo)spin-orbit coupling.
	
	We extract the QGT within an angular region $10\degree<\alpha<45\degree$ of $k-$space (see Fig.~S4). Fig.~\ref{fig:transm_exp_QGT}(d) shows that non-zero components exist for both the higher and lower energy modes in Figs.~\ref{fig:transm_exp_QGT}(a),(b), with clear features around the diagonal of the Brillouin zone. We find positive values for the quantum metric components $g_{xx}$ and $g_{yy}$, with $g_{xx}\approx g_{yy}\approx-g_{xy}$.  
	Remarkably, we also find a non-zero Berry curvature, which is not expected in our square lattice geometry. The Berry curvature is much smaller than the quantum metric components. Using the symmetry of the square lattice (see Fig.~S1 for the properties of SLR bands, and further discussion in Ref.~\cite{CuerdaPRB2023}), we find that the quantum metric components are symmetric with respect to the diagonal of the Brillouin zone, and likewise the Berry curvature is antisymmetric. 
	
	
	We provide a simple two-band model (derived in the Supplemental Material~\cite{SupplementalMat}) to intuitively interpret the experimental QGT results. In our model, the two bands correspond to the two polarization directions of the plasmonic-photonic modes, and the dispersion of the SLR modes is encoded in a simplified form:
	\begin{equation}\label{2bandHam}
		\hat{H}=\epsilon(\mathbf{k})I_{2\times 2}+\mathbf{\Omega}(\mathbf{k})\cdot\boldsymbol{\sigma},
	\end{equation}
	where
	\begin{align}
		\epsilon(\mathbf{k})&=\dfrac{1}{2}(E_{-1,0}(\mathbf{k})+E_{0,-1}(\mathbf{k}))\label{epk},\\
		\Omega_{x}(\mathbf{k})&=\dfrac{1}{2k^2}\left(k_{x}^{2}-k_{y}^{2}\right)\left(E_{0,-1}(\mathbf{k})-E_{-1,0}(\mathbf{k})\right),\label{omega_x}\\
		\Omega_y(\mathbf{k})&=\dfrac{g}{2}\sqrt{k_x^2 + k_y^2}\label{omega_y}\textrm{,}
	\end{align} 
	and the energies $E_{-1,0}$ and $E_{0,-1}$ are defined in Eq.~\eqref{energy_ela}. In the most general non-Hermitian case, the Hamiltonian \eqref{2bandHam}-\eqref{omega_y} has two sets of eigenvectors, right $|R_{\pm}\rangle$ and left $|L_{\pm}\rangle$, for each of the complex-valued eigenenergies $E_{\pm}(\mathbf{k})$ that define the two bands. Complete expressions of these and the QGT are found in the Supplemental Material~\cite{SupplementalMat}. The term in Eq.~\eqref{omega_y} models the band splitting at the diagonal of the Brillouin zone, and $g=g'+ig''$ controls its size, both in the energy band structure and in the inherent losses of each mode. This minimal model is introduced to investigate degeneracy removal at the diagonal of the Brillouin zone, and it is limited to its vicinity. Nevertheless, Ref.~\cite{CuerdaPRB2023} shows by detailed T-matrix calculations that the heuristic two-band model introduced here gives the same qualitative behaviour as a microscopic description of the system.
	
\begin{figure}[t!]
		\includegraphics[width=0.99\columnwidth]{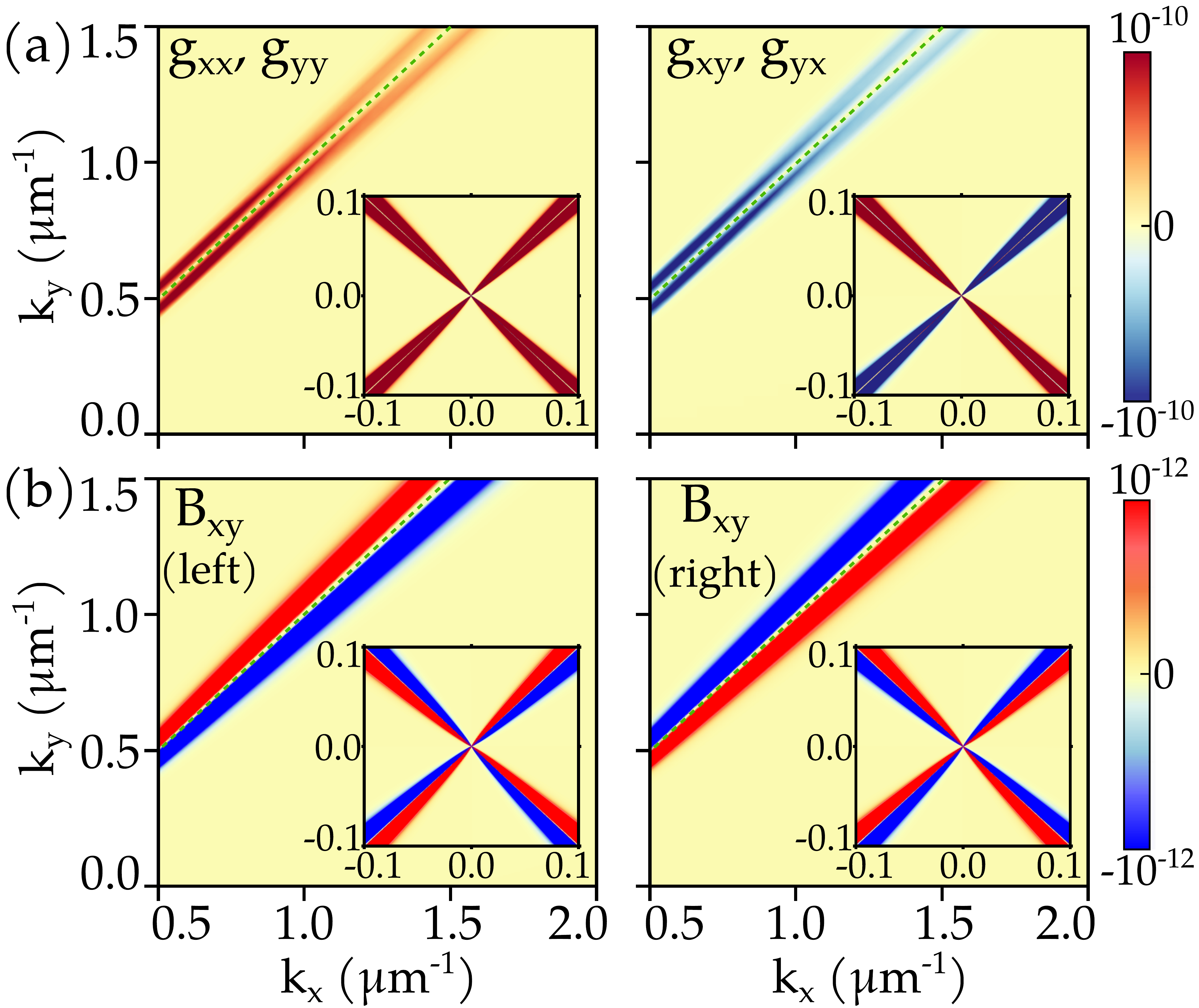}
		\caption{\label{fig:qgt_2band} QGT obtained with the two-band model. Main panels show the same area of the Brillouin zone as the experiments in Fig.~\ref{fig:transm_exp_QGT}(d), and the insets display neighboring regions to the $\Gamma$-point. The green dashed line marks the diagonal of the Brillouin zone. (a) Quantum metric components. We find that $g_{xx}=g_{yy}=-g_{xy}$ and  $g_{xy}=g_{yx}$. (b) Berry curvature calculated with the right and left eigenvectors of the Hamiltonian. We found that $\mathfrak{B}_{xy}^{n}=-\mathfrak{B}_{yx}^{n}$. Colorscale units are m$^2$.} 
	\end{figure}

	The quantum metric obtained with the two-band model (see Fig.~\ref{fig:qgt_2band}(a)) presents an excellent qualitative agreement with the experimental results. In both cases, the quantum metric components are non-zero around the diagonal of the Brillouin zone, and the sign of the components coincides with Fig.~\ref{fig:transm_exp_QGT}(d). The quantum metric depends essentially on the energy band splitting, and all components are zero for $g=0$. For a typical coupling constant that fulfills $g''\approx 10^{-2}g'$ (as in the simulations of Ref.~\cite{CuerdaPRB2023}), the quantum metric is the same as in the Hermitian limit ($g''=0$), with negligible non-Hermitian corrections. 
	
	The Berry curvature, in contrast, strongly depends on the losses, and both the right and left forms of the Berry curvature (see Supplemental Material~\cite{SupplementalMat}) become zero for $g''=0$. This, together with our analysis in Ref.~\cite{CuerdaPRB2023}, proves that the Berry curvature is of non-Hermitian nature. Fig.~\ref{fig:qgt_2band}(b) shows the left and right Berry curvature that exhibit an excellent qualitative agreement with the experimental results, with an antisymmetric distribution at each diagonal of the Brillouin zone. The difference in magnitude between the Berry curvature and the quantum metric components is $10^{-2}$ in the two-band model, differing from the experimental ratio. However, microscopic simulations show good agreement with the experiments~\cite{CuerdaPRB2023}.

	We remark that the Hamiltonian in Eq.~\eqref{2bandHam} has the form of a $\mathbf{k}$-dependent effective magnetic field $\mathbf{\Omega}$ coupled to a pseudospin, or a spin-orbit coupling Hamiltonian in an electronic system~\cite{KavokinPRL2005}. In our description, the polarization of light takes the role of a pseudospin, and the effective magnetic field is given by the TE/TM mode structure of a square plasmonic lattice and the band splitting at the diagonal of the Brillouin zone. The origin of the TE/TM splitting in our case is different from the one in microcavity polariton systems (see Supplemental Material~\cite{SupplementalMat}). Polarization-selective phenomena are observed in tight-binding systems~\cite{PoddubnyACSPhot2014}, but in our lattice the splitting along the diagonal arises due to long-range radiative interactions.
	
	The two-band model shows (see Supplemental Material~\cite{SupplementalMat}) that exactly at the diagonal ($\Omega_x =0$) the quantum metric and Berry curvature always vanish, as we also see in the T-matrix simulations and experiments. Around the diagonal (in a region where $\Omega_x \sim \Omega_y' , \Omega_y ''$), in the Hermitian case, a $k$-dependent rotation of the eigenstates in the horizontal/vertical polarization plane appears which leads to a finite quantum metric. With losses, a net LCP/RCP polarization may emerge (although not resolved by our experiments) and contribute to the Berry curvature. Intriguingly, Berry curvature can arise also due to the interplay of losses and the spin-orbit coupling, which is likely the main origin in our case. 


	The distribution of quantum metric and Berry curvature found here is unprecedented, following entire high-symmetry lines instead of accumulating around certain points in $k$-space. Whereas non-Hermitian quantum metric~\cite{LiaoPRL2021} has been recently reported, our results yield, to our knowledge, the pioneering observation of Berry curvature induced by losses in optical systems.
	
	Splittings of degenerate bands at high-symmetry points are often found in periodic systems when time-reversal symmetry is broken by (effective) magnetic field; in such cases the bandgap opening may give rise to separate bands with non-zero Chern numbers~\cite{RaghuPRA2008,HaldanePRL2008}. Here, we do not find a full bandgap in the whole Brillouin zone. Thus, time-reversal symmetry breaking by losses in our case causes non-trivial quantum geometry, but not topology. The simple two-band model we introduced can be easily adapted to other configurations, e.g.~different lattice geometries~\cite{WuPRL2015}, and will be powerful in designing topologically non-trivial structures.  
	
	
	Our work lays the foundation for fertile paths of future research in long-range coupled photonic lattice systems. We anticipate that further intriguing non-Hermitian phenomena are to be found around other (than $\Gamma$) high-symmetry points \cite{GuoPRL2019,JuarezACSPHot2022}, or with symmetry breaking e.g.~with multiparticle or chiral unit cells~\cite{HeilmannACSPhot2022,SalernoPRL2022,GoerlitzerAdvMat2020}. Magnetic nanoparticles may also break the time-reversal symmetry~\cite{KatajaNatComm2015,FreireNatPhot2022,MaccaferriAPL2023}. Finally, the above non-Hermitian effects were also found when having gain instead of losses: all these research lines may be combined with lasing or condensation phenomena. In analogy to fermionic systems~\cite{TormaNatRevPhys2022}, we expect the QGT to be relevant for bulk response in interacting, non-linear photonic and polaritonic systems.

	\begin{acknowledgments}
		We acknowledge useful discussions with Mikko Rosenberg and Marek Ne{\v{c}}ada. This work was supported
		by the Academy of Finland under Project No.~349313, Project No.~318937
		(PROFI), and the Academy of Finland Flagship Programme in
		Photonics Research and Innovation (PREIN) Project
		No. 320167, as well as by the Jane and Aatos Erkko Foundation and the Technology Industries of Finland Centennial Foundation as part of the Future Makers funding program. J.C.~acknowledges former support by the Academy of Finland under project No.~325608. J.M.T.~acknowledges financial support by the Magnus Ehrnrooth Foundation. Part of the research was performed at the OtaNano Nanofab cleanroom (Micronova Nanofabrication Centre), supported by Aalto University.
	\end{acknowledgments}

	
	%


\begin{thebibliography}{90}%
		\makeatletter
		\providecommand \@ifxundefined [1]{%
			\@ifx{#1\undefined}
		}%
		\providecommand \@ifnum [1]{%
			\ifnum #1\expandafter \@firstoftwo
			\else \expandafter \@secondoftwo
			\fi
		}%
		\providecommand \@ifx [1]{%
			\ifx #1\expandafter \@firstoftwo
			\else \expandafter \@secondoftwo
			\fi
		}%
		\providecommand \natexlab [1]{#1}%
		\providecommand \enquote  [1]{``#1''}%
		\providecommand \bibnamefont  [1]{#1}%
		\providecommand \bibfnamefont [1]{#1}%
		\providecommand \citenamefont [1]{#1}%
		\providecommand \href@noop [0]{\@secondoftwo}%
		\providecommand \href [0]{\begingroup \@sanitize@url \@href}%
		\providecommand \@href[1]{\@@startlink{#1}\@@href}%
		\providecommand \@@href[1]{\endgroup#1\@@endlink}%
		\providecommand \@sanitize@url [0]{\catcode `\\12\catcode `\$12\catcode
			`\&12\catcode `\#12\catcode `\^12\catcode `\_12\catcode `\%12\relax}%
		\providecommand \@@startlink[1]{}%
		\providecommand \@@endlink[0]{}%
		\providecommand \url  [0]{\begingroup\@sanitize@url \@url }%
		\providecommand \@url [1]{\endgroup\@href {#1}{\urlprefix }}%
		\providecommand \urlprefix  [0]{URL }%
		\providecommand \Eprint [0]{\href }%
		\providecommand \doibase [0]{https://doi.org/}%
		\providecommand \selectlanguage [0]{\@gobble}%
		\providecommand \bibinfo  [0]{\@secondoftwo}%
		\providecommand \bibfield  [0]{\@secondoftwo}%
		\providecommand \translation [1]{[#1]}%
		\providecommand \BibitemOpen [0]{}%
		\providecommand \bibitemStop [0]{}%
		\providecommand \bibitemNoStop [0]{.\EOS\space}%
		\providecommand \EOS [0]{\spacefactor3000\relax}%
		\providecommand \BibitemShut  [1]{\csname bibitem#1\endcsname}%
		\let\auto@bib@innerbib\@empty
		\bibitem [{\citenamefont {Klitzing}\ \emph {et~al.}(1980)\citenamefont
			{Klitzing}, \citenamefont {Dorda},\ and\ \citenamefont
			{Pepper}}]{KlitzingPRL1980}%
		\BibitemOpen
		\bibfield  {author} {\bibinfo {author} {\bibfnamefont {K.~v.}\ \bibnamefont
				{Klitzing}}, \bibinfo {author} {\bibfnamefont {G.}~\bibnamefont {Dorda}},\
			and\ \bibinfo {author} {\bibfnamefont {M.}~\bibnamefont {Pepper}},\
		}\bibfield  {title} {\bibinfo {title} {New method for high-accuracy
				determination of the fine-structure constant based on quantized {H}all
				resistance},\ }\href {https://doi.org/10.1103/PhysRevLett.45.494} {\bibfield
			{journal} {\bibinfo  {journal} {Phys. Rev. Lett.}\ }\textbf {\bibinfo
				{volume} {45}},\ \bibinfo {pages} {494} (\bibinfo {year} {1980})}\BibitemShut
		{NoStop}%
		\bibitem [{\citenamefont {Thouless}\ \emph {et~al.}(1982)\citenamefont
			{Thouless}, \citenamefont {Kohmoto}, \citenamefont {Nightingale},\ and\
			\citenamefont {den Nijs}}]{ThoulessPRL1982}%
		\BibitemOpen
		\bibfield  {author} {\bibinfo {author} {\bibfnamefont {D.~J.}\ \bibnamefont
				{Thouless}}, \bibinfo {author} {\bibfnamefont {M.}~\bibnamefont {Kohmoto}},
			\bibinfo {author} {\bibfnamefont {M.~P.}\ \bibnamefont {Nightingale}},\ and\
			\bibinfo {author} {\bibfnamefont {M.}~\bibnamefont {den Nijs}},\ }\bibfield
		{title} {\bibinfo {title} {Quantized {H}all conductance in a two-dimensional
				periodic potential},\ }\href {https://doi.org/10.1103/PhysRevLett.49.405}
		{\bibfield  {journal} {\bibinfo  {journal} {Phys. Rev. Lett.}\ }\textbf
			{\bibinfo {volume} {49}},\ \bibinfo {pages} {405} (\bibinfo {year}
			{1982})}\BibitemShut {NoStop}%
		\bibitem [{\citenamefont {Bernevig}\ and\ \citenamefont
			{Hughes}(2013)}]{BernevigSuperconductors}%
		\BibitemOpen
		\bibfield  {author} {\bibinfo {author} {\bibfnamefont {B.~A.}\ \bibnamefont
				{Bernevig}}\ and\ \bibinfo {author} {\bibfnamefont {T.~L.}\ \bibnamefont
				{Hughes}},\ }\href@noop {} {\emph {\bibinfo {title} {Topological {I}nsulators
					and {T}opological {S}uperconductors}}}\ (\bibinfo  {publisher} {Princeton
			University Press, Princeton, NJ},\ \bibinfo {year} {2013})\BibitemShut
		{NoStop}%
		\bibitem [{\citenamefont {Haldane}(1988)}]{HaldanePRL1988}%
		\BibitemOpen
		\bibfield  {author} {\bibinfo {author} {\bibfnamefont {F.~D.~M.}\
				\bibnamefont {Haldane}},\ }\bibfield  {title} {\bibinfo {title} {Model for a
				{Q}uantum {H}all {E}ffect without {L}andau {L}evels: {C}ondensed-{M}atter
				{R}ealization of the ``{P}arity {A}nomaly''},\ }\href
		{https://doi.org/10.1103/PhysRevLett.61.2015} {\bibfield  {journal} {\bibinfo
				{journal} {Phys. Rev. Lett.}\ }\textbf {\bibinfo {volume} {61}},\ \bibinfo
			{pages} {2015} (\bibinfo {year} {1988})}\BibitemShut {NoStop}%
		\bibitem [{\citenamefont {Kane}\ and\ \citenamefont
			{Mele}(2005)}]{Kane1PRL2005}%
		\BibitemOpen
		\bibfield  {author} {\bibinfo {author} {\bibfnamefont {C.~L.}\ \bibnamefont
				{Kane}}\ and\ \bibinfo {author} {\bibfnamefont {E.~J.}\ \bibnamefont
				{Mele}},\ }\bibfield  {title} {\bibinfo {title} {${Z}_{2}$ {T}opological
				{O}rder and the {Q}uantum {S}pin {H}all {E}ffect},\ }\href
		{https://doi.org/10.1103/PhysRevLett.95.146802} {\bibfield  {journal}
			{\bibinfo  {journal} {Phys. Rev. Lett.}\ }\textbf {\bibinfo {volume} {95}},\
			\bibinfo {pages} {146802} (\bibinfo {year} {2005})}\BibitemShut {NoStop}%
		\bibitem [{\citenamefont {Price}\ \emph {et~al.}(2022)\citenamefont {Price},
			\citenamefont {Chong}, \citenamefont {Khanikaev}, \citenamefont {Schomerus},
			\citenamefont {Maczewsky}, \citenamefont {Kremer}, \citenamefont {Heinrich},
			\citenamefont {Szameit}, \citenamefont {Zilberberg}, \citenamefont {Yang},
			\citenamefont {Zhang}, \citenamefont {Alù}, \citenamefont {Thomale},
			\citenamefont {Carusotto}, \citenamefont {St-Jean}, \citenamefont {Amo},
			\citenamefont {Dutt}, \citenamefont {Yuan}, \citenamefont {Fan},
			\citenamefont {Yin}, \citenamefont {Peng}, \citenamefont {Ozawa},\ and\
			\citenamefont {Blanco-Redondo}}]{PriceJPhysPhot2022}%
		\BibitemOpen
		\bibfield  {author} {\bibinfo {author} {\bibfnamefont {H.}~\bibnamefont
				{Price}}, \bibinfo {author} {\bibfnamefont {Y.}~\bibnamefont {Chong}},
			\bibinfo {author} {\bibfnamefont {A.}~\bibnamefont {Khanikaev}}, \bibinfo
			{author} {\bibfnamefont {H.}~\bibnamefont {Schomerus}}, \bibinfo {author}
			{\bibfnamefont {L.~J.}\ \bibnamefont {Maczewsky}}, \bibinfo {author}
			{\bibfnamefont {M.}~\bibnamefont {Kremer}}, \bibinfo {author} {\bibfnamefont
				{M.}~\bibnamefont {Heinrich}}, \bibinfo {author} {\bibfnamefont
				{A.}~\bibnamefont {Szameit}}, \bibinfo {author} {\bibfnamefont
				{O.}~\bibnamefont {Zilberberg}}, \bibinfo {author} {\bibfnamefont
				{Y.}~\bibnamefont {Yang}}, \bibinfo {author} {\bibfnamefont {B.}~\bibnamefont
				{Zhang}}, \bibinfo {author} {\bibfnamefont {A.}~\bibnamefont {Alù}},
			\bibinfo {author} {\bibfnamefont {R.}~\bibnamefont {Thomale}}, \bibinfo
			{author} {\bibfnamefont {I.}~\bibnamefont {Carusotto}}, \bibinfo {author}
			{\bibfnamefont {P.}~\bibnamefont {St-Jean}}, \bibinfo {author} {\bibfnamefont
				{A.}~\bibnamefont {Amo}}, \bibinfo {author} {\bibfnamefont {A.}~\bibnamefont
				{Dutt}}, \bibinfo {author} {\bibfnamefont {L.}~\bibnamefont {Yuan}}, \bibinfo
			{author} {\bibfnamefont {S.}~\bibnamefont {Fan}}, \bibinfo {author}
			{\bibfnamefont {X.}~\bibnamefont {Yin}}, \bibinfo {author} {\bibfnamefont
				{C.}~\bibnamefont {Peng}}, \bibinfo {author} {\bibfnamefont {T.}~\bibnamefont
				{Ozawa}},\ and\ \bibinfo {author} {\bibfnamefont {A.}~\bibnamefont
				{Blanco-Redondo}},\ }\bibfield  {title} {\bibinfo {title} {Roadmap on
				topological photonics},\ }\href {https://doi.org/10.1088/2515-7647/ac4ee4}
		{\bibfield  {journal} {\bibinfo  {journal} {J. Phys. Photonics}\ }\textbf
			{\bibinfo {volume} {4}},\ \bibinfo {pages} {032501} (\bibinfo {year}
			{2022})}\BibitemShut {NoStop}%
		\bibitem [{\citenamefont {Ota}\ \emph {et~al.}(2020)\citenamefont {Ota},
			\citenamefont {Takata}, \citenamefont {Ozawa}, \citenamefont {Amo},
			\citenamefont {Jia}, \citenamefont {Kante}, \citenamefont {Notomi},
			\citenamefont {Arakawa},\ and\ \citenamefont
			{Iwamoto}}]{OtaNanophotonics2020}%
		\BibitemOpen
		\bibfield  {author} {\bibinfo {author} {\bibfnamefont {Y.}~\bibnamefont
				{Ota}}, \bibinfo {author} {\bibfnamefont {K.}~\bibnamefont {Takata}},
			\bibinfo {author} {\bibfnamefont {T.}~\bibnamefont {Ozawa}}, \bibinfo
			{author} {\bibfnamefont {A.}~\bibnamefont {Amo}}, \bibinfo {author}
			{\bibfnamefont {Z.}~\bibnamefont {Jia}}, \bibinfo {author} {\bibfnamefont
				{B.}~\bibnamefont {Kante}}, \bibinfo {author} {\bibfnamefont
				{M.}~\bibnamefont {Notomi}}, \bibinfo {author} {\bibfnamefont
				{Y.}~\bibnamefont {Arakawa}},\ and\ \bibinfo {author} {\bibfnamefont
				{S.}~\bibnamefont {Iwamoto}},\ }\bibfield  {title} {\bibinfo {title} {Active
				topological photonics},\ }\href
		{https://doi.org/doi:10.1515/nanoph-2019-0376} {\bibfield  {journal}
			{\bibinfo  {journal} {Nanophotonics}\ }\textbf {\bibinfo {volume} {9}},\
			\bibinfo {pages} {547} (\bibinfo {year} {2020})}\BibitemShut {NoStop}%
		\bibitem [{\citenamefont {Ozawa}\ \emph {et~al.}(2019)\citenamefont {Ozawa},
			\citenamefont {Price}, \citenamefont {Amo}, \citenamefont {Goldman},
			\citenamefont {Hafezi}, \citenamefont {Lu}, \citenamefont {Rechtsman},
			\citenamefont {Schuster}, \citenamefont {Simon}, \citenamefont {Zilberberg},\
			and\ \citenamefont {Carusotto}}]{OzawaRevModPhys2019}%
		\BibitemOpen
		\bibfield  {author} {\bibinfo {author} {\bibfnamefont {T.}~\bibnamefont
				{Ozawa}}, \bibinfo {author} {\bibfnamefont {H.~M.}\ \bibnamefont {Price}},
			\bibinfo {author} {\bibfnamefont {A.}~\bibnamefont {Amo}}, \bibinfo {author}
			{\bibfnamefont {N.}~\bibnamefont {Goldman}}, \bibinfo {author} {\bibfnamefont
				{M.}~\bibnamefont {Hafezi}}, \bibinfo {author} {\bibfnamefont
				{L.}~\bibnamefont {Lu}}, \bibinfo {author} {\bibfnamefont {M.~C.}\
				\bibnamefont {Rechtsman}}, \bibinfo {author} {\bibfnamefont {D.}~\bibnamefont
				{Schuster}}, \bibinfo {author} {\bibfnamefont {J.}~\bibnamefont {Simon}},
			\bibinfo {author} {\bibfnamefont {O.}~\bibnamefont {Zilberberg}},\ and\
			\bibinfo {author} {\bibfnamefont {I.}~\bibnamefont {Carusotto}},\ }\bibfield
		{title} {\bibinfo {title} {Topological photonics},\ }\href
		{https://doi.org/10.1103/RevModPhys.91.015006} {\bibfield  {journal}
			{\bibinfo  {journal} {Rev. Mod. Phys.}\ }\textbf {\bibinfo {volume} {91}},\
			\bibinfo {pages} {015006} (\bibinfo {year} {2019})}\BibitemShut {NoStop}%
		\bibitem [{\citenamefont {Khanikaev}\ and\ \citenamefont
			{Shvets}(2017)}]{KhanikaevNatPhot2017}%
		\BibitemOpen
		\bibfield  {author} {\bibinfo {author} {\bibfnamefont {A.}~\bibnamefont
				{Khanikaev}}\ and\ \bibinfo {author} {\bibfnamefont {G.}~\bibnamefont
				{Shvets}},\ }\bibfield  {title} {\bibinfo {title} {Two-dimensional
				topological photonics},\ }\href {https://doi.org/10.1038/s41566-017-0048-5}
		{\bibfield  {journal} {\bibinfo  {journal} {Nat. Photon.}\ }\textbf {\bibinfo
				{volume} {11}},\ \bibinfo {pages} {763–773} (\bibinfo {year}
			{2017})}\BibitemShut {NoStop}%
		\bibitem [{\citenamefont {Sun}\ \emph {et~al.}(2017)\citenamefont {Sun},
			\citenamefont {He}, \citenamefont {Liu}, \citenamefont {Lu}, \citenamefont
			{Zhu},\ and\ \citenamefont {Chen}}]{SunProgQuantElec2017}%
		\BibitemOpen
		\bibfield  {author} {\bibinfo {author} {\bibfnamefont {X.-C.}\ \bibnamefont
				{Sun}}, \bibinfo {author} {\bibfnamefont {C.}~\bibnamefont {He}}, \bibinfo
			{author} {\bibfnamefont {X.-P.}\ \bibnamefont {Liu}}, \bibinfo {author}
			{\bibfnamefont {M.-H.}\ \bibnamefont {Lu}}, \bibinfo {author} {\bibfnamefont
				{S.-N.}\ \bibnamefont {Zhu}},\ and\ \bibinfo {author} {\bibfnamefont {Y.-F.}\
				\bibnamefont {Chen}},\ }\bibfield  {title} {\bibinfo {title} {Two-dimensional
				topological photonic systems},\ }\href
		{https://doi.org/https://doi.org/10.1016/j.pquantelec.2017.07.004} {\bibfield
			{journal} {\bibinfo  {journal} {Prog. Quantum. Electron.}\ }\textbf
			{\bibinfo {volume} {55}},\ \bibinfo {pages} {52} (\bibinfo {year}
			{2017})}\BibitemShut {NoStop}%
		\bibitem [{\citenamefont {Lu}\ \emph {et~al.}(2014)\citenamefont {Lu},
			\citenamefont {Joannopoulos},\ and\ \citenamefont
			{Soljačić}}]{LuNatPhot2014}%
		\BibitemOpen
		\bibfield  {author} {\bibinfo {author} {\bibfnamefont {L.}~\bibnamefont
				{Lu}}, \bibinfo {author} {\bibfnamefont {J.}~\bibnamefont {Joannopoulos}},\
			and\ \bibinfo {author} {\bibfnamefont {M.}~\bibnamefont {Soljačić}},\
		}\bibfield  {title} {\bibinfo {title} {Topological photonics},\ }\href
		{https://doi.org/10.1038/nphoton.2014.248} {\bibfield  {journal} {\bibinfo
				{journal} {Nat. Photon.}\ }\textbf {\bibinfo {volume} {8}},\ \bibinfo {pages}
			{821–829} (\bibinfo {year} {2014})}\BibitemShut {NoStop}%
		\bibitem [{\citenamefont {Xu}\ \emph {et~al.}(2017)\citenamefont {Xu},
			\citenamefont {Wang},\ and\ \citenamefont {Duan}}]{XuPRL2017}%
		\BibitemOpen
		\bibfield  {author} {\bibinfo {author} {\bibfnamefont {Y.}~\bibnamefont
				{Xu}}, \bibinfo {author} {\bibfnamefont {S.-T.}\ \bibnamefont {Wang}},\ and\
			\bibinfo {author} {\bibfnamefont {L.-M.}\ \bibnamefont {Duan}},\ }\bibfield
		{title} {\bibinfo {title} {Weyl exceptional rings in a three-dimensional
				dissipative cold atomic gas},\ }\href
		{https://doi.org/10.1103/PhysRevLett.118.045701} {\bibfield  {journal}
			{\bibinfo  {journal} {Phys. Rev. Lett.}\ }\textbf {\bibinfo {volume} {118}},\
			\bibinfo {pages} {045701} (\bibinfo {year} {2017})}\BibitemShut {NoStop}%
		\bibitem [{\citenamefont {Smirnova}\ \emph {et~al.}(2020)\citenamefont
			{Smirnova}, \citenamefont {Leykam}, \citenamefont {Chong},\ and\
			\citenamefont {Kivshar}}]{SmirnovaAPR2020}%
		\BibitemOpen
		\bibfield  {author} {\bibinfo {author} {\bibfnamefont {D.}~\bibnamefont
				{Smirnova}}, \bibinfo {author} {\bibfnamefont {D.}~\bibnamefont {Leykam}},
			\bibinfo {author} {\bibfnamefont {Y.}~\bibnamefont {Chong}},\ and\ \bibinfo
			{author} {\bibfnamefont {Y.}~\bibnamefont {Kivshar}},\ }\bibfield  {title}
		{\bibinfo {title} {Nonlinear topological photonics},\ }\href
		{https://doi.org/10.1063/1.5142397} {\bibfield  {journal} {\bibinfo
				{journal} {Appl. Phys. Rev.}\ }\textbf {\bibinfo {volume} {7}},\ \bibinfo
			{pages} {021306} (\bibinfo {year} {2020})}\BibitemShut {NoStop}%
		\bibitem [{\citenamefont {Provost}\ and\ \citenamefont
			{Vallee}(1980)}]{Provost1980}%
		\BibitemOpen
		\bibfield  {author} {\bibinfo {author} {\bibfnamefont {J.~P.}\ \bibnamefont
				{Provost}}\ and\ \bibinfo {author} {\bibfnamefont {G.}~\bibnamefont
				{Vallee}},\ }\bibfield  {title} {\bibinfo {title} {Riemannian structure on
				manifolds of quantum states},\ }\href
		{https://doi.org/https://doi.org/10.1007/BF02193559} {\bibfield  {journal}
			{\bibinfo  {journal} {Commun. Math. Phys.}\ }\textbf {\bibinfo {volume}
				{76}},\ \bibinfo {pages} {289–301} (\bibinfo {year} {1980})}\BibitemShut
		{NoStop}%
		\bibitem [{\citenamefont {Peotta}\ and\ \citenamefont
			{T{\"o}rm{\"a}}(2015)}]{PeottaNatComm2015}%
		\BibitemOpen
		\bibfield  {author} {\bibinfo {author} {\bibfnamefont {S.}~\bibnamefont
				{Peotta}}\ and\ \bibinfo {author} {\bibfnamefont {P.}~\bibnamefont
				{T{\"o}rm{\"a}}},\ }\bibfield  {title} {\bibinfo {title} {Superfluidity in
				topologically nontrivial flat bands},\ }\href
		{https://doi.org/10.1038/ncomms9944} {\bibfield  {journal} {\bibinfo
				{journal} {Nat. Commun.}\ }\textbf {\bibinfo {volume} {6}},\ \bibinfo {pages}
			{8944} (\bibinfo {year} {2015})}\BibitemShut {NoStop}%
		\bibitem [{\citenamefont {Liang}\ \emph {et~al.}(2017)\citenamefont {Liang},
			\citenamefont {Vanhala}, \citenamefont {Peotta}, \citenamefont {Siro},
			\citenamefont {Harju},\ and\ \citenamefont {T\"orm\"a}}]{LiangPRB2017}%
		\BibitemOpen
		\bibfield  {author} {\bibinfo {author} {\bibfnamefont {L.}~\bibnamefont
				{Liang}}, \bibinfo {author} {\bibfnamefont {T.~I.}\ \bibnamefont {Vanhala}},
			\bibinfo {author} {\bibfnamefont {S.}~\bibnamefont {Peotta}}, \bibinfo
			{author} {\bibfnamefont {T.}~\bibnamefont {Siro}}, \bibinfo {author}
			{\bibfnamefont {A.}~\bibnamefont {Harju}},\ and\ \bibinfo {author}
			{\bibfnamefont {P.}~\bibnamefont {T\"orm\"a}},\ }\bibfield  {title} {\bibinfo
			{title} {Band geometry, {B}erry curvature, and superfluid weight},\ }\href
		{https://doi.org/10.1103/PhysRevB.95.024515} {\bibfield  {journal} {\bibinfo
				{journal} {Phys. Rev. B}\ }\textbf {\bibinfo {volume} {95}},\ \bibinfo
			{pages} {024515} (\bibinfo {year} {2017})}\BibitemShut {NoStop}%
		\bibitem [{\citenamefont {Huhtinen}\ \emph {et~al.}(2022)\citenamefont
			{Huhtinen}, \citenamefont {Herzog-Arbeitman}, \citenamefont {Chew},
			\citenamefont {Bernevig},\ and\ \citenamefont {T\"orm\"a}}]{HuhtinenPRB2022}%
		\BibitemOpen
		\bibfield  {author} {\bibinfo {author} {\bibfnamefont {K.-E.}\ \bibnamefont
				{Huhtinen}}, \bibinfo {author} {\bibfnamefont {J.}~\bibnamefont
				{Herzog-Arbeitman}}, \bibinfo {author} {\bibfnamefont {A.}~\bibnamefont
				{Chew}}, \bibinfo {author} {\bibfnamefont {B.~A.}\ \bibnamefont {Bernevig}},\
			and\ \bibinfo {author} {\bibfnamefont {P.}~\bibnamefont {T\"orm\"a}},\
		}\bibfield  {title} {\bibinfo {title} {Revisiting flat band
				superconductivity: Dependence on minimal quantum metric and band touchings},\
		}\href {https://doi.org/10.1103/PhysRevB.106.014518} {\bibfield  {journal}
			{\bibinfo  {journal} {Phys. Rev. B}\ }\textbf {\bibinfo {volume} {106}},\
			\bibinfo {pages} {014518} (\bibinfo {year} {2022})}\BibitemShut {NoStop}%
		\bibitem [{\citenamefont {T{\"o}rm{\"a}}\ \emph {et~al.}(2022)\citenamefont
			{T{\"o}rm{\"a}}, \citenamefont {Peotta},\ and\ \citenamefont
			{Bernevig}}]{TormaNatRevPhys2022}%
		\BibitemOpen
		\bibfield  {author} {\bibinfo {author} {\bibfnamefont {P.}~\bibnamefont
				{T{\"o}rm{\"a}}}, \bibinfo {author} {\bibfnamefont {S.}~\bibnamefont
				{Peotta}},\ and\ \bibinfo {author} {\bibfnamefont {B.~A.}\ \bibnamefont
				{Bernevig}},\ }\bibfield  {title} {\bibinfo {title} {Superconductivity,
				superfluidity and quantum geometry in twisted multilayer systems},\ }\href
		{https://doi.org/10.1038/s42254-022-00466-y} {\bibfield  {journal} {\bibinfo
				{journal} {Nat. Rev. Phys.}\ }\textbf {\bibinfo {volume} {4}},\ \bibinfo
			{pages} {528–542} (\bibinfo {year} {2022})}\BibitemShut {NoStop}%
		\bibitem [{\citenamefont {Pi\'echon}\ \emph {et~al.}(2016)\citenamefont
			{Pi\'echon}, \citenamefont {Raoux}, \citenamefont {Fuchs},\ and\
			\citenamefont {Montambaux}}]{PiechonPRB2016}%
		\BibitemOpen
		\bibfield  {author} {\bibinfo {author} {\bibfnamefont {F.}~\bibnamefont
				{Pi\'echon}}, \bibinfo {author} {\bibfnamefont {A.}~\bibnamefont {Raoux}},
			\bibinfo {author} {\bibfnamefont {J.-N.}\ \bibnamefont {Fuchs}},\ and\
			\bibinfo {author} {\bibfnamefont {G.}~\bibnamefont {Montambaux}},\ }\bibfield
		{title} {\bibinfo {title} {Geometric orbital susceptibility: Quantum metric
				without {B}erry curvature},\ }\href
		{https://doi.org/10.1103/PhysRevB.94.134423} {\bibfield  {journal} {\bibinfo
				{journal} {Phys. Rev. B}\ }\textbf {\bibinfo {volume} {94}},\ \bibinfo
			{pages} {134423} (\bibinfo {year} {2016})}\BibitemShut {NoStop}%
		\bibitem [{\citenamefont {Braun}\ \emph {et~al.}(2018)\citenamefont {Braun},
			\citenamefont {Adesso}, \citenamefont {Benatti}, \citenamefont {Floreanini},
			\citenamefont {Marzolino}, \citenamefont {Mitchell},\ and\ \citenamefont
			{Pirandola}}]{BraunRevModPhys2018}%
		\BibitemOpen
		\bibfield  {author} {\bibinfo {author} {\bibfnamefont {D.}~\bibnamefont
				{Braun}}, \bibinfo {author} {\bibfnamefont {G.}~\bibnamefont {Adesso}},
			\bibinfo {author} {\bibfnamefont {F.}~\bibnamefont {Benatti}}, \bibinfo
			{author} {\bibfnamefont {R.}~\bibnamefont {Floreanini}}, \bibinfo {author}
			{\bibfnamefont {U.}~\bibnamefont {Marzolino}}, \bibinfo {author}
			{\bibfnamefont {M.~W.}\ \bibnamefont {Mitchell}},\ and\ \bibinfo {author}
			{\bibfnamefont {S.}~\bibnamefont {Pirandola}},\ }\bibfield  {title} {\bibinfo
			{title} {Quantum-enhanced measurements without entanglement},\ }\href
		{https://doi.org/10.1103/RevModPhys.90.035006} {\bibfield  {journal}
			{\bibinfo  {journal} {Rev. Mod. Phys.}\ }\textbf {\bibinfo {volume} {90}},\
			\bibinfo {pages} {035006} (\bibinfo {year} {2018})}\BibitemShut {NoStop}%
		\bibitem [{\citenamefont {Zheng}\ \emph {et~al.}(2022)\citenamefont {Zheng},
			\citenamefont {Xu}, \citenamefont {Ma}, \citenamefont {Li}, \citenamefont
			{Dong}, \citenamefont {Zhang}, \citenamefont {Wang}, \citenamefont {Sun},
			\citenamefont {Wu}, \citenamefont {Zhao}, \citenamefont {Li}, \citenamefont
			{Lan}, \citenamefont {Tan},\ and\ \citenamefont
			{Yu}}]{ZhengChinPhysLett2022}%
		\BibitemOpen
		\bibfield  {author} {\bibinfo {author} {\bibfnamefont {W.}~\bibnamefont
				{Zheng}}, \bibinfo {author} {\bibfnamefont {J.}~\bibnamefont {Xu}}, \bibinfo
			{author} {\bibfnamefont {Z.}~\bibnamefont {Ma}}, \bibinfo {author}
			{\bibfnamefont {Y.}~\bibnamefont {Li}}, \bibinfo {author} {\bibfnamefont
				{Y.}~\bibnamefont {Dong}}, \bibinfo {author} {\bibfnamefont {Y.}~\bibnamefont
				{Zhang}}, \bibinfo {author} {\bibfnamefont {X.}~\bibnamefont {Wang}},
			\bibinfo {author} {\bibfnamefont {G.}~\bibnamefont {Sun}}, \bibinfo {author}
			{\bibfnamefont {P.}~\bibnamefont {Wu}}, \bibinfo {author} {\bibfnamefont
				{J.}~\bibnamefont {Zhao}}, \bibinfo {author} {\bibfnamefont {S.}~\bibnamefont
				{Li}}, \bibinfo {author} {\bibfnamefont {D.}~\bibnamefont {Lan}}, \bibinfo
			{author} {\bibfnamefont {X.}~\bibnamefont {Tan}},\ and\ \bibinfo {author}
			{\bibfnamefont {Y.}~\bibnamefont {Yu}},\ }\bibfield  {title} {\bibinfo
			{title} {Measuring quantum geometric tensor of non-abelian system in
				superconducting circuits},\ }\href
		{https://doi.org/10.1088/0256-307X/39/10/100202} {\bibfield  {journal}
			{\bibinfo  {journal} {Chin. Phys. Lett.}\ }\textbf {\bibinfo {volume} {39}},\
			\bibinfo {pages} {100202} (\bibinfo {year} {2022})}\BibitemShut {NoStop}%
		\bibitem [{\citenamefont {Yu}\ \emph {et~al.}(2019)\citenamefont {Yu},
			\citenamefont {Yang}, \citenamefont {Gong}, \citenamefont {Cao},
			\citenamefont {Lu}, \citenamefont {Liu}, \citenamefont {Zhang}, \citenamefont
			{Plenio}, \citenamefont {Jelezko}, \citenamefont {Ozawa}, \citenamefont
			{Goldman},\ and\ \citenamefont {Cai}}]{YuNatSciRev2019}%
		\BibitemOpen
		\bibfield  {author} {\bibinfo {author} {\bibfnamefont {M.}~\bibnamefont
				{Yu}}, \bibinfo {author} {\bibfnamefont {P.}~\bibnamefont {Yang}}, \bibinfo
			{author} {\bibfnamefont {M.}~\bibnamefont {Gong}}, \bibinfo {author}
			{\bibfnamefont {Q.}~\bibnamefont {Cao}}, \bibinfo {author} {\bibfnamefont
				{Q.}~\bibnamefont {Lu}}, \bibinfo {author} {\bibfnamefont {H.}~\bibnamefont
				{Liu}}, \bibinfo {author} {\bibfnamefont {S.}~\bibnamefont {Zhang}}, \bibinfo
			{author} {\bibfnamefont {M.~B.}\ \bibnamefont {Plenio}}, \bibinfo {author}
			{\bibfnamefont {F.}~\bibnamefont {Jelezko}}, \bibinfo {author} {\bibfnamefont
				{T.}~\bibnamefont {Ozawa}}, \bibinfo {author} {\bibfnamefont
				{N.}~\bibnamefont {Goldman}},\ and\ \bibinfo {author} {\bibfnamefont
				{J.}~\bibnamefont {Cai}},\ }\bibfield  {title} {\bibinfo {title}
			{Experimental measurement of the quantum geometric tensor using coupled
				qubits in diamond},\ }\href {https://doi.org/10.1093/nsr/nwz193} {\bibfield
			{journal} {\bibinfo  {journal} {Nat. Sci. Rev.}\ }\textbf {\bibinfo {volume}
				{7}},\ \bibinfo {pages} {254–260} (\bibinfo {year} {2019})}\BibitemShut
		{NoStop}%
		\bibitem [{\citenamefont {Yi}\ \emph {et~al.}(2023)\citenamefont {Yi},
			\citenamefont {Yu}, \citenamefont {Yuan}, \citenamefont {Jiao}, \citenamefont
			{Yang}, \citenamefont {Jiang}, \citenamefont {Zhang}, \citenamefont {Chen},\
			and\ \citenamefont {Pan}}]{YiArxiv2023}%
		\BibitemOpen
		\bibfield  {author} {\bibinfo {author} {\bibfnamefont {C.-R.}\ \bibnamefont
				{Yi}}, \bibinfo {author} {\bibfnamefont {J.}~\bibnamefont {Yu}}, \bibinfo
			{author} {\bibfnamefont {H.}~\bibnamefont {Yuan}}, \bibinfo {author}
			{\bibfnamefont {R.-H.}\ \bibnamefont {Jiao}}, \bibinfo {author}
			{\bibfnamefont {Y.-M.}\ \bibnamefont {Yang}}, \bibinfo {author}
			{\bibfnamefont {X.}~\bibnamefont {Jiang}}, \bibinfo {author} {\bibfnamefont
				{J.-Y.}\ \bibnamefont {Zhang}}, \bibinfo {author} {\bibfnamefont
				{S.}~\bibnamefont {Chen}},\ and\ \bibinfo {author} {\bibfnamefont {J.-W.}\
				\bibnamefont {Pan}},\ }\href@noop {} {\bibinfo {title} {Extracting the
				quantum geometric tensor of an optical {R}aman lattice by {B}loch state
				tomography}} (\bibinfo {year} {2023}),\ \Eprint
		{https://arxiv.org/abs/2301.06090} {arXiv:2301.06090 [cond-mat.quant-gas]}
		\BibitemShut {NoStop}%
		\bibitem [{\citenamefont {Gianfrate}\ \emph {et~al.}(2020)\citenamefont
			{Gianfrate}, \citenamefont {Bleu}, \citenamefont {Dominici}, \citenamefont
			{Ardizzone}, \citenamefont {de~Giorgi}, \citenamefont {Ballarini},
			\citenamefont {Lerario}, \citenamefont {West}, \citenamefont {Pfeiffer},
			\citenamefont {Solnyshkov}, \citenamefont {Sanvitto},\ and\ \citenamefont
			{Malpuech}}]{GianfrateNature2020}%
		\BibitemOpen
		\bibfield  {author} {\bibinfo {author} {\bibfnamefont {A.}~\bibnamefont
				{Gianfrate}}, \bibinfo {author} {\bibfnamefont {O.}~\bibnamefont {Bleu}},
			\bibinfo {author} {\bibfnamefont {L.}~\bibnamefont {Dominici}}, \bibinfo
			{author} {\bibfnamefont {V.}~\bibnamefont {Ardizzone}}, \bibinfo {author}
			{\bibfnamefont {M.}~\bibnamefont {de~Giorgi}}, \bibinfo {author}
			{\bibfnamefont {D.}~\bibnamefont {Ballarini}}, \bibinfo {author}
			{\bibfnamefont {G.}~\bibnamefont {Lerario}}, \bibinfo {author} {\bibfnamefont
				{K.~W.}\ \bibnamefont {West}}, \bibinfo {author} {\bibfnamefont {L.~N.}\
				\bibnamefont {Pfeiffer}}, \bibinfo {author} {\bibfnamefont {D.~D.}\
				\bibnamefont {Solnyshkov}}, \bibinfo {author} {\bibfnamefont
				{D.}~\bibnamefont {Sanvitto}},\ and\ \bibinfo {author} {\bibfnamefont
				{G.}~\bibnamefont {Malpuech}},\ }\bibfield  {title} {\bibinfo {title}
			{Measurement of the quantum geometric tensor and of the anomalous {H}all
				drift},\ }\href {https://doi.org/10.1038/s41586-020-1989-2} {\bibfield
			{journal} {\bibinfo  {journal} {Nature}\ }\textbf {\bibinfo {volume} {578}},\
			\bibinfo {pages} {381–385} (\bibinfo {year} {2020})}\BibitemShut {NoStop}%
		\bibitem [{\citenamefont {Ren}\ \emph {et~al.}(2021)\citenamefont {Ren},
			\citenamefont {Liao}, \citenamefont {Li}, \citenamefont {Li}, \citenamefont
			{Bleu}, \citenamefont {Malpuech}, \citenamefont {Yao}, \citenamefont {Fu},\
			and\ \citenamefont {Solnyshkov}}]{RenNatComm2021}%
		\BibitemOpen
		\bibfield  {author} {\bibinfo {author} {\bibfnamefont {J.}~\bibnamefont
				{Ren}}, \bibinfo {author} {\bibfnamefont {Q.}~\bibnamefont {Liao}}, \bibinfo
			{author} {\bibfnamefont {F.}~\bibnamefont {Li}}, \bibinfo {author}
			{\bibfnamefont {Y.}~\bibnamefont {Li}}, \bibinfo {author} {\bibfnamefont
				{O.}~\bibnamefont {Bleu}}, \bibinfo {author} {\bibfnamefont {G.}~\bibnamefont
				{Malpuech}}, \bibinfo {author} {\bibfnamefont {J.}~\bibnamefont {Yao}},
			\bibinfo {author} {\bibfnamefont {H.}~\bibnamefont {Fu}},\ and\ \bibinfo
			{author} {\bibfnamefont {D.}~\bibnamefont {Solnyshkov}},\ }\bibfield  {title}
		{\bibinfo {title} {Nontrivial band geometry in an optically active system},\
		}\href {https://doi.org/10.1038/s41467-020-20845-2} {\bibfield  {journal}
			{\bibinfo  {journal} {Nat. Commun.}\ }\textbf {\bibinfo {volume} {12}},\
			\bibinfo {pages} {689} (\bibinfo {year} {2021})}\BibitemShut {NoStop}%
		\bibitem [{\citenamefont {El-Ganainy}\ \emph {et~al.}(2018)\citenamefont
			{El-Ganainy}, \citenamefont {Makris}, \citenamefont {Khajavikhan},
			\citenamefont {Musslimani}, \citenamefont {Rotter},\ and\ \citenamefont
			{Christodoulides}}]{ElGanainyNatPhys2018}%
		\BibitemOpen
		\bibfield  {author} {\bibinfo {author} {\bibfnamefont {R.}~\bibnamefont
				{El-Ganainy}}, \bibinfo {author} {\bibfnamefont {K.~G.}\ \bibnamefont
				{Makris}}, \bibinfo {author} {\bibfnamefont {M.}~\bibnamefont {Khajavikhan}},
			\bibinfo {author} {\bibfnamefont {Z.~H.}\ \bibnamefont {Musslimani}},
			\bibinfo {author} {\bibfnamefont {S.}~\bibnamefont {Rotter}},\ and\ \bibinfo
			{author} {\bibfnamefont {D.~N.}\ \bibnamefont {Christodoulides}},\ }\bibfield
		{title} {\bibinfo {title} {Non-{H}ermitian physics and {PT} symmetry},\
		}\href {https://doi.org/10.1038/nphys4323} {\bibfield  {journal} {\bibinfo
				{journal} {Nat. Phys.}\ }\textbf {\bibinfo {volume} {14}},\ \bibinfo {pages}
			{11} (\bibinfo {year} {2018})}\BibitemShut {NoStop}%
		\bibitem [{\citenamefont {Zhao}\ \emph {et~al.}(2019)\citenamefont {Zhao},
			\citenamefont {Qiao}, \citenamefont {Wu}, \citenamefont {Midya},
			\citenamefont {Longhi},\ and\ \citenamefont {Feng}}]{ZhaoScience2019}%
		\BibitemOpen
		\bibfield  {author} {\bibinfo {author} {\bibfnamefont {H.}~\bibnamefont
				{Zhao}}, \bibinfo {author} {\bibfnamefont {X.}~\bibnamefont {Qiao}}, \bibinfo
			{author} {\bibfnamefont {T.}~\bibnamefont {Wu}}, \bibinfo {author}
			{\bibfnamefont {B.}~\bibnamefont {Midya}}, \bibinfo {author} {\bibfnamefont
				{S.}~\bibnamefont {Longhi}},\ and\ \bibinfo {author} {\bibfnamefont
				{L.}~\bibnamefont {Feng}},\ }\bibfield  {title} {\bibinfo {title}
			{Non-{H}ermitian topological light steering},\ }\href
		{https://doi.org/10.1126/science.aay1064} {\bibfield  {journal} {\bibinfo
				{journal} {Science}\ }\textbf {\bibinfo {volume} {365}},\ \bibinfo {pages}
			{1163} (\bibinfo {year} {2019})}\BibitemShut {NoStop}%
		\bibitem [{\citenamefont {Moiseyev}(2011)}]{Moiseyev2011}%
		\BibitemOpen
		\bibfield  {author} {\bibinfo {author} {\bibfnamefont {N.}~\bibnamefont
				{Moiseyev}},\ }\href {https://doi.org/10.1017/CBO9780511976186} {\emph
			{\bibinfo {title} {Non-{H}ermitian Quantum Mechanics}}}\ (\bibinfo
		{publisher} {Cambridge University Press},\ \bibinfo {year}
		{2011})\BibitemShut {NoStop}%
		\bibitem [{\citenamefont {Bergholtz}\ \emph {et~al.}(2021)\citenamefont
			{Bergholtz}, \citenamefont {Budich},\ and\ \citenamefont
			{Kunst}}]{BergholtzRevModPhys2021}%
		\BibitemOpen
		\bibfield  {author} {\bibinfo {author} {\bibfnamefont {E.~J.}\ \bibnamefont
				{Bergholtz}}, \bibinfo {author} {\bibfnamefont {J.~C.}\ \bibnamefont
				{Budich}},\ and\ \bibinfo {author} {\bibfnamefont {F.~K.}\ \bibnamefont
				{Kunst}},\ }\bibfield  {title} {\bibinfo {title} {Exceptional topology of
				non-{H}ermitian systems},\ }\href
		{https://doi.org/10.1103/RevModPhys.93.015005} {\bibfield  {journal}
			{\bibinfo  {journal} {Rev. Mod. Phys.}\ }\textbf {\bibinfo {volume} {93}},\
			\bibinfo {pages} {015005} (\bibinfo {year} {2021})}\BibitemShut {NoStop}%
		\bibitem [{\citenamefont {Ding}\ \emph {et~al.}(2022)\citenamefont {Ding},
			\citenamefont {Fang},\ and\ \citenamefont {Ma}}]{DingNatRevPhys2022}%
		\BibitemOpen
		\bibfield  {author} {\bibinfo {author} {\bibfnamefont {K.}~\bibnamefont
				{Ding}}, \bibinfo {author} {\bibfnamefont {C.}~\bibnamefont {Fang}},\ and\
			\bibinfo {author} {\bibfnamefont {G.}~\bibnamefont {Ma}},\ }\bibfield
		{title} {\bibinfo {title} {Non-{H}ermitian topology and exceptional-point
				geometries},\ }\href {https://doi.org/10.1038/s42254-022-00516-5} {\bibfield
			{journal} {\bibinfo  {journal} {Nat. Rev. Phys.}\ }\textbf {\bibinfo {volume}
				{4}},\ \bibinfo {pages} {745–760} (\bibinfo {year} {2022})}\BibitemShut
		{NoStop}%
		\bibitem [{\citenamefont {Okuma}\ and\ \citenamefont
			{Sato}(2023)}]{OkumaAnnuRevCondMat2023}%
		\BibitemOpen
		\bibfield  {author} {\bibinfo {author} {\bibfnamefont {N.}~\bibnamefont
				{Okuma}}\ and\ \bibinfo {author} {\bibfnamefont {M.}~\bibnamefont {Sato}},\
		}\bibfield  {title} {\bibinfo {title} {Non-{H}ermitian topological phenomena:
				A review},\ }\href {https://doi.org/10.1146/annurev-conmatphys-040521-033133}
		{\bibfield  {journal} {\bibinfo  {journal} {Annu. Rev. Condens. Matter
					Phys.}\ }\textbf {\bibinfo {volume} {14}},\ \bibinfo {pages} {83} (\bibinfo
			{year} {2023})}\BibitemShut {NoStop}%
		\bibitem [{\citenamefont {Nasari}\ \emph {et~al.}(2023)\citenamefont {Nasari},
			\citenamefont {Pyrialakos}, \citenamefont {Christodoulides},\ and\
			\citenamefont {Khajavikhan}}]{NasariOptMatExp2023}%
		\BibitemOpen
		\bibfield  {author} {\bibinfo {author} {\bibfnamefont {H.}~\bibnamefont
				{Nasari}}, \bibinfo {author} {\bibfnamefont {G.~G.}\ \bibnamefont
				{Pyrialakos}}, \bibinfo {author} {\bibfnamefont {D.~N.}\ \bibnamefont
				{Christodoulides}},\ and\ \bibinfo {author} {\bibfnamefont {M.}~\bibnamefont
				{Khajavikhan}},\ }\bibfield  {title} {\bibinfo {title} {Non-{H}ermitian
				topological photonics},\ }\href {https://doi.org/10.1364/OME.483361}
		{\bibfield  {journal} {\bibinfo  {journal} {Opt. Mater. Express}\ }\textbf
			{\bibinfo {volume} {13}},\ \bibinfo {pages} {870} (\bibinfo {year}
			{2023})}\BibitemShut {NoStop}%
		\bibitem [{\citenamefont {Liu}\ \emph {et~al.}(2023)\citenamefont {Liu},
			\citenamefont {Lai}, \citenamefont {Wang}, \citenamefont {Cheng},
			\citenamefont {Tian},\ and\ \citenamefont {Chen}}]{LiuNanophotonics2023}%
		\BibitemOpen
		\bibfield  {author} {\bibinfo {author} {\bibfnamefont {H.}~\bibnamefont
				{Liu}}, \bibinfo {author} {\bibfnamefont {P.}~\bibnamefont {Lai}}, \bibinfo
			{author} {\bibfnamefont {H.}~\bibnamefont {Wang}}, \bibinfo {author}
			{\bibfnamefont {H.}~\bibnamefont {Cheng}}, \bibinfo {author} {\bibfnamefont
				{J.}~\bibnamefont {Tian}},\ and\ \bibinfo {author} {\bibfnamefont
				{S.}~\bibnamefont {Chen}},\ }\bibfield  {title} {\bibinfo {title}
			{Topological phases and non-{H}ermitian topology in photonic artificial
				microstructures},\ }\bibfield  {journal} {\bibinfo  {journal}
			{Nanophotonics}\ }\href {https://doi.org/doi:10.1515/nanoph-2022-0778}
		{doi:10.1515/nanoph-2022-0778} (\bibinfo {year} {2023})\BibitemShut {NoStop}%
		\bibitem [{\citenamefont {Wang}\ and\ \citenamefont
			{Chong}(2023)}]{WangJOSAB2023}%
		\BibitemOpen
		\bibfield  {author} {\bibinfo {author} {\bibfnamefont {Q.}~\bibnamefont
				{Wang}}\ and\ \bibinfo {author} {\bibfnamefont {Y.~D.}\ \bibnamefont
				{Chong}},\ }\bibfield  {title} {\bibinfo {title} {Non-{H}ermitian photonic
				lattices: tutorial},\ }\href {https://doi.org/10.1364/JOSAB.481963}
		{\bibfield  {journal} {\bibinfo  {journal} {J. Opt. Soc. Am. B}\ }\textbf
			{\bibinfo {volume} {40}},\ \bibinfo {pages} {1443} (\bibinfo {year}
			{2023})}\BibitemShut {NoStop}%
		\bibitem [{\citenamefont {Gong}\ \emph {et~al.}(2018)\citenamefont {Gong},
			\citenamefont {Ashida}, \citenamefont {Kawabata}, \citenamefont {Takasan},
			\citenamefont {Higashikawa},\ and\ \citenamefont {Ueda}}]{GongPRX2018}%
		\BibitemOpen
		\bibfield  {author} {\bibinfo {author} {\bibfnamefont {Z.}~\bibnamefont
				{Gong}}, \bibinfo {author} {\bibfnamefont {Y.}~\bibnamefont {Ashida}},
			\bibinfo {author} {\bibfnamefont {K.}~\bibnamefont {Kawabata}}, \bibinfo
			{author} {\bibfnamefont {K.}~\bibnamefont {Takasan}}, \bibinfo {author}
			{\bibfnamefont {S.}~\bibnamefont {Higashikawa}},\ and\ \bibinfo {author}
			{\bibfnamefont {M.}~\bibnamefont {Ueda}},\ }\bibfield  {title} {\bibinfo
			{title} {Topological phases of non-{H}ermitian systems},\ }\href
		{https://doi.org/10.1103/PhysRevX.8.031079} {\bibfield  {journal} {\bibinfo
				{journal} {Phys. Rev. X}\ }\textbf {\bibinfo {volume} {8}},\ \bibinfo {pages}
			{031079} (\bibinfo {year} {2018})}\BibitemShut {NoStop}%
		\bibitem [{\citenamefont {Kawabata}\ \emph {et~al.}(2019)\citenamefont
			{Kawabata}, \citenamefont {Shiozaki}, \citenamefont {Ueda},\ and\
			\citenamefont {Sato}}]{KawabataPRX2019}%
		\BibitemOpen
		\bibfield  {author} {\bibinfo {author} {\bibfnamefont {K.}~\bibnamefont
				{Kawabata}}, \bibinfo {author} {\bibfnamefont {K.}~\bibnamefont {Shiozaki}},
			\bibinfo {author} {\bibfnamefont {M.}~\bibnamefont {Ueda}},\ and\ \bibinfo
			{author} {\bibfnamefont {M.}~\bibnamefont {Sato}},\ }\bibfield  {title}
		{\bibinfo {title} {Symmetry and topology in non-{H}ermitian physics},\ }\href
		{https://doi.org/10.1103/PhysRevX.9.041015} {\bibfield  {journal} {\bibinfo
				{journal} {Phys. Rev. X}\ }\textbf {\bibinfo {volume} {9}},\ \bibinfo {pages}
			{041015} (\bibinfo {year} {2019})}\BibitemShut {NoStop}%
		\bibitem [{\citenamefont {Yao}\ and\ \citenamefont {Wang}(2018)}]{YaoPRL2018}%
		\BibitemOpen
		\bibfield  {author} {\bibinfo {author} {\bibfnamefont {S.}~\bibnamefont
				{Yao}}\ and\ \bibinfo {author} {\bibfnamefont {Z.}~\bibnamefont {Wang}},\
		}\bibfield  {title} {\bibinfo {title} {Edge states and topological invariants
				of non-{H}ermitian systems},\ }\href
		{https://doi.org/10.1103/PhysRevLett.121.086803} {\bibfield  {journal}
			{\bibinfo  {journal} {Phys. Rev. Lett.}\ }\textbf {\bibinfo {volume} {121}},\
			\bibinfo {pages} {086803} (\bibinfo {year} {2018})}\BibitemShut {NoStop}%
		\bibitem [{\citenamefont {Doppler}\ \emph {et~al.}(2016)\citenamefont
			{Doppler}, \citenamefont {Mailybaev}, \citenamefont {B\"ohm}, \citenamefont
			{Kuhl}, \citenamefont {Girschik}, \citenamefont {Libisch}, \citenamefont
			{Milburn}, \citenamefont {Rabl}, \citenamefont {Moiseyev},\ and\
			\citenamefont {Rotter}}]{DopplerNature2016}%
		\BibitemOpen
		\bibfield  {author} {\bibinfo {author} {\bibfnamefont {J.}~\bibnamefont
				{Doppler}}, \bibinfo {author} {\bibfnamefont {A.~A.}\ \bibnamefont
				{Mailybaev}}, \bibinfo {author} {\bibfnamefont {J.}~\bibnamefont {B\"ohm}},
			\bibinfo {author} {\bibfnamefont {U.}~\bibnamefont {Kuhl}}, \bibinfo {author}
			{\bibfnamefont {A.}~\bibnamefont {Girschik}}, \bibinfo {author}
			{\bibfnamefont {F.}~\bibnamefont {Libisch}}, \bibinfo {author} {\bibfnamefont
				{T.~J.}\ \bibnamefont {Milburn}}, \bibinfo {author} {\bibfnamefont
				{P.}~\bibnamefont {Rabl}}, \bibinfo {author} {\bibfnamefont {N.}~\bibnamefont
				{Moiseyev}},\ and\ \bibinfo {author} {\bibfnamefont {S.}~\bibnamefont
				{Rotter}},\ }\bibfield  {title} {\bibinfo {title} {Dynamically encircling an
				exceptional point for asymmetric mode switching},\ }\href
		{https://doi.org/10.1038/nature18605} {\bibfield  {journal} {\bibinfo
				{journal} {Nature}\ }\textbf {\bibinfo {volume} {537}},\ \bibinfo {pages}
			{76} (\bibinfo {year} {2016})}\BibitemShut {NoStop}%
		\bibitem [{\citenamefont {Chen}\ \emph {et~al.}(2017)\citenamefont {Chen},
			\citenamefont {\"Ozdemir}, \citenamefont {Zhao}, \citenamefont {Wiersig},\
			and\ \citenamefont {Yang}}]{ChenNature2017}%
		\BibitemOpen
		\bibfield  {author} {\bibinfo {author} {\bibfnamefont {W.}~\bibnamefont
				{Chen}}, \bibinfo {author} {\bibfnamefont {{\c S}.~K.}\ \bibnamefont
				{\"Ozdemir}}, \bibinfo {author} {\bibfnamefont {G.}~\bibnamefont {Zhao}},
			\bibinfo {author} {\bibfnamefont {J.}~\bibnamefont {Wiersig}},\ and\ \bibinfo
			{author} {\bibfnamefont {L.}~\bibnamefont {Yang}},\ }\bibfield  {title}
		{\bibinfo {title} {Exceptional points enhance sensing in an optical
				microcavity},\ }\href {https://doi.org/10.1038/nature23281} {\bibfield
			{journal} {\bibinfo  {journal} {Nature}\ }\textbf {\bibinfo {volume} {548}},\
			\bibinfo {pages} {192} (\bibinfo {year} {2017})}\BibitemShut {NoStop}%
		\bibitem [{\citenamefont {Wiersig}(2014)}]{WiersigPRL2014}%
		\BibitemOpen
		\bibfield  {author} {\bibinfo {author} {\bibfnamefont {J.}~\bibnamefont
				{Wiersig}},\ }\bibfield  {title} {\bibinfo {title} {Enhancing the sensitivity
				of frequency and energy splitting detection by using exceptional points:
				Application to microcavity sensors for single-particle detection},\ }\href
		{https://doi.org/10.1103/PhysRevLett.112.203901} {\bibfield  {journal}
			{\bibinfo  {journal} {Phys. Rev. Lett.}\ }\textbf {\bibinfo {volume} {112}},\
			\bibinfo {pages} {203901} (\bibinfo {year} {2014})}\BibitemShut {NoStop}%
		\bibitem [{\citenamefont {Miri}\ and\ \citenamefont
			{Alù}(2019)}]{MiriScience2019}%
		\BibitemOpen
		\bibfield  {author} {\bibinfo {author} {\bibfnamefont {M.-A.}\ \bibnamefont
				{Miri}}\ and\ \bibinfo {author} {\bibfnamefont {A.}~\bibnamefont {Alù}},\
		}\bibfield  {title} {\bibinfo {title} {Exceptional points in optics and
				photonics},\ }\href {https://doi.org/10.1126/science.aar7709} {\bibfield
			{journal} {\bibinfo  {journal} {Science}\ }\textbf {\bibinfo {volume}
				{363}},\ \bibinfo {pages} {eaar7709} (\bibinfo {year} {2019})}\BibitemShut
		{NoStop}%
		\bibitem [{\citenamefont {Özdemir}\ \emph {et~al.}(2019)\citenamefont
			{Özdemir}, \citenamefont {Rotter},\ and\ \citenamefont
			{Nori}}]{OzdemirNatMat2019}%
		\BibitemOpen
		\bibfield  {author} {\bibinfo {author} {\bibfnamefont {{\c S}.~K.}\
				\bibnamefont {Özdemir}}, \bibinfo {author} {\bibfnamefont {S.}~\bibnamefont
				{Rotter}},\ and\ \bibinfo {author} {\bibfnamefont {F.}~\bibnamefont {Nori}},\
		}\bibfield  {title} {\bibinfo {title} {Parity–{T}ime symmetry and
				exceptional points in photonics},\ }\href
		{https://doi.org/10.1038/s41563-019-0304-9} {\bibfield  {journal} {\bibinfo
				{journal} {Nat. Mater.}\ }\textbf {\bibinfo {volume} {18}},\ \bibinfo {pages}
			{783} (\bibinfo {year} {2019})}\BibitemShut {NoStop}%
		\bibitem [{\citenamefont {Denner}\ \emph {et~al.}(2021)\citenamefont {Denner},
			\citenamefont {Skurativska}, \citenamefont {Schindler}, \citenamefont
			{Fischer}, \citenamefont {Thomale}, \citenamefont {Bzdušek},\ and\
			\citenamefont {Neupert}}]{DennerNatComm2021}%
		\BibitemOpen
		\bibfield  {author} {\bibinfo {author} {\bibfnamefont {M.~M.}\ \bibnamefont
				{Denner}}, \bibinfo {author} {\bibfnamefont {A.}~\bibnamefont {Skurativska}},
			\bibinfo {author} {\bibfnamefont {F.}~\bibnamefont {Schindler}}, \bibinfo
			{author} {\bibfnamefont {M.~H.}\ \bibnamefont {Fischer}}, \bibinfo {author}
			{\bibfnamefont {R.}~\bibnamefont {Thomale}}, \bibinfo {author} {\bibfnamefont
				{T.}~\bibnamefont {Bzdušek}},\ and\ \bibinfo {author} {\bibfnamefont
				{T.}~\bibnamefont {Neupert}},\ }\bibfield  {title} {\bibinfo {title}
			{Exceptional topological insulators},\ }\href
		{https://doi.org/10.1038/s41467-021-25947-z} {\bibfield  {journal} {\bibinfo
				{journal} {Nat. Commun.}\ }\textbf {\bibinfo {volume} {12}},\ \bibinfo
			{pages} {5681} (\bibinfo {year} {2021})}\BibitemShut {NoStop}%
		\bibitem [{\citenamefont {Solnyshkov}\ \emph {et~al.}(2021)\citenamefont
			{Solnyshkov}, \citenamefont {Leblanc}, \citenamefont {Bessonart},
			\citenamefont {Nalitov}, \citenamefont {Ren}, \citenamefont {Liao},
			\citenamefont {Li},\ and\ \citenamefont {Malpuech}}]{SolnyshkovPRB2021}%
		\BibitemOpen
		\bibfield  {author} {\bibinfo {author} {\bibfnamefont {D.~D.}\ \bibnamefont
				{Solnyshkov}}, \bibinfo {author} {\bibfnamefont {C.}~\bibnamefont {Leblanc}},
			\bibinfo {author} {\bibfnamefont {L.}~\bibnamefont {Bessonart}}, \bibinfo
			{author} {\bibfnamefont {A.}~\bibnamefont {Nalitov}}, \bibinfo {author}
			{\bibfnamefont {J.}~\bibnamefont {Ren}}, \bibinfo {author} {\bibfnamefont
				{Q.}~\bibnamefont {Liao}}, \bibinfo {author} {\bibfnamefont {F.}~\bibnamefont
				{Li}},\ and\ \bibinfo {author} {\bibfnamefont {G.}~\bibnamefont {Malpuech}},\
		}\bibfield  {title} {\bibinfo {title} {Quantum metric and wave packets at
				exceptional points in non-{H}ermitian systems},\ }\href
		{https://doi.org/10.1103/PhysRevB.103.125302} {\bibfield  {journal} {\bibinfo
				{journal} {Phys. Rev. B}\ }\textbf {\bibinfo {volume} {103}},\ \bibinfo
			{pages} {125302} (\bibinfo {year} {2021})}\BibitemShut {NoStop}%
		\bibitem [{\citenamefont {Shen}\ \emph {et~al.}(2018)\citenamefont {Shen},
			\citenamefont {Zhen},\ and\ \citenamefont {Fu}}]{ShenPRL2018}%
		\BibitemOpen
		\bibfield  {author} {\bibinfo {author} {\bibfnamefont {H.}~\bibnamefont
				{Shen}}, \bibinfo {author} {\bibfnamefont {B.}~\bibnamefont {Zhen}},\ and\
			\bibinfo {author} {\bibfnamefont {L.}~\bibnamefont {Fu}},\ }\bibfield
		{title} {\bibinfo {title} {Topological band theory for non-{H}ermitian
				{H}amiltonians},\ }\href {https://doi.org/10.1103/PhysRevLett.120.146402}
		{\bibfield  {journal} {\bibinfo  {journal} {Phys. Rev. Lett.}\ }\textbf
			{\bibinfo {volume} {120}},\ \bibinfo {pages} {146402} (\bibinfo {year}
			{2018})}\BibitemShut {NoStop}%
		\bibitem [{\citenamefont {Leykam}\ \emph {et~al.}(2017)\citenamefont {Leykam},
			\citenamefont {Bliokh}, \citenamefont {Huang}, \citenamefont {Chong},\ and\
			\citenamefont {Nori}}]{LeykamPRL2017}%
		\BibitemOpen
		\bibfield  {author} {\bibinfo {author} {\bibfnamefont {D.}~\bibnamefont
				{Leykam}}, \bibinfo {author} {\bibfnamefont {K.}~\bibnamefont {Bliokh}},
			\bibinfo {author} {\bibfnamefont {C.}~\bibnamefont {Huang}}, \bibinfo
			{author} {\bibfnamefont {Y.~D.}\ \bibnamefont {Chong}},\ and\ \bibinfo
			{author} {\bibfnamefont {F.}~\bibnamefont {Nori}},\ }\bibfield  {title}
		{\bibinfo {title} {Edge modes, degeneracies, and topological numbers in
				non-{H}ermitian systems},\ }\href
		{https://doi.org/10.1103/PhysRevLett.118.040401} {\bibfield  {journal}
			{\bibinfo  {journal} {Phys. Rev. Lett.}\ }\textbf {\bibinfo {volume} {118}},\
			\bibinfo {pages} {040401} (\bibinfo {year} {2017})}\BibitemShut {NoStop}%
		\bibitem [{\citenamefont {Su}\ \emph {et~al.}(2021)\citenamefont {Su},
			\citenamefont {Estrecho}, \citenamefont {Biegańska}, \citenamefont {Huang},
			\citenamefont {Wurdack}, \citenamefont {Pieczarka}, \citenamefont {Truscott},
			\citenamefont {Liew}, \citenamefont {Ostrovskaya},\ and\ \citenamefont
			{Xiong}}]{SuSciAdv2021}%
		\BibitemOpen
		\bibfield  {author} {\bibinfo {author} {\bibfnamefont {R.}~\bibnamefont
				{Su}}, \bibinfo {author} {\bibfnamefont {E.}~\bibnamefont {Estrecho}},
			\bibinfo {author} {\bibfnamefont {D.}~\bibnamefont {Biegańska}}, \bibinfo
			{author} {\bibfnamefont {Y.}~\bibnamefont {Huang}}, \bibinfo {author}
			{\bibfnamefont {M.}~\bibnamefont {Wurdack}}, \bibinfo {author} {\bibfnamefont
				{M.}~\bibnamefont {Pieczarka}}, \bibinfo {author} {\bibfnamefont {A.~G.}\
				\bibnamefont {Truscott}}, \bibinfo {author} {\bibfnamefont {T.~C.~H.}\
				\bibnamefont {Liew}}, \bibinfo {author} {\bibfnamefont {E.~A.}\ \bibnamefont
				{Ostrovskaya}},\ and\ \bibinfo {author} {\bibfnamefont {Q.}~\bibnamefont
				{Xiong}},\ }\bibfield  {title} {\bibinfo {title} {Direct measurement of a
				non-{H}ermitian topological invariant in a hybrid light-matter system},\
		}\href {https://doi.org/10.1126/sciadv.abj8905} {\bibfield  {journal}
			{\bibinfo  {journal} {Sci. Adv.}\ }\textbf {\bibinfo {volume} {7}},\ \bibinfo
			{pages} {eabj8905} (\bibinfo {year} {2021})}\BibitemShut {NoStop}%
		\bibitem [{\citenamefont {Liao}\ \emph {et~al.}(2021)\citenamefont {Liao},
			\citenamefont {Leblanc}, \citenamefont {Ren}, \citenamefont {Li},
			\citenamefont {Li}, \citenamefont {Solnyshkov}, \citenamefont {Malpuech},
			\citenamefont {Yao},\ and\ \citenamefont {Fu}}]{LiaoPRL2021}%
		\BibitemOpen
		\bibfield  {author} {\bibinfo {author} {\bibfnamefont {Q.}~\bibnamefont
				{Liao}}, \bibinfo {author} {\bibfnamefont {C.}~\bibnamefont {Leblanc}},
			\bibinfo {author} {\bibfnamefont {J.}~\bibnamefont {Ren}}, \bibinfo {author}
			{\bibfnamefont {F.}~\bibnamefont {Li}}, \bibinfo {author} {\bibfnamefont
				{Y.}~\bibnamefont {Li}}, \bibinfo {author} {\bibfnamefont {D.~D.}\
				\bibnamefont {Solnyshkov}}, \bibinfo {author} {\bibfnamefont
				{G.}~\bibnamefont {Malpuech}}, \bibinfo {author} {\bibfnamefont
				{J.}~\bibnamefont {Yao}},\ and\ \bibinfo {author} {\bibfnamefont
				{H.}~\bibnamefont {Fu}},\ }\bibfield  {title} {\bibinfo {title} {Experimental
				measurement of the divergent quantum metric of an exceptional point},\ }\href
		{https://doi.org/10.1103/PhysRevLett.127.107402} {\bibfield  {journal}
			{\bibinfo  {journal} {Phys. Rev. Lett.}\ }\textbf {\bibinfo {volume} {127}},\
			\bibinfo {pages} {107402} (\bibinfo {year} {2021})}\BibitemShut {NoStop}%
		\bibitem [{\citenamefont {Zhou}\ \emph {et~al.}(2013)\citenamefont {Zhou},
			\citenamefont {Dridi}, \citenamefont {Suh}, \citenamefont {Kim},
			\citenamefont {Co}, \citenamefont {Wasielewski}, \citenamefont {Schatz},\
			and\ \citenamefont {Odom}}]{ZhouNatNano2013}%
		\BibitemOpen
		\bibfield  {author} {\bibinfo {author} {\bibfnamefont {W.}~\bibnamefont
				{Zhou}}, \bibinfo {author} {\bibfnamefont {M.}~\bibnamefont {Dridi}},
			\bibinfo {author} {\bibfnamefont {J.~Y.}\ \bibnamefont {Suh}}, \bibinfo
			{author} {\bibfnamefont {C.~H.}\ \bibnamefont {Kim}}, \bibinfo {author}
			{\bibfnamefont {D.~T.}\ \bibnamefont {Co}}, \bibinfo {author} {\bibfnamefont
				{M.~R.}\ \bibnamefont {Wasielewski}}, \bibinfo {author} {\bibfnamefont
				{G.~C.}\ \bibnamefont {Schatz}},\ and\ \bibinfo {author} {\bibfnamefont
				{T.~W.}\ \bibnamefont {Odom}},\ }\bibfield  {title} {\bibinfo {title} {Lasing
				action in strongly coupled plasmonic nanocavity arrays},\ }\href
		{https://doi.org/10.1038/nnano.2013.99} {\bibfield  {journal} {\bibinfo
				{journal} {Nat. Nanotechnol.}\ }\textbf {\bibinfo {volume} {8}},\ \bibinfo
			{pages} {506–511} (\bibinfo {year} {2013})}\BibitemShut {NoStop}%
		\bibitem [{\citenamefont {Schokker}\ and\ \citenamefont
			{Koenderink}(2016)}]{SchokkerOptica2016}%
		\BibitemOpen
		\bibfield  {author} {\bibinfo {author} {\bibfnamefont {A.~H.}\ \bibnamefont
				{Schokker}}\ and\ \bibinfo {author} {\bibfnamefont {A.~F.}\ \bibnamefont
				{Koenderink}},\ }\bibfield  {title} {\bibinfo {title} {Lasing in
				quasi-periodic and aperiodic plasmon lattices},\ }\href
		{https://doi.org/10.1364/OPTICA.3.000686} {\bibfield  {journal} {\bibinfo
				{journal} {Optica}\ }\textbf {\bibinfo {volume} {3}},\ \bibinfo {pages} {686}
			(\bibinfo {year} {2016})}\BibitemShut {NoStop}%
		\bibitem [{\citenamefont {De~Giorgi}\ \emph {et~al.}(2018)\citenamefont
			{De~Giorgi}, \citenamefont {Ramezani}, \citenamefont {Todisco}, \citenamefont
			{Halpin}, \citenamefont {Caputo}, \citenamefont {Fieramosca}, \citenamefont
			{G\'omez-Rivas},\ and\ \citenamefont {Sanvitto}}]{DeGiorgiACSPhot2018}%
		\BibitemOpen
		\bibfield  {author} {\bibinfo {author} {\bibfnamefont {M.}~\bibnamefont
				{De~Giorgi}}, \bibinfo {author} {\bibfnamefont {M.}~\bibnamefont {Ramezani}},
			\bibinfo {author} {\bibfnamefont {F.}~\bibnamefont {Todisco}}, \bibinfo
			{author} {\bibfnamefont {A.}~\bibnamefont {Halpin}}, \bibinfo {author}
			{\bibfnamefont {D.}~\bibnamefont {Caputo}}, \bibinfo {author} {\bibfnamefont
				{A.}~\bibnamefont {Fieramosca}}, \bibinfo {author} {\bibfnamefont
				{J.}~\bibnamefont {G\'omez-Rivas}},\ and\ \bibinfo {author} {\bibfnamefont
				{D.}~\bibnamefont {Sanvitto}},\ }\bibfield  {title} {\bibinfo {title}
			{Interaction and coherence of a plasmon–exciton polariton condensate},\
		}\href {https://doi.org/10.1021/acsphotonics.8b00662} {\bibfield  {journal}
			{\bibinfo  {journal} {ACS Photonics}\ }\textbf {\bibinfo {volume} {5}},\
			\bibinfo {pages} {3666} (\bibinfo {year} {2018})}\BibitemShut {NoStop}%
		\bibitem [{\citenamefont {Wang}\ \emph {et~al.}(2018)\citenamefont {Wang},
			\citenamefont {Ramezani}, \citenamefont {Väkeväinen}, \citenamefont
			{Törmä}, \citenamefont {G\'{o}mez-Rivas},\ and\ \citenamefont
			{Odom}}]{WangMatToday2018}%
		\BibitemOpen
		\bibfield  {author} {\bibinfo {author} {\bibfnamefont {W.}~\bibnamefont
				{Wang}}, \bibinfo {author} {\bibfnamefont {M.}~\bibnamefont {Ramezani}},
			\bibinfo {author} {\bibfnamefont {A.~I.}\ \bibnamefont {Väkeväinen}},
			\bibinfo {author} {\bibfnamefont {P.}~\bibnamefont {Törmä}}, \bibinfo
			{author} {\bibfnamefont {J.}~\bibnamefont {G\'{o}mez-Rivas}},\ and\ \bibinfo
			{author} {\bibfnamefont {T.~W.}\ \bibnamefont {Odom}},\ }\bibfield  {title}
		{\bibinfo {title} {The rich photonic world of plasmonic nanoparticle
				arrays},\ }\href
		{https://doi.org/https://doi.org/10.1016/j.mattod.2017.09.002} {\bibfield
			{journal} {\bibinfo  {journal} {Mater. Today}\ }\textbf {\bibinfo {volume}
				{21}},\ \bibinfo {pages} {303–314} (\bibinfo {year} {2018})}\BibitemShut
		{NoStop}%
		\bibitem [{\citenamefont {Wu}\ \emph {et~al.}(2020)\citenamefont {Wu},
			\citenamefont {Ha}, \citenamefont {Shendre}, \citenamefont {Durmusoglu},
			\citenamefont {Koh}, \citenamefont {Abujetas}, \citenamefont {Sánchez-Gil},
			\citenamefont {Paniagua-Domínguez}, \citenamefont {Demir},\ and\
			\citenamefont {Kuznetsov}}]{WuNanoLett2020}%
		\BibitemOpen
		\bibfield  {author} {\bibinfo {author} {\bibfnamefont {M.}~\bibnamefont
				{Wu}}, \bibinfo {author} {\bibfnamefont {S.~T.}\ \bibnamefont {Ha}}, \bibinfo
			{author} {\bibfnamefont {S.}~\bibnamefont {Shendre}}, \bibinfo {author}
			{\bibfnamefont {E.~G.}\ \bibnamefont {Durmusoglu}}, \bibinfo {author}
			{\bibfnamefont {W.-K.}\ \bibnamefont {Koh}}, \bibinfo {author} {\bibfnamefont
				{D.~R.}\ \bibnamefont {Abujetas}}, \bibinfo {author} {\bibfnamefont {J.~A.}\
				\bibnamefont {Sánchez-Gil}}, \bibinfo {author} {\bibfnamefont
				{R.}~\bibnamefont {Paniagua-Domínguez}}, \bibinfo {author} {\bibfnamefont
				{H.~V.}\ \bibnamefont {Demir}},\ and\ \bibinfo {author} {\bibfnamefont
				{A.~I.}\ \bibnamefont {Kuznetsov}},\ }\bibfield  {title} {\bibinfo {title}
			{Room-temperature lasing in colloidal nanoplatelets via mie-resonant bound
				states in the continuum},\ }\href
		{https://doi.org/10.1021/acs.nanolett.0c01975} {\bibfield  {journal}
			{\bibinfo  {journal} {Nano Lett.}\ }\textbf {\bibinfo {volume} {20}},\
			\bibinfo {pages} {6005} (\bibinfo {year} {2020})}\BibitemShut {NoStop}%
		\bibitem [{\citenamefont {Ramezani}\ \emph {et~al.}(2017)\citenamefont
			{Ramezani}, \citenamefont {Halpin}, \citenamefont
			{Fern\'{a}ndez-Dom\'{i}nguez}, \citenamefont {Feist}, \citenamefont
			{Rodriguez}, \citenamefont {Garc\'{i}a-Vidal},\ and\ \citenamefont
			{G\'{o}mez-Rivas}}]{RamezaniOptica2017}%
		\BibitemOpen
		\bibfield  {author} {\bibinfo {author} {\bibfnamefont {M.}~\bibnamefont
				{Ramezani}}, \bibinfo {author} {\bibfnamefont {A.}~\bibnamefont {Halpin}},
			\bibinfo {author} {\bibfnamefont {A.~I.}\ \bibnamefont
				{Fern\'{a}ndez-Dom\'{i}nguez}}, \bibinfo {author} {\bibfnamefont
				{J.}~\bibnamefont {Feist}}, \bibinfo {author} {\bibfnamefont {S.~R.-K.}\
				\bibnamefont {Rodriguez}}, \bibinfo {author} {\bibfnamefont {F.~J.}\
				\bibnamefont {Garc\'{i}a-Vidal}},\ and\ \bibinfo {author} {\bibfnamefont
				{J.}~\bibnamefont {G\'{o}mez-Rivas}},\ }\bibfield  {title} {\bibinfo {title}
			{Plasmon-exciton-polariton lasing},\ }\href
		{https://doi.org/10.1364/OPTICA.4.000031} {\bibfield  {journal} {\bibinfo
				{journal} {Optica}\ }\textbf {\bibinfo {volume} {4}},\ \bibinfo {pages} {31}
			(\bibinfo {year} {2017})}\BibitemShut {NoStop}%
		\bibitem [{\citenamefont {Hakala}\ \emph {et~al.}(2018)\citenamefont {Hakala},
			\citenamefont {Moilanen}, \citenamefont {V{\"a}kev{\"a}inen}, \citenamefont
			{Guo}, \citenamefont {Martikainen}, \citenamefont {Daskalakis}, \citenamefont
			{Rekola}, \citenamefont {Julku},\ and\ \citenamefont
			{T{\"o}rm{\"a}}}]{HakalaNatPhys2018}%
		\BibitemOpen
		\bibfield  {author} {\bibinfo {author} {\bibfnamefont {T.~K.}\ \bibnamefont
				{Hakala}}, \bibinfo {author} {\bibfnamefont {A.~J.}\ \bibnamefont
				{Moilanen}}, \bibinfo {author} {\bibfnamefont {A.~I.}\ \bibnamefont
				{V{\"a}kev{\"a}inen}}, \bibinfo {author} {\bibfnamefont {R.}~\bibnamefont
				{Guo}}, \bibinfo {author} {\bibfnamefont {J.-P.}\ \bibnamefont
				{Martikainen}}, \bibinfo {author} {\bibfnamefont {K.~S.}\ \bibnamefont
				{Daskalakis}}, \bibinfo {author} {\bibfnamefont {H.~T.}\ \bibnamefont
				{Rekola}}, \bibinfo {author} {\bibfnamefont {A.}~\bibnamefont {Julku}},\ and\
			\bibinfo {author} {\bibfnamefont {P.}~\bibnamefont {T{\"o}rm{\"a}}},\
		}\bibfield  {title} {\bibinfo {title} {Bose--{E}instein {C}ondensation in a
				{P}lasmonic {L}attice},\ }\href {https://doi.org/10.1038/s41567-018-0109-9}
		{\bibfield  {journal} {\bibinfo  {journal} {Nat. Phys.}\ }\textbf {\bibinfo
				{volume} {14}},\ \bibinfo {pages} {739–744} (\bibinfo {year}
			{2018})}\BibitemShut {NoStop}%
		\bibitem [{\citenamefont {V{\"{a}}kev{\"{a}}inen}\ \emph
			{et~al.}(2020)\citenamefont {V{\"{a}}kev{\"{a}}inen}, \citenamefont
			{Moilanen}, \citenamefont {Ne{\v{c}}ada}, \citenamefont {Hakala},
			\citenamefont {Daskalakis},\ and\ \citenamefont
			{T{\"{o}}rm{\"{a}}}}]{VakevainenNatComm2020}%
		\BibitemOpen
		\bibfield  {author} {\bibinfo {author} {\bibfnamefont {A.~I.}\ \bibnamefont
				{V{\"{a}}kev{\"{a}}inen}}, \bibinfo {author} {\bibfnamefont {A.~J.}\
				\bibnamefont {Moilanen}}, \bibinfo {author} {\bibfnamefont {M.}~\bibnamefont
				{Ne{\v{c}}ada}}, \bibinfo {author} {\bibfnamefont {T.~K.}\ \bibnamefont
				{Hakala}}, \bibinfo {author} {\bibfnamefont {K.~D.}\ \bibnamefont
				{Daskalakis}},\ and\ \bibinfo {author} {\bibfnamefont {P.}~\bibnamefont
				{T{\"{o}}rm{\"{a}}}},\ }\bibfield  {title} {\bibinfo {title} {{Sub-picosecond
					thermalization dynamics in condensation of strongly coupled lattice
					plasmons}},\ }\href {https://doi.org/10.1038/s41467-020-16906-1} {\bibfield
			{journal} {\bibinfo  {journal} {Nat. Commun.}\ }\textbf {\bibinfo {volume}
				{11}},\ \bibinfo {pages} {3139} (\bibinfo {year} {2020})}\BibitemShut
		{NoStop}%
		\bibitem [{\citenamefont {Koshelev}\ and\ \citenamefont
			{Kivshar}(2021)}]{KoshelevACSPhot2021}%
		\BibitemOpen
		\bibfield  {author} {\bibinfo {author} {\bibfnamefont {K.}~\bibnamefont
				{Koshelev}}\ and\ \bibinfo {author} {\bibfnamefont {Y.}~\bibnamefont
				{Kivshar}},\ }\bibfield  {title} {\bibinfo {title} {Dielectric resonant
				metaphotonics},\ }\href {https://doi.org/10.1021/acsphotonics.0c01315}
		{\bibfield  {journal} {\bibinfo  {journal} {ACS Photonics}\ }\textbf
			{\bibinfo {volume} {8}},\ \bibinfo {pages} {102} (\bibinfo {year}
			{2021})}\BibitemShut {NoStop}%
		\bibitem [{\citenamefont {Guan}\ \emph {et~al.}(2022)\citenamefont {Guan},
			\citenamefont {Park}, \citenamefont {Deng}, \citenamefont {Tan},
			\citenamefont {Hu},\ and\ \citenamefont {Odom}}]{GuanChemRev2022}%
		\BibitemOpen
		\bibfield  {author} {\bibinfo {author} {\bibfnamefont {J.}~\bibnamefont
				{Guan}}, \bibinfo {author} {\bibfnamefont {J.-E.}\ \bibnamefont {Park}},
			\bibinfo {author} {\bibfnamefont {S.}~\bibnamefont {Deng}}, \bibinfo {author}
			{\bibfnamefont {M.~J.~H.}\ \bibnamefont {Tan}}, \bibinfo {author}
			{\bibfnamefont {J.}~\bibnamefont {Hu}},\ and\ \bibinfo {author}
			{\bibfnamefont {T.~W.}\ \bibnamefont {Odom}},\ }\bibfield  {title} {\bibinfo
			{title} {Light–matter interactions in hybrid material metasurfaces},\
		}\href {https://doi.org/10.1021/acs.chemrev.2c00011} {\bibfield  {journal}
			{\bibinfo  {journal} {Chem. Rev.}\ }\textbf {\bibinfo {volume} {122}},\
			\bibinfo {pages} {15177} (\bibinfo {year} {2022})}\BibitemShut {NoStop}%
		\bibitem [{\citenamefont {Castellanos}\ \emph {et~al.}(2023)\citenamefont
			{Castellanos}, \citenamefont {Ramezani}, \citenamefont {Murai},\ and\
			\citenamefont {Gómez~Rivas}}]{CastellanosAdvOptMat2023}%
		\BibitemOpen
		\bibfield  {author} {\bibinfo {author} {\bibfnamefont {G.~W.}\ \bibnamefont
				{Castellanos}}, \bibinfo {author} {\bibfnamefont {M.}~\bibnamefont
				{Ramezani}}, \bibinfo {author} {\bibfnamefont {S.}~\bibnamefont {Murai}},\
			and\ \bibinfo {author} {\bibfnamefont {J.}~\bibnamefont {Gómez~Rivas}},\
		}\bibfield  {title} {\bibinfo {title} {Non-equilibrium {B}ose–{E}instein
				condensation of exciton-polaritons in silicon metasurfaces},\ }\href
		{https://doi.org/https://doi.org/10.1002/adom.202202305} {\bibfield
			{journal} {\bibinfo  {journal} {Adv. Opt. Mater.}\ }\textbf {\bibinfo
				{volume} {11}},\ \bibinfo {pages} {2202305} (\bibinfo {year}
			{2023})}\BibitemShut {NoStop}%
		\bibitem [{\citenamefont {Rider}\ \emph {et~al.}(2022)\citenamefont {Rider},
			\citenamefont {Buendía}, \citenamefont {Abujetas}, \citenamefont {Huidobro},
			\citenamefont {Sánchez-Gil},\ and\ \citenamefont
			{Giannini}}]{RiderACSPhot2022}%
		\BibitemOpen
		\bibfield  {author} {\bibinfo {author} {\bibfnamefont {M.~S.}\ \bibnamefont
				{Rider}}, \bibinfo {author} {\bibfnamefont {A.}~\bibnamefont {Buendía}},
			\bibinfo {author} {\bibfnamefont {D.~R.}\ \bibnamefont {Abujetas}}, \bibinfo
			{author} {\bibfnamefont {P.~A.}\ \bibnamefont {Huidobro}}, \bibinfo {author}
			{\bibfnamefont {J.~A.}\ \bibnamefont {Sánchez-Gil}},\ and\ \bibinfo {author}
			{\bibfnamefont {V.}~\bibnamefont {Giannini}},\ }\bibfield  {title} {\bibinfo
			{title} {Advances and prospects in topological nanoparticle photonics},\
		}\href {https://doi.org/10.1021/acsphotonics.1c01874} {\bibfield  {journal}
			{\bibinfo  {journal} {ACS Photonics}\ }\textbf {\bibinfo {volume} {9}},\
			\bibinfo {pages} {1483} (\bibinfo {year} {2022})}\BibitemShut {NoStop}%
		\bibitem [{\citenamefont {Taskinen}\ \emph {et~al.}(2021)\citenamefont
			{Taskinen}, \citenamefont {Kliuiev}, \citenamefont {Moilanen},\ and\
			\citenamefont {T{\"{o}}rm{\"{a}}}}]{TaskinenNanoLett2021}%
		\BibitemOpen
		\bibfield  {author} {\bibinfo {author} {\bibfnamefont {J.~M.}\ \bibnamefont
				{Taskinen}}, \bibinfo {author} {\bibfnamefont {P.}~\bibnamefont {Kliuiev}},
			\bibinfo {author} {\bibfnamefont {A.~J.}\ \bibnamefont {Moilanen}},\ and\
			\bibinfo {author} {\bibfnamefont {P.}~\bibnamefont {T{\"{o}}rm{\"{a}}}},\
		}\bibfield  {title} {\bibinfo {title} {Polarization and phase textures in
				lattice plasmon condensates},\ }\href
		{https://doi.org/10.1021/acs.nanolett.1c01395} {\bibfield  {journal}
			{\bibinfo  {journal} {Nano Lett.}\ }\textbf {\bibinfo {volume} {21}},\
			\bibinfo {pages} {5262–5268} (\bibinfo {year} {2021})}\BibitemShut
		{NoStop}%
		\bibitem [{\citenamefont {Ha}\ \emph {et~al.}(2018)\citenamefont {Ha},
			\citenamefont {Fu}, \citenamefont {Emani}, \citenamefont {Pan}, \citenamefont
			{Bakker}, \citenamefont {Paniagua-Domínguez},\ and\ \citenamefont
			{Kuznetsov}}]{HaNatNano2018}%
		\BibitemOpen
		\bibfield  {author} {\bibinfo {author} {\bibfnamefont {S.~T.}\ \bibnamefont
				{Ha}}, \bibinfo {author} {\bibfnamefont {Y.~H.}\ \bibnamefont {Fu}}, \bibinfo
			{author} {\bibfnamefont {N.~K.}\ \bibnamefont {Emani}}, \bibinfo {author}
			{\bibfnamefont {Z.}~\bibnamefont {Pan}}, \bibinfo {author} {\bibfnamefont
				{R.~M.}\ \bibnamefont {Bakker}}, \bibinfo {author} {\bibfnamefont
				{R.}~\bibnamefont {Paniagua-Domínguez}},\ and\ \bibinfo {author}
			{\bibfnamefont {A.~I.}\ \bibnamefont {Kuznetsov}},\ }\bibfield  {title}
		{\bibinfo {title} {Directional lasing in resonant semiconductor nanoantenna
				arrays},\ }\href {https://doi.org/10.1038/s41565-018-0245-5} {\bibfield
			{journal} {\bibinfo  {journal} {Nat. Nanotechnol.}\ }\textbf {\bibinfo
				{volume} {13}},\ \bibinfo {pages} {1042–1047} (\bibinfo {year}
			{2018})}\BibitemShut {NoStop}%
		\bibitem [{\citenamefont {Murai}\ \emph {et~al.}(2020)\citenamefont {Murai},
			\citenamefont {Abujetas}, \citenamefont {Castellanos}, \citenamefont
			{Sánchez-Gil}, \citenamefont {Zhang},\ and\ \citenamefont
			{Rivas}}]{MuraiACSPhot2020}%
		\BibitemOpen
		\bibfield  {author} {\bibinfo {author} {\bibfnamefont {S.}~\bibnamefont
				{Murai}}, \bibinfo {author} {\bibfnamefont {D.~R.}\ \bibnamefont {Abujetas}},
			\bibinfo {author} {\bibfnamefont {G.~W.}\ \bibnamefont {Castellanos}},
			\bibinfo {author} {\bibfnamefont {J.~A.}\ \bibnamefont {Sánchez-Gil}},
			\bibinfo {author} {\bibfnamefont {F.}~\bibnamefont {Zhang}},\ and\ \bibinfo
			{author} {\bibfnamefont {J.~G.}\ \bibnamefont {Rivas}},\ }\bibfield  {title}
		{\bibinfo {title} {Bound states in the continuum in the visible emerging from
				out-of-plane magnetic dipoles},\ }\href
		{https://doi.org/10.1021/acsphotonics.0c00723} {\bibfield  {journal}
			{\bibinfo  {journal} {ACS Photonics}\ }\textbf {\bibinfo {volume} {7}},\
			\bibinfo {pages} {2204} (\bibinfo {year} {2020})}\BibitemShut {NoStop}%
		\bibitem [{\citenamefont {Heilmann}\ \emph {et~al.}(2022)\citenamefont
			{Heilmann}, \citenamefont {Salerno}, \citenamefont {Cuerda}, \citenamefont
			{Hakala},\ and\ \citenamefont {T{\"{o}}rm{\"{a}}}}]{HeilmannACSPhot2022}%
		\BibitemOpen
		\bibfield  {author} {\bibinfo {author} {\bibfnamefont {R.}~\bibnamefont
				{Heilmann}}, \bibinfo {author} {\bibfnamefont {G.}~\bibnamefont {Salerno}},
			\bibinfo {author} {\bibfnamefont {J.}~\bibnamefont {Cuerda}}, \bibinfo
			{author} {\bibfnamefont {T.~K.}\ \bibnamefont {Hakala}},\ and\ \bibinfo
			{author} {\bibfnamefont {P.}~\bibnamefont {T{\"{o}}rm{\"{a}}}},\ }\bibfield
		{title} {\bibinfo {title} {Quasi-{BIC} mode lasing in a quadrumer plasmonic
				lattice},\ }\href {https://doi.org/10.1021/acsphotonics.1c01416} {\bibfield
			{journal} {\bibinfo  {journal} {ACS Photonics}\ }\textbf {\bibinfo {volume}
				{9}},\ \bibinfo {pages} {224–232} (\bibinfo {year} {2022})}\BibitemShut
		{NoStop}%
		\bibitem [{\citenamefont {Liu}\ \emph {et~al.}(2021)\citenamefont {Liu},
			\citenamefont {Liu}, \citenamefont {Shi},\ and\ \citenamefont
			{Kivshar}}]{LiuNanophotonics2021}%
		\BibitemOpen
		\bibfield  {author} {\bibinfo {author} {\bibfnamefont {W.}~\bibnamefont
				{Liu}}, \bibinfo {author} {\bibfnamefont {W.}~\bibnamefont {Liu}}, \bibinfo
			{author} {\bibfnamefont {L.}~\bibnamefont {Shi}},\ and\ \bibinfo {author}
			{\bibfnamefont {Y.}~\bibnamefont {Kivshar}},\ }\bibfield  {title} {\bibinfo
			{title} {Topological polarization singularities in metaphotonics},\ }\href
		{https://doi.org/doi:10.1515/nanoph-2020-0654} {\bibfield  {journal}
			{\bibinfo  {journal} {Nanophotonics}\ }\textbf {\bibinfo {volume} {10}},\
			\bibinfo {pages} {1469} (\bibinfo {year} {2021})}\BibitemShut {NoStop}%
		\bibitem [{\citenamefont {Salerno}\ \emph {et~al.}(2022)\citenamefont
			{Salerno}, \citenamefont {Heilmann}, \citenamefont {Arjas}, \citenamefont
			{Aronen}, \citenamefont {Martikainen},\ and\ \citenamefont
			{T\"orm\"a}}]{SalernoPRL2022}%
		\BibitemOpen
		\bibfield  {author} {\bibinfo {author} {\bibfnamefont {G.}~\bibnamefont
				{Salerno}}, \bibinfo {author} {\bibfnamefont {R.}~\bibnamefont {Heilmann}},
			\bibinfo {author} {\bibfnamefont {K.}~\bibnamefont {Arjas}}, \bibinfo
			{author} {\bibfnamefont {K.}~\bibnamefont {Aronen}}, \bibinfo {author}
			{\bibfnamefont {J.-P.}\ \bibnamefont {Martikainen}},\ and\ \bibinfo {author}
			{\bibfnamefont {P.}~\bibnamefont {T\"orm\"a}},\ }\bibfield  {title} {\bibinfo
			{title} {Loss-driven topological transitions in lasing},\ }\href
		{https://doi.org/10.1103/PhysRevLett.129.173901} {\bibfield  {journal}
			{\bibinfo  {journal} {Phys. Rev. Lett.}\ }\textbf {\bibinfo {volume} {129}},\
			\bibinfo {pages} {173901} (\bibinfo {year} {2022})}\BibitemShut {NoStop}%
		\bibitem [{\citenamefont {Kravets}\ \emph {et~al.}(2018)\citenamefont
			{Kravets}, \citenamefont {Kabashin}, \citenamefont {Barnes},\ and\
			\citenamefont {Grigorenko}}]{KravetsChemRev2018}%
		\BibitemOpen
		\bibfield  {author} {\bibinfo {author} {\bibfnamefont {V.~G.}\ \bibnamefont
				{Kravets}}, \bibinfo {author} {\bibfnamefont {A.~V.}\ \bibnamefont
				{Kabashin}}, \bibinfo {author} {\bibfnamefont {W.~L.}\ \bibnamefont
				{Barnes}},\ and\ \bibinfo {author} {\bibfnamefont {A.~N.}\ \bibnamefont
				{Grigorenko}},\ }\bibfield  {title} {\bibinfo {title} {Plasmonic surface
				lattice resonances: A review of properties and applications},\ }\href
		{https://doi.org/10.1021/acs.chemrev.8b00243} {\bibfield  {journal} {\bibinfo
				{journal} {Chem. Rev.}\ }\textbf {\bibinfo {volume} {118}},\ \bibinfo
			{pages} {5912} (\bibinfo {year} {2018})}\BibitemShut {NoStop}%
		\bibitem [{\citenamefont {Augui\'e}\ and\ \citenamefont
			{Barnes}(2008)}]{AuguiePRL2008}%
		\BibitemOpen
		\bibfield  {author} {\bibinfo {author} {\bibfnamefont {B.}~\bibnamefont
				{Augui\'e}}\ and\ \bibinfo {author} {\bibfnamefont {W.~L.}\ \bibnamefont
				{Barnes}},\ }\bibfield  {title} {\bibinfo {title} {Collective resonances in
				gold nanoparticle arrays},\ }\href
		{https://doi.org/10.1103/PhysRevLett.101.143902} {\bibfield  {journal}
			{\bibinfo  {journal} {Phys. Rev. Lett.}\ }\textbf {\bibinfo {volume} {101}},\
			\bibinfo {pages} {143902} (\bibinfo {year} {2008})}\BibitemShut {NoStop}%
		\bibitem [{\citenamefont {Rodriguez}\ \emph {et~al.}(2011)\citenamefont
			{Rodriguez}, \citenamefont {Abass}, \citenamefont {Maes}, \citenamefont
			{Janssen}, \citenamefont {Vecchi},\ and\ \citenamefont
			{G\'omez~Rivas}}]{RodriguezPRX2011}%
		\BibitemOpen
		\bibfield  {author} {\bibinfo {author} {\bibfnamefont {S.~R.~K.}\
				\bibnamefont {Rodriguez}}, \bibinfo {author} {\bibfnamefont {A.}~\bibnamefont
				{Abass}}, \bibinfo {author} {\bibfnamefont {B.}~\bibnamefont {Maes}},
			\bibinfo {author} {\bibfnamefont {O.~T.~A.}\ \bibnamefont {Janssen}},
			\bibinfo {author} {\bibfnamefont {G.}~\bibnamefont {Vecchi}},\ and\ \bibinfo
			{author} {\bibfnamefont {J.}~\bibnamefont {G\'omez~Rivas}},\ }\bibfield
		{title} {\bibinfo {title} {Coupling bright and dark plasmonic lattice
				resonances},\ }\href {https://doi.org/10.1103/PhysRevX.1.021019} {\bibfield
			{journal} {\bibinfo  {journal} {Phys. Rev. X}\ }\textbf {\bibinfo {volume}
				{1}},\ \bibinfo {pages} {021019} (\bibinfo {year} {2011})}\BibitemShut
		{NoStop}%
		\bibitem [{\citenamefont {Guo}\ \emph {et~al.}(2017)\citenamefont {Guo},
			\citenamefont {Hakala},\ and\ \citenamefont {T{\"o}rm{\"a}}}]{GuoPRB2017}%
		\BibitemOpen
		\bibfield  {author} {\bibinfo {author} {\bibfnamefont {R.}~\bibnamefont
				{Guo}}, \bibinfo {author} {\bibfnamefont {T.~K.}\ \bibnamefont {Hakala}},\
			and\ \bibinfo {author} {\bibfnamefont {P.}~\bibnamefont {T{\"o}rm{\"a}}},\
		}\bibfield  {title} {\bibinfo {title} {Geometry dependence of surface lattice
				resonances in plasmonic nanoparticle arrays},\ }\href
		{https://doi.org/10.1103/PhysRevB.95.155423} {\bibfield  {journal} {\bibinfo
				{journal} {Phys. Rev. B}\ }\textbf {\bibinfo {volume} {95}},\ \bibinfo
			{pages} {155423} (\bibinfo {year} {2017})}\BibitemShut {NoStop}%
		\bibitem [{\citenamefont {Knudson}\ \emph {et~al.}(2019)\citenamefont
			{Knudson}, \citenamefont {Li}, \citenamefont {Wang}, \citenamefont {Wang},
			\citenamefont {Schaller},\ and\ \citenamefont {Odom}}]{KnudsonACSNano2019}%
		\BibitemOpen
		\bibfield  {author} {\bibinfo {author} {\bibfnamefont {M.~P.}\ \bibnamefont
				{Knudson}}, \bibinfo {author} {\bibfnamefont {R.}~\bibnamefont {Li}},
			\bibinfo {author} {\bibfnamefont {D.}~\bibnamefont {Wang}}, \bibinfo {author}
			{\bibfnamefont {W.}~\bibnamefont {Wang}}, \bibinfo {author} {\bibfnamefont
				{R.~D.}\ \bibnamefont {Schaller}},\ and\ \bibinfo {author} {\bibfnamefont
				{T.~W.}\ \bibnamefont {Odom}},\ }\bibfield  {title} {\bibinfo {title}
			{Polarization-dependent lasing behavior from low-symmetry nanocavity
				arrays},\ }\href {https://doi.org/10.1021/acsnano.9b01142} {\bibfield
			{journal} {\bibinfo  {journal} {ACS Nano}\ }\textbf {\bibinfo {volume}
				{13}},\ \bibinfo {pages} {7435} (\bibinfo {year} {2019})}\BibitemShut
		{NoStop}%
		\bibitem [{\citenamefont {Weick}\ \emph {et~al.}(2013)\citenamefont {Weick},
			\citenamefont {Woollacott}, \citenamefont {Barnes}, \citenamefont {Hess},\
			and\ \citenamefont {Mariani}}]{WeickPRL2013}%
		\BibitemOpen
		\bibfield  {author} {\bibinfo {author} {\bibfnamefont {G.}~\bibnamefont
				{Weick}}, \bibinfo {author} {\bibfnamefont {C.}~\bibnamefont {Woollacott}},
			\bibinfo {author} {\bibfnamefont {W.~L.}\ \bibnamefont {Barnes}}, \bibinfo
			{author} {\bibfnamefont {O.}~\bibnamefont {Hess}},\ and\ \bibinfo {author}
			{\bibfnamefont {E.}~\bibnamefont {Mariani}},\ }\bibfield  {title} {\bibinfo
			{title} {Dirac-like plasmons in honeycomb lattices of metallic
				nanoparticles},\ }\href {https://doi.org/10.1103/PhysRevLett.110.106801}
		{\bibfield  {journal} {\bibinfo  {journal} {Phys. Rev. Lett.}\ }\textbf
			{\bibinfo {volume} {110}},\ \bibinfo {pages} {106801} (\bibinfo {year}
			{2013})}\BibitemShut {NoStop}%
		\bibitem [{\citenamefont {Kuznetsov}\ \emph {et~al.}(2016)\citenamefont
			{Kuznetsov}, \citenamefont {Miroshnichenko}, \citenamefont {Brongersma},
			\citenamefont {Kivshar},\ and\ \citenamefont
			{Luk’yanchuk}}]{KuznetsovScience2016}%
		\BibitemOpen
		\bibfield  {author} {\bibinfo {author} {\bibfnamefont {A.~I.}\ \bibnamefont
				{Kuznetsov}}, \bibinfo {author} {\bibfnamefont {A.~E.}\ \bibnamefont
				{Miroshnichenko}}, \bibinfo {author} {\bibfnamefont {M.~L.}\ \bibnamefont
				{Brongersma}}, \bibinfo {author} {\bibfnamefont {Y.~S.}\ \bibnamefont
				{Kivshar}},\ and\ \bibinfo {author} {\bibfnamefont {B.}~\bibnamefont
				{Luk’yanchuk}},\ }\bibfield  {title} {\bibinfo {title} {Optically resonant
				dielectric nanostructures},\ }\href {https://doi.org/10.1126/science.aag2472}
		{\bibfield  {journal} {\bibinfo  {journal} {Science}\ }\textbf {\bibinfo
				{volume} {354}},\ \bibinfo {pages} {aag2472} (\bibinfo {year}
			{2016})}\BibitemShut {NoStop}%
		\bibitem [{\citenamefont {Krasnok}\ and\ \citenamefont
			{Alù}(2020)}]{KrasnokIEEE2020}%
		\BibitemOpen
		\bibfield  {author} {\bibinfo {author} {\bibfnamefont {A.}~\bibnamefont
				{Krasnok}}\ and\ \bibinfo {author} {\bibfnamefont {A.}~\bibnamefont {Alù}},\
		}\bibfield  {title} {\bibinfo {title} {Active nanophotonics},\ }\href
		{https://doi.org/10.1109/JPROC.2020.2985048} {\bibfield  {journal} {\bibinfo
				{journal} {Proc. IEEE}\ }\textbf {\bibinfo {volume} {108}},\ \bibinfo {pages}
			{628} (\bibinfo {year} {2020})}\BibitemShut {NoStop}%
		\bibitem [{Sup()}]{SupplementalMat}%
		\BibitemOpen
		\href@noop {} {}\bibinfo {note} {See Supplemental Material at
			http://link.aps.org/ supplemental/10.1103/PhysRevLett.XX.XXXXXX for further
			information about the empty lattice band formation, sample fabrication,
			finite element simulations, experimental setup and transmission measurements,
			derivation of the two band model and resulting analytical expressions, origin
			of Berry curvature and Stokes vector analysis, comparison of transmission and
			eigenmode calculations, and discussion on other two band models in the
			available literature.}\BibitemShut {Stop}%
		\bibitem [{\citenamefont {Moerland}\ \emph {et~al.}(2017)\citenamefont
			{Moerland}, \citenamefont {Hakala}, \citenamefont {Martikainen},
			\citenamefont {Rekola}, \citenamefont {V{\"a}kev{\"a}inen},\ and\
			\citenamefont {T{\"o}rm{\"a}}}]{QuantumPlasmonics}%
		\BibitemOpen
		\bibfield  {author} {\bibinfo {author} {\bibfnamefont {R.~J.}\ \bibnamefont
				{Moerland}}, \bibinfo {author} {\bibfnamefont {T.~K.}\ \bibnamefont
				{Hakala}}, \bibinfo {author} {\bibfnamefont {J.-P.}\ \bibnamefont
				{Martikainen}}, \bibinfo {author} {\bibfnamefont {H.~T.}\ \bibnamefont
				{Rekola}}, \bibinfo {author} {\bibfnamefont {A.~I.}\ \bibnamefont
				{V{\"a}kev{\"a}inen}},\ and\ \bibinfo {author} {\bibfnamefont
				{P.}~\bibnamefont {T{\"o}rm{\"a}}},\ }\bibfield  {title} {\bibinfo {title}
			{Strong coupling between organic molecules and plasmonic nanostructures},\
		}in\ \href {https://doi.org/10.1007/978-3-319-45820-5_6} {\emph {\bibinfo
				{booktitle} {Quantum Plasmonics}}},\ \bibinfo {editor} {edited by\ \bibinfo
			{editor} {\bibfnamefont {S.~I.}\ \bibnamefont {Bozhevolnyi}}, \bibinfo
			{editor} {\bibfnamefont {L.}~\bibnamefont {Mart{\'i}n-Moreno}},\ and\
			\bibinfo {editor} {\bibfnamefont {F.~J.}\ \bibnamefont {Garc{\'i}a-Vidal}}}\
		(\bibinfo  {publisher} {Springer International Publishing},\ \bibinfo
		{address} {Cham},\ \bibinfo {year} {2017})\ pp.\ \bibinfo {pages}
		{121--150}\BibitemShut {NoStop}%
		\bibitem [{\citenamefont {Bleu}\ \emph {et~al.}(2018)\citenamefont {Bleu},
			\citenamefont {Solnyshkov},\ and\ \citenamefont {Malpuech}}]{BleuPRB2018}%
		\BibitemOpen
		\bibfield  {author} {\bibinfo {author} {\bibfnamefont {O.}~\bibnamefont
				{Bleu}}, \bibinfo {author} {\bibfnamefont {D.~D.}\ \bibnamefont
				{Solnyshkov}},\ and\ \bibinfo {author} {\bibfnamefont {G.}~\bibnamefont
				{Malpuech}},\ }\bibfield  {title} {\bibinfo {title} {Measuring the quantum
				geometric tensor in two-dimensional photonic and exciton-polariton systems},\
		}\href {https://doi.org/10.1103/PhysRevB.97.195422} {\bibfield  {journal}
			{\bibinfo  {journal} {Phys. Rev. B}\ }\textbf {\bibinfo {volume} {97}},\
			\bibinfo {pages} {195422} (\bibinfo {year} {2018})}\BibitemShut {NoStop}%
		\bibitem [{\citenamefont {Zhen}\ \emph {et~al.}(2015)\citenamefont {Zhen},
			\citenamefont {Hsu}, \citenamefont {Igarashi}, \citenamefont {Lu},
			\citenamefont {Kaminer}, \citenamefont {Pick}, \citenamefont {Chua},
			\citenamefont {Joannopoulos},\ and\ \citenamefont
			{Soljačić}}]{ZhenNature2015}%
		\BibitemOpen
		\bibfield  {author} {\bibinfo {author} {\bibfnamefont {B.}~\bibnamefont
				{Zhen}}, \bibinfo {author} {\bibfnamefont {C.-W.}\ \bibnamefont {Hsu}},
			\bibinfo {author} {\bibfnamefont {Y.}~\bibnamefont {Igarashi}}, \bibinfo
			{author} {\bibfnamefont {L.}~\bibnamefont {Lu}}, \bibinfo {author}
			{\bibfnamefont {I.}~\bibnamefont {Kaminer}}, \bibinfo {author} {\bibfnamefont
				{A.}~\bibnamefont {Pick}}, \bibinfo {author} {\bibfnamefont {S.-L.}\
				\bibnamefont {Chua}}, \bibinfo {author} {\bibfnamefont {J.~D.}\ \bibnamefont
				{Joannopoulos}},\ and\ \bibinfo {author} {\bibfnamefont {M.}~\bibnamefont
				{Soljačić}},\ }\bibfield  {title} {\bibinfo {title} {Spawning rings of
				exceptional points out of {D}irac cones},\ }\href
		{https://doi.org/10.1038/nature14889} {\bibfield  {journal} {\bibinfo
				{journal} {Nature}\ }\textbf {\bibinfo {volume} {525}},\ \bibinfo {pages}
			{354} (\bibinfo {year} {2015})}\BibitemShut {NoStop}%
		\bibitem [{\citenamefont {Cuerda}\ \emph {et~al.}()\citenamefont {Cuerda},
			\citenamefont {Taskinen}, \citenamefont {Källman}, \citenamefont {Grabitz},\
			and\ \citenamefont {Törmä}}]{CuerdaPRB2023}%
		\BibitemOpen
		\bibfield  {author} {\bibinfo {author} {\bibfnamefont {J.}~\bibnamefont
				{Cuerda}}, \bibinfo {author} {\bibfnamefont {J.~M.}\ \bibnamefont
				{Taskinen}}, \bibinfo {author} {\bibfnamefont {N.}~\bibnamefont {Källman}},
			\bibinfo {author} {\bibfnamefont {L.}~\bibnamefont {Grabitz}},\ and\ \bibinfo
			{author} {\bibfnamefont {P.}~\bibnamefont {Törmä}},\ }\bibfield  {title}
		{\bibinfo {title} {Pseudospin-orbit coupling and non-{H}ermitian effects in the
				{Q}uantum {G}eometric {T}ensor of a plasmonic lattice, accepted for publication in {P}hysical
				{R}eview {B} (jointly submitted with this manuscript)},\ }\href
		{https://arxiv.org/abs/2305.13244} {\bibinfo  {journal} {arXiv:2305.13244
				}\ }\BibitemShut {NoStop}%
		\bibitem [{\citenamefont {Kavokin}\ \emph {et~al.}(2005)\citenamefont
			{Kavokin}, \citenamefont {Malpuech},\ and\ \citenamefont
			{Glazov}}]{KavokinPRL2005}%
		\BibitemOpen
		\bibfield  {journal} {  }\bibfield  {author} {\bibinfo {author} {\bibfnamefont
				{A.}~\bibnamefont {Kavokin}}, \bibinfo {author} {\bibfnamefont
				{G.}~\bibnamefont {Malpuech}},\ and\ \bibinfo {author} {\bibfnamefont
				{M.}~\bibnamefont {Glazov}},\ }\bibfield  {title} {\bibinfo {title} {Optical
				{S}pin {H}all {E}ffect},\ }\href
		{https://doi.org/10.1103/PhysRevLett.95.136601} {\bibfield  {journal}
			{\bibinfo  {journal} {Phys. Rev. Lett.}\ }\textbf {\bibinfo {volume} {95}},\
			\bibinfo {pages} {136601} (\bibinfo {year} {2005})}\BibitemShut {NoStop}%
		\bibitem [{\citenamefont {Poddubny}\ \emph {et~al.}(2014)\citenamefont
			{Poddubny}, \citenamefont {Miroshnichenko}, \citenamefont {Slobozhanyuk},\
			and\ \citenamefont {Kivshar}}]{PoddubnyACSPhot2014}%
		\BibitemOpen
		\bibfield  {author} {\bibinfo {author} {\bibfnamefont {A.}~\bibnamefont
				{Poddubny}}, \bibinfo {author} {\bibfnamefont {A.}~\bibnamefont
				{Miroshnichenko}}, \bibinfo {author} {\bibfnamefont {A.}~\bibnamefont
				{Slobozhanyuk}},\ and\ \bibinfo {author} {\bibfnamefont {Y.}~\bibnamefont
				{Kivshar}},\ }\bibfield  {title} {\bibinfo {title} {Topological {M}ajorana
				states in zigzag chains of plasmonic nanoparticles},\ }\href
		{https://doi.org/10.1021/ph4000949} {\bibfield  {journal} {\bibinfo
				{journal} {ACS Photonics}\ }\textbf {\bibinfo {volume} {1}},\ \bibinfo
			{pages} {101} (\bibinfo {year} {2014})}\BibitemShut {NoStop}%
		\bibitem [{\citenamefont {Raghu}\ and\ \citenamefont
			{Haldane}(2008)}]{RaghuPRA2008}%
		\BibitemOpen
		\bibfield  {author} {\bibinfo {author} {\bibfnamefont {S.}~\bibnamefont
				{Raghu}}\ and\ \bibinfo {author} {\bibfnamefont {F.~D.~M.}\ \bibnamefont
				{Haldane}},\ }\bibfield  {title} {\bibinfo {title} {Analogs of
				quantum-{H}all-effect edge states in photonic crystals},\ }\href
		{https://doi.org/10.1103/PhysRevA.78.033834} {\bibfield  {journal} {\bibinfo
				{journal} {Phys. Rev. A}\ }\textbf {\bibinfo {volume} {78}},\ \bibinfo
			{pages} {033834} (\bibinfo {year} {2008})}\BibitemShut {NoStop}%
		\bibitem [{\citenamefont {Haldane}\ and\ \citenamefont
			{Raghu}(2008)}]{HaldanePRL2008}%
		\BibitemOpen
		\bibfield  {author} {\bibinfo {author} {\bibfnamefont {F.~D.~M.}\
				\bibnamefont {Haldane}}\ and\ \bibinfo {author} {\bibfnamefont
				{S.}~\bibnamefont {Raghu}},\ }\bibfield  {title} {\bibinfo {title} {Possible
				realization of directional optical waveguides in photonic crystals with
				broken time-reversal symmetry},\ }\href
		{https://doi.org/10.1103/PhysRevLett.100.013904} {\bibfield  {journal}
			{\bibinfo  {journal} {Phys. Rev. Lett.}\ }\textbf {\bibinfo {volume} {100}},\
			\bibinfo {pages} {013904} (\bibinfo {year} {2008})}\BibitemShut {NoStop}%
		\bibitem [{\citenamefont {Wu}\ and\ \citenamefont {Hu}(2015)}]{WuPRL2015}%
		\BibitemOpen
		\bibfield  {author} {\bibinfo {author} {\bibfnamefont {L.-H.}\ \bibnamefont
				{Wu}}\ and\ \bibinfo {author} {\bibfnamefont {X.}~\bibnamefont {Hu}},\
		}\bibfield  {title} {\bibinfo {title} {Scheme for achieving a topological
				photonic crystal by using dielectric material},\ }\href
		{https://doi.org/10.1103/PhysRevLett.114.223901} {\bibfield  {journal}
			{\bibinfo  {journal} {Phys. Rev. Lett.}\ }\textbf {\bibinfo {volume} {114}},\
			\bibinfo {pages} {223901} (\bibinfo {year} {2015})}\BibitemShut {NoStop}%
		\bibitem [{\citenamefont {Guo}\ \emph {et~al.}(2019)\citenamefont {Guo},
			\citenamefont {Ne\ifmmode~\check{c}\else \v{c}\fi{}ada}, \citenamefont
			{Hakala}, \citenamefont {V\"akev\"ainen},\ and\ \citenamefont
			{T\"orm\"a}}]{GuoPRL2019}%
		\BibitemOpen
		\bibfield  {author} {\bibinfo {author} {\bibfnamefont {R.}~\bibnamefont
				{Guo}}, \bibinfo {author} {\bibfnamefont {M.}~\bibnamefont
				{Ne\ifmmode~\check{c}\else \v{c}\fi{}ada}}, \bibinfo {author} {\bibfnamefont
				{T.~K.}\ \bibnamefont {Hakala}}, \bibinfo {author} {\bibfnamefont {A.~I.}\
				\bibnamefont {V\"akev\"ainen}},\ and\ \bibinfo {author} {\bibfnamefont
				{P.}~\bibnamefont {T\"orm\"a}},\ }\bibfield  {title} {\bibinfo {title}
			{Lasing at {$K$} points of a honeycomb plasmonic lattice},\ }\href
		{https://doi.org/10.1103/PhysRevLett.122.013901} {\bibfield  {journal}
			{\bibinfo  {journal} {Phys. Rev. Lett.}\ }\textbf {\bibinfo {volume} {122}},\
			\bibinfo {pages} {013901} (\bibinfo {year} {2019})}\BibitemShut {NoStop}%
		\bibitem [{\citenamefont {Juarez}\ \emph {et~al.}(2022)\citenamefont {Juarez},
			\citenamefont {Li}, \citenamefont {Guan}, \citenamefont {Reese},
			\citenamefont {Schaller},\ and\ \citenamefont {Odom}}]{JuarezACSPHot2022}%
		\BibitemOpen
		\bibfield  {author} {\bibinfo {author} {\bibfnamefont {X.~G.}\ \bibnamefont
				{Juarez}}, \bibinfo {author} {\bibfnamefont {R.}~\bibnamefont {Li}}, \bibinfo
			{author} {\bibfnamefont {J.}~\bibnamefont {Guan}}, \bibinfo {author}
			{\bibfnamefont {T.}~\bibnamefont {Reese}}, \bibinfo {author} {\bibfnamefont
				{R.~D.}\ \bibnamefont {Schaller}},\ and\ \bibinfo {author} {\bibfnamefont
				{T.~W.}\ \bibnamefont {Odom}},\ }\bibfield  {title} {\bibinfo {title}
			{M-point lasing in hexagonal and honeycomb plasmonic lattices},\ }\href
		{https://doi.org/10.1021/acsphotonics.1c01618} {\bibfield  {journal}
			{\bibinfo  {journal} {ACS Photonics}\ }\textbf {\bibinfo {volume} {9}},\
			\bibinfo {pages} {52} (\bibinfo {year} {2022})}\BibitemShut {NoStop}%
		\bibitem [{\citenamefont {Goerlitzer}\ \emph {et~al.}(2020)\citenamefont
			{Goerlitzer}, \citenamefont {Mohammadi}, \citenamefont {Nechayev},
			\citenamefont {Volk}, \citenamefont {Rey}, \citenamefont {Banzer},
			\citenamefont {Karg},\ and\ \citenamefont {Vogel}}]{GoerlitzerAdvMat2020}%
		\BibitemOpen
		\bibfield  {author} {\bibinfo {author} {\bibfnamefont {E.~S.~A.}\
				\bibnamefont {Goerlitzer}}, \bibinfo {author} {\bibfnamefont
				{R.}~\bibnamefont {Mohammadi}}, \bibinfo {author} {\bibfnamefont
				{S.}~\bibnamefont {Nechayev}}, \bibinfo {author} {\bibfnamefont
				{K.}~\bibnamefont {Volk}}, \bibinfo {author} {\bibfnamefont {M.}~\bibnamefont
				{Rey}}, \bibinfo {author} {\bibfnamefont {P.}~\bibnamefont {Banzer}},
			\bibinfo {author} {\bibfnamefont {M.}~\bibnamefont {Karg}},\ and\ \bibinfo
			{author} {\bibfnamefont {N.}~\bibnamefont {Vogel}},\ }\bibfield  {title}
		{\bibinfo {title} {Chiral surface lattice resonances},\ }\href
		{https://doi.org/https://doi.org/10.1002/adma.202001330} {\bibfield
			{journal} {\bibinfo  {journal} {Adv. Mater.}\ }\textbf {\bibinfo {volume}
				{32}},\ \bibinfo {pages} {2001330} (\bibinfo {year} {2020})}\BibitemShut
		{NoStop}%
		\bibitem [{\citenamefont {Kataja}\ \emph {et~al.}(2015)\citenamefont {Kataja},
			\citenamefont {Hakala}, \citenamefont {Julku}, \citenamefont {Huttunen},
			\citenamefont {van Dijken},\ and\ \citenamefont
			{T{\"o}rm{\"a}}}]{KatajaNatComm2015}%
		\BibitemOpen
		\bibfield  {author} {\bibinfo {author} {\bibfnamefont {M.}~\bibnamefont
				{Kataja}}, \bibinfo {author} {\bibfnamefont {T.~K.}\ \bibnamefont {Hakala}},
			\bibinfo {author} {\bibfnamefont {A.}~\bibnamefont {Julku}}, \bibinfo
			{author} {\bibfnamefont {M.~J.}\ \bibnamefont {Huttunen}}, \bibinfo {author}
			{\bibfnamefont {S.}~\bibnamefont {van Dijken}},\ and\ \bibinfo {author}
			{\bibfnamefont {P.}~\bibnamefont {T{\"o}rm{\"a}}},\ }\bibfield  {title}
		{\bibinfo {title} {Surface lattice resonances and magneto-optical response in
				magnetic nanoparticle arrays},\ }\href {https://doi.org/10.1038/ncomms8072}
		{\bibfield  {journal} {\bibinfo  {journal} {Nat. Commun.}\ }\textbf {\bibinfo
				{volume} {6}},\ \bibinfo {pages} {7072} (\bibinfo {year} {2015})}\BibitemShut
		{NoStop}%
		\bibitem [{\citenamefont {Freire-Fernández}\ \emph {et~al.}(2022)\citenamefont
			{Freire-Fernández}, \citenamefont {Cuerda}, \citenamefont {Daskalakis},
			\citenamefont {Perumbilavil}, \citenamefont {Martikainen}, \citenamefont
			{Arjas}, \citenamefont {Törmä},\ and\ \citenamefont {van
				Dijken}}]{FreireNatPhot2022}%
		\BibitemOpen
		\bibfield  {author} {\bibinfo {author} {\bibfnamefont {F.}~\bibnamefont
				{Freire-Fernández}}, \bibinfo {author} {\bibfnamefont {J.}~\bibnamefont
				{Cuerda}}, \bibinfo {author} {\bibfnamefont {K.~S.}\ \bibnamefont
				{Daskalakis}}, \bibinfo {author} {\bibfnamefont {S.}~\bibnamefont
				{Perumbilavil}}, \bibinfo {author} {\bibfnamefont {J.-P.}\ \bibnamefont
				{Martikainen}}, \bibinfo {author} {\bibfnamefont {K.}~\bibnamefont {Arjas}},
			\bibinfo {author} {\bibfnamefont {P.}~\bibnamefont {Törmä}},\ and\ \bibinfo
			{author} {\bibfnamefont {S.}~\bibnamefont {van Dijken}},\ }\bibfield  {title}
		{\bibinfo {title} {Magnetic on–off switching of a plasmonic laser},\ }\href
		{https://doi.org/10.1038/s41566-021-00922-8} {\bibfield  {journal} {\bibinfo
				{journal} {Nat. Photon.}\ }\textbf {\bibinfo {volume} {16}},\ \bibinfo
			{pages} {27} (\bibinfo {year} {2022})}\BibitemShut {NoStop}%
		\bibitem [{\citenamefont {Maccaferri}\ \emph {et~al.}(2023)\citenamefont
			{Maccaferri}, \citenamefont {Gabbani}, \citenamefont {Pineider},
			\citenamefont {Kaihara}, \citenamefont {Tapani},\ and\ \citenamefont
			{Vavassori}}]{MaccaferriAPL2023}%
		\BibitemOpen
		\bibfield  {author} {\bibinfo {author} {\bibfnamefont {N.}~\bibnamefont
				{Maccaferri}}, \bibinfo {author} {\bibfnamefont {A.}~\bibnamefont {Gabbani}},
			\bibinfo {author} {\bibfnamefont {F.}~\bibnamefont {Pineider}}, \bibinfo
			{author} {\bibfnamefont {T.}~\bibnamefont {Kaihara}}, \bibinfo {author}
			{\bibfnamefont {T.}~\bibnamefont {Tapani}},\ and\ \bibinfo {author}
			{\bibfnamefont {P.}~\bibnamefont {Vavassori}},\ }\bibfield  {title} {\bibinfo
			{title} {Magnetoplasmonics in confined geometries: Current challenges and
				future opportunities},\ }\href {https://doi.org/10.1063/5.0136941} {\bibfield
			{journal} {\bibinfo  {journal} {Appl. Phys. Lett.}\ }\textbf {\bibinfo
				{volume} {122}},\ \bibinfo {pages} {120502} (\bibinfo {year}
			{2023})}\BibitemShut {NoStop}%
	\end{thebibliography}
\end{document}


\renewcommand{\theequation}{S\arabic{equation}}
	\renewcommand{\thefigure}{S\arabic{figure}}
	
	\preprint{APS/123-QED}
	
	\title{Supplemental Material:\\Observation of Quantum metric and non-Hermitian Berry curvature in a plasmonic lattice}
	
	\author{Javier Cuerda}
	\email{javier.cuerda@aalto.fi}
	\author{Jani M. Taskinen}
	\author{Nicki Källman}
	\author{Leo Grabitz}
	\author{Päivi Törmä}%
	\email{paivi.torma@aalto.fi}
	\affiliation{%
		Department of Applied Physics, Aalto University School of Science, Aalto FI-00076, Finland
	}%

	\date{\today}
	
	\setlength{\parskip}{0pt} 


	\maketitle

	\section{Empty lattice band degeneracy at the diagonal of the Brillouin zone}\label{sec:ela}
\begin{figure*}
		\includegraphics[width=0.9\textwidth]{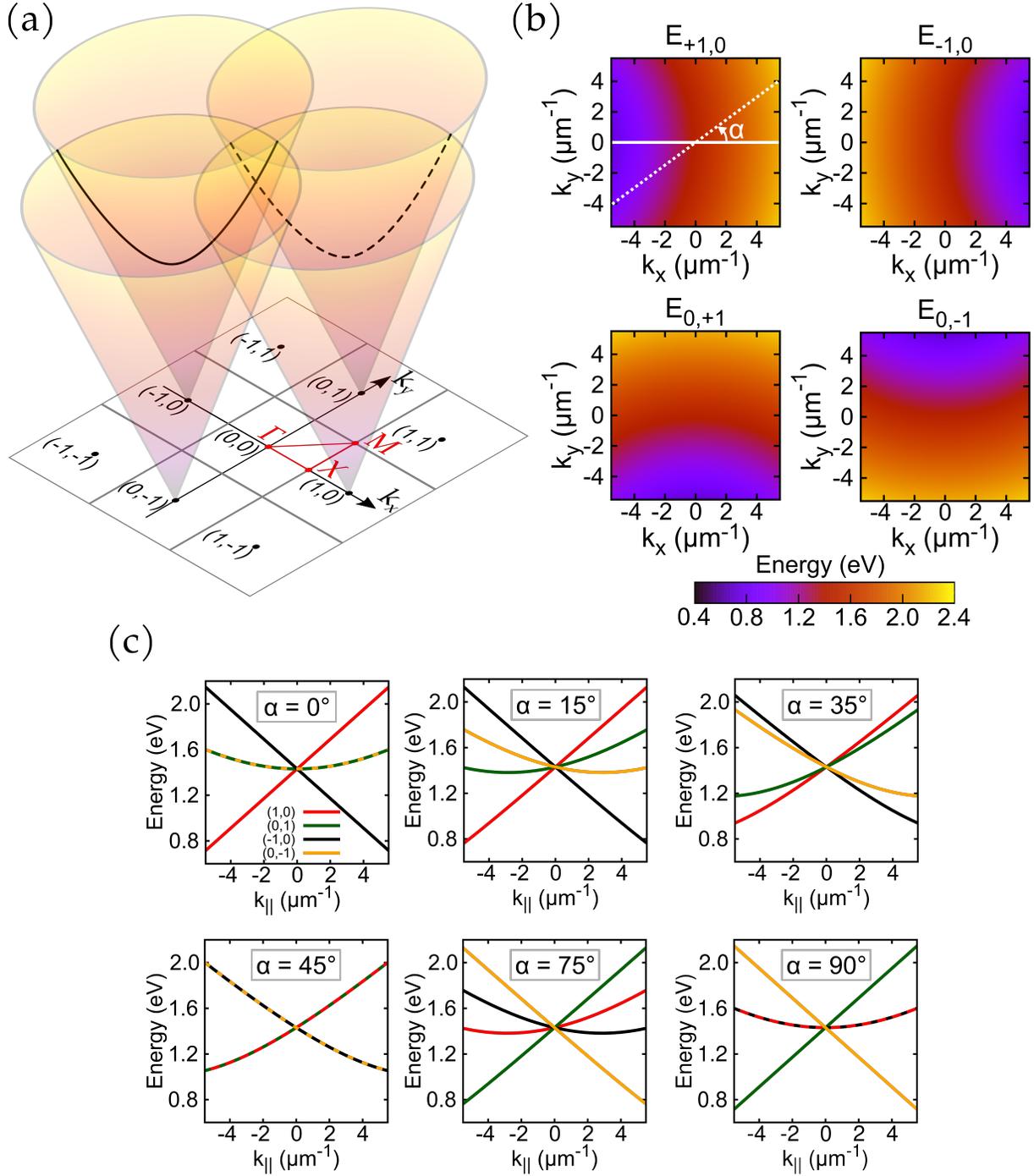}
		\caption{\label{fig:ela}Band formation in the empty lattice approximation and degeneracy in the diagonal of the first Brillouin zone. (a) Diffraction order positions of the lattice in $k-$space labeled as $(m,n)$, where $\mathbf{G}_{mn}=mG_{x}\hat{\mathbf{k}}_{x}+nG_{y}\hat{\mathbf{k}}_{y}$ is the reciprocal vector, and $G_{x,y}\equiv G=2\pi/p$. The first Brillouin zone is defined by the center square, and relevant high-symmetry points and trajectories are shown in red. Empty lattice dispersion bands emerge from the intersection of trajectories in reciprocal space with the diffraction cones centered at each reciprocal vector. The pairs of diffraction cones $(0,-1)$, $(-1,0)$ and $(0,+1)$, $(+1,0)$ are equidistant from the diagonal of the first Brillouin zone $\Gamma-M$, thus these four empty lattice modes become two doubly degenerate bands (dashed and solid black lines) along that trajectory, as explained in the following panels. (b) Energy distribution of empty lattice modes $(\pm 1,0)$ and $(0,\pm 1)$ within the first Brillouin zone. (c) Empty lattice band formation within the first Brillouin zone, along straight trajectories in $k-$space that cross through the $\Gamma$-point at an angle $\alpha$, as defined by the white dotted line in panel (b).}
	\end{figure*}

	The band structure of SLR modes in the two-dimensional $k-$space is mainly ruled by their photonic component related to the lattice diffraction orders, and the main effect of single nanoparticle LSPRs is to couple the diffraction orders by scattering and thus open bandgaps at high-symmetry points of the Brillouin zone \cite{GuoPRL2019,RodriguezPRX2011,JuarezACSPHot2022}. Therefore, as a first approach to understanding the lattice modes involved in the QGT extraction, we use the empty lattice approximation that contains information on the lattice geometry but not on the metallic nanoparticles; see Fig.~\ref{fig:ela}.
	
	We examine the optical response of an infinite square lattice with a period of $p=570$ nm embedded in a host with a refractive index of $n_{h}=1.52$. The formation of energy bands in a plasmonic lattice relies on the long-range radiative coupling between nanoparticles, hence nearest-neighbor approximation between unit cells cannot be assumed. Instead, we model the lattice considering the interplay of diffraction orders in the first Brillouin zone that gives rise to the energy band dispersions of the system. The band structure is characterized by light cones centered at the reciprocal vector $\mathbf{G}_{mn}$ of each diffraction order $(m,n)$ of the lattice (see Fig.~\ref{fig:ela}(a))~\cite{GuoPRB2017,QuantumPlasmonics}. 
	The folding of bands within the first Brillouin zone, according to the intuitive picture in Fig.~\ref{fig:ela}(a), gives rise to four empty lattice dispersions shown in Fig.~\ref{fig:ela}(b) as given by Eq.~(1) in the main text. The dispersion profiles in Fig.~\ref{fig:ela}(b) are related with each other through the four-fold symmetry of the Brillouin zone, and we identify four high-symmetry axes in the $(k_{x},k_{y})$ plane: horizontal ($k_{y}=0$), vertical ($k_{x}=0$), diagonal ($k_{x}=k_{y}$) and antidiagonal ($k_{x}=-k_{y}$). Therefore we may calculate the modes only within the irreducible Brillouin zone, defined by the angle range $0\degree\leq\alpha\leq45\degree$ in Fig.~\ref{fig:ela}(b); the empty lattice modes in the rest of the Brillouin zone are then found by simple symmetry rules. As explained in the main text, we may exploit these symmetry relations to extrapolate the results of the QGT components in Fig.~2(d) to the whole Brillouin zone. Further discussion on this is found in Ref.~\cite{CuerdaPRB2023}.
	
	Fig.~\ref{fig:ela}(c) shows the empty lattice bands for various cross sections of the energy profiles in Fig.~\ref{fig:ela}(b) with trajectories in $k-$space across the $\Gamma$-point defined as $\mathbf{k}=k_{||}(\cos\alpha,\sin\alpha)$. For $\mathbf{k}$ pointing along the high-symmetry axis $k_{y}=0$ (that is, for $\alpha=0\degree$), two diffraction orders $(m,n)=(\pm1,0)$ present a linear dispersion whereas $(0,\pm 1)$ become degenerate with a parabolic dependence on $k_{||}$. For $k_{x}=0$ ($\alpha=90\degree$), this behavior holds due to the symmetry of the square lattice, but now modes $(0,\pm 1)$ are linear, and $(\pm1,0)$ are parabolic and degenerate. The behavior for $\alpha=0\degree$ and $90\degree$ is prototypical of square plasmonic lattices and has been widely studied \cite{KravetsChemRev2018,GuoPRB2017,AuguiePRL2008,SchokkerPRB2017,RodriguezPRX2011}. More generally, since the diagonal of the Brillouin zone is a high-symmetry axis, same dispersions for the angle $\alpha<45\degree$ are found for $\alpha'=90-\alpha$ with empty lattice modes transforming as follows: $(0,\pm1)\rightarrow(\pm 1,0)$ and $(\pm 1,0)\rightarrow(0,\pm1)$ for $\alpha\rightarrow\alpha'$ (an example of this are cases $\alpha=15\degree$ and $\alpha'=75\degree$ in Fig.~\ref{fig:ela}(c)). Accordingly, the modes $(0,\pm1)$ are found to be degenerate with $(\pm1,0)$, respectively, for $\alpha=45\degree$, that is, the diagonal of the first Brillouin zone.
	
	\section{Plasmonic sample fabrication}\label{appendix:samples}
	
	The plasmonic samples consist of cylindrical gold nanoparticles arranged in a square array geometry (see Fig.~1(c) in the main text). The periodicity is set to 570~nm, and the nominal radius and height of the particles are 75~nm and 50~nm, respectively. The SLR modes of the plasmonic structures are captured using a transmission setup, which allows us to image polarization-resolved crosscuts of the sample dispersion in ($k_{||},E$)-space. The 2D reciprocal space is reconstructed by measuring different trajectories through the $\Gamma$-point as a function of the angle $\alpha$ (defined in Fig.~2(a)). We measure an angle range of $-1 \degree \leq \alpha \leq 46\degree$ to guarantee the inclusion of the diagonal, and the symmetry relations mentioned in Section \ref{sec:ela} are utilized to construct a complete dataset around the $\Gamma$-point. 
	
	The nanoparticles are fabricated on a $75.6\times 25.0 \times 1.0$~mm$^{3}$ glass substrate whose refractive index is $n_h = 1.52$. A positive resist (PMMA~A4) is spin coated on the substrate at 3000~RPM and solidified on a hot plate at 175~$^\circ$C for 120~s. A 10~nm layer of aluminum is evaporated on the resist as a conduction layer for electron beam lithography. After the lithography process, the aluminum layer is removed by a 90~s immersion in a 50~\% solution of AZ~351~B developer, and the patterned resist layer is developed in a solution of 1:3 MIBK:IPA for 25~s. A 2~nm adhesion layer of titanium and a 50~nm layer of gold are evaporated on the sample, and finally the excess resist and metal films are removed in an overnight acetone bath. The size of the resulting arrays is 200~$\times$~200~µm$^2$. The fabrication process excluding the final step is carried out in cleanroom conditions. The samples are prepared for the transmission measurements by immersing them in index-matching oil between the substrate and a second glass slide.
	\section{Finite element simulation of the modes at the diagonal and at the $\Gamma$-point}\label{supp:FEM}
	Here we provide realistic numerical simulations and a thorough explanation that support the qualitative physics outlined in Figs.~1(a) and 2(c). A plasmonic lattice such as the one studied in this work sustains modes called surface lattice resonances (SLRs). SLRs are Bloch waves because the system is periodic, therefore the electric field $\mathbf{E}$ has a spatial dependence within a unit cell such that
	\begin{equation}\label{sup:wavefunction}
		\mathbf{E}(\mathbf{r})=\mathbf{A}(\mathbf{r})e^{i\mathbf{k}_{mn}\cdot\mathbf{r}}\textrm{,} 
	\end{equation}
	where $\mathbf{k}_{mn}=\mathbf{k}_{||}+\mathbf{G}_{mn}$. The wavevector $\mathbf{k}_{||}$ points along the lattice plane and $\mathbf{G}_{mn}=mG_{x}\hat{\mathbf{x}}+nG_{y}\hat{\mathbf{y}}$ is the reciprocal vector of the lattice, with $G_{x,y}=2\pi/p$ and $p$ is the square lattice period. 
	
	From Eq.~\eqref{sup:wavefunction}, we can explain the main features of the SLR modes illustrated in Fig.~1(a). SLRs come about as a combination of the single particle response, acting as a small (dipolar) antenna, and the diffraction orders imposed by the lattice periodicity. Importantly, the orientation of the dipoles depends on the polarization. While the field distribution of the dipoles is encoded in the complex-valued amplitude $\mathbf{A}(\mathbf{r})$ in Eq.~\eqref{sup:wavefunction}, the phase $\varphi(\mathbf{r})\equiv\mathbf{k}_{mn}\cdot\mathbf{r}$ that is presented in Fig.~1(a) describes the diffraction orders of the lattice. For $y-$polarized incoming far-field radiation (indicated as a purple arrow in the left panel of Fig.~1(a)), the dipoles radiate in the $x-$direction and only the degenerate $(m,n)=(\pm1,0)$ diffraction orders can be excited since $(0,\pm1)$ propagate orthogonally to the radiative direction (see Ref.~\cite{GuoPRB2017} for more details). Therefore, at the $\Gamma$-point, $k_{||}=0$, the dipoles oscillate in phase with each other, forming a standing wave:
	\begin{equation}\label{sup:standingwave}
		\mathbf{E}(\mathbf{r})=\mathbf{A}(\mathbf{r})(e^{iG_{x}x}+e^{-iG_{x}x})=2\mathbf{A}(\mathbf{r})\cos(G_{x}x)\textrm{.}
	\end{equation}
	\begin{figure}[htp!]
		\centering
		\includegraphics[width=0.355\textwidth]{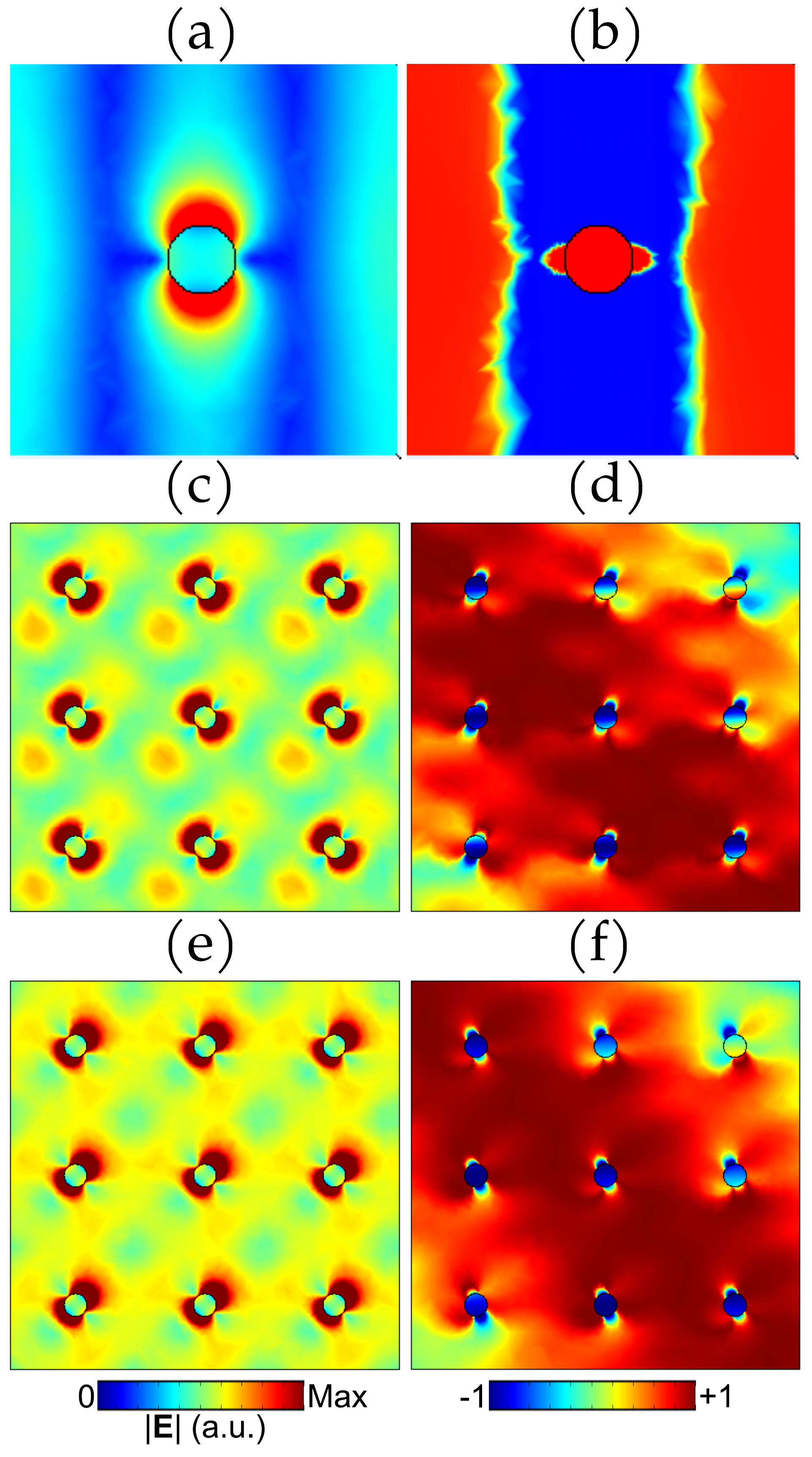}
		\caption{\label{fig:FEM} 
			Finite element simulations showing the in-plane $E$-field and phase of the considered SLR modes at (a),(b) the $\Gamma$-point (schematically shown in Fig.~1(a) of the main text) and (c)-(f) at the diagonal of the Brillouin zone (schematic in Fig.~2(c) of the main paper). A realistic three-dimensional structure is simulated, and the results are shown at the lattice plane (at the mid-height of the nanoparticles). In panels (c)-(f), several unit cells are simulated to observe the interplay of the feedback given by the diffraction orders with the dipolar fields at the individual nanoparticle level, and to allow the buildup of the phase shown schematically in Fig.~2(c) of the main paper. (a) Electric field norm $|\mathbf{E}|$ of the SLR mode. (b) Phase of the in-plane electric field $\Phi(E_{y})\equiv E_{y}/|E_{y}|$. As illustrated by Fig.~1(a), the panels (a),(b) above correspond to both the TE and the TM mode at the $\Gamma$-point, since they are the $\pi/2$-rotated version of one another by the symmetry of the square lattice. Note that the actual phase profile of (b) is very different from the one that is shown by color in Fig.~1(a) of the main paper - the latter is only a schematic illustration. (c) Field profile ($|\mathbf{E}|$) of the TE-polarized mode and (d) phase $\Phi_{0,-1}$ of this mode. (d) Field profile ($|\mathbf{E}|$) of the TM-polarized mode and (e) phase $\Phi_{-1,0}$ of this mode. The two modes in (c) and in (e) are not identical upon rotation. The corresponding phase shown in (d) and (f) are quite similar in the large scale
			because they mainly depend on the $\mathbf{k}_{||}$, which is the same in both cases – however,
			around the nanoparticles there is a crucial difference between the modes (for point-like objects instead of finite size
			nanoparticles, this difference would disappear). Periodic boundary conditions are set in (c)-(f) such that a parallel wavevector is pointing towards the main diagonal of the Brillouin zone: $k_{||}=(k_{x},k_{y})$, with $k_{x}=k_{y}=k_{0}$ and $k_{0}=1$ $\mu$m$^{-1}$.}
		
	\end{figure}
	Fig.~\ref{fig:FEM}(a), showing full numerical finite element (FEM) simulations at the $\Gamma$-point, reveals a field profile which is a combination of the dipole excitation at the single particle level and intensity maxima corresponding to the standing wave in Eq.~\eqref{sup:standingwave} and Fig.~1(a). The phase distribution $\varphi(\mathbf{r})$ in real space may be calculated as follows: $\Phi(\mathbf{r})\equiv e^{i\varphi(\mathbf{r})}=E_{y}/|E_{y}|$ (shown in Fig.~\ref{fig:FEM}(b)), where $E_{y}$ is the $y$-component of the electric field $\mathbf{E}$. Both Figs.~\ref{fig:FEM}(a) and (b) show good agreement with the schematic picture in Fig.~1(a).
	
	Next, we analyze the SLR modes away from the $\Gamma$-point, at the main diagonal of the Brillouin zone: $\mathbf{k}_{||}=(k_{x},k_{y})$, with $k_{x}=k_{y}=k_{0}$ and $k_{0}=1$ $\mu$m$^{-1}$. As explained in the main text, our SLR modes are formed from the empty lattice modes $(m,n)=(0,-1)$ and $(-1,0)$, thus the field profile in Eq.~\eqref{sup:wavefunction} has the following form:
	\begin{align}
		\mathbf{E}_{(-1,0)}(\mathbf{r})&=\mathbf{A}_{1}(\mathbf{r})e^{-iG_{x}x}e^{i\mathbf{k}_{||}\cdot\mathbf{r}},\label{sup:femdiag1}\\
		\mathbf{E}_{(0,-1)}(\mathbf{r})&=\mathbf{A}_{2}(\mathbf{r})e^{-iG_{y}y}e^{i\mathbf{k}_{||}\cdot\mathbf{r}},\label{sup:femdiag2}
	\end{align}
	with complex-valued vectorial amplitudes $\mathbf{A}_{1,2}(\mathbf{r})$. 
	
	Figs.~\ref{fig:FEM}(c),(e) show the electric field norm $|\mathbf{E}|$ from the FEM numerical simulations corresponding to the two modes found at the diagonal of the Brillouin zone. The electric field is characterized by dipolar patterns that agree well with Fig.~2(c) of the main text. Additional features in the electric field norm result (as revealed by comparison with Eqs.~\eqref{sup:femdiag1}, \eqref{sup:femdiag2}) from the relevant diffraction orders and from the $\mathbf{k}_{||}$-vector. 
	
	Given the expressions \eqref{sup:femdiag1} and \eqref{sup:femdiag2} of the considered modes, the numerical counterpart of the phase $\varphi'(\mathbf{r})\equiv\mathbf{k}_{||}\cdot\mathbf{r}$ illustrated in Fig.~2(c) is defined by the phase-dependent spatial distribution $\Phi_{mn}(\mathbf{r})\equiv e^{i\varphi'(\mathbf{r})}=e^{i\mathbf{G}_{mn}\cdot\mathbf{r}}\Phi(\mathbf{r})$. Figs.~\ref{fig:FEM}(d) and (f) show $\Phi_{0,-1}(\mathbf{r})$ and $\Phi_{-1,0}(\mathbf{r})$, that correspond to the field profiles of Figs.~\ref{fig:FEM}(c) and (e), respectively, and $\Phi(\mathbf{r})=E_{x}/|E_{x}|$. The field profiles are clearly different. The phase distributions have a similar large scale structure as it is determined by the same non-zero momentum in both cases, but there are relevant differences close to the nanoparticles. Indeed these small features, which would be absent for point-like particles, make the two modes to be different even after $\pi/2$-rotation and lift the degeneracy. All these findings are consistent with the mechanism proposed in Fig.~2(c) of the main text.
	\vspace{1cm}
	\section{Experimental setup and transmission measurements}\label{supp:exp_setup_transm}
	
\begin{figure}[htp!]
		\includegraphics[width=1\columnwidth]{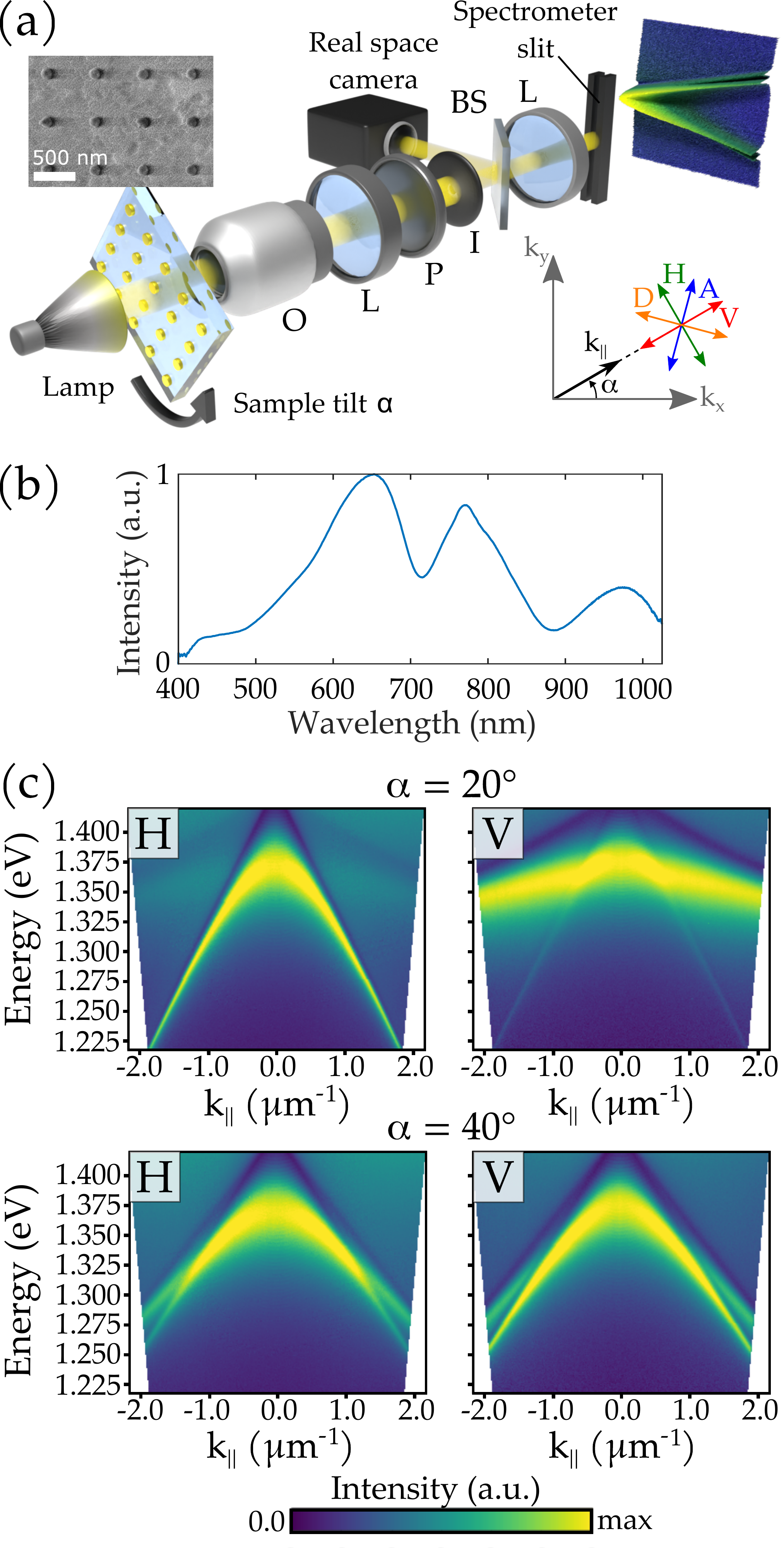}
		\caption{\label{fig:transm_exp} Experimental setup and results of the transmission measurements. (a) Schematic representation of our experimental setup for polarization-resolved transmission measurements. Here, O stands for objective, L for lens, P for polarizer, I for iris, and BS for beamsplitter. Linear polarizations are defined with respect to the sample tilt angle $\alpha$, see lower right inset. A scanning electron micrograph of the experimental sample is provided. (b) Normalized intensity spectrum of the halogen light source used in the transmission measurements. (c) Results of the transmission measurements for different sample tilt angles for horizontal (H) and vertical (V) polarization filters. The color scale values are calculated as $1 - T$, where $T$ is transmission.} 
		\vspace*{-2mm}
	\end{figure}

	A diagram of the experimental setup used in the transmission experiments is shown in Fig.~\ref{fig:transm_exp}(a). The sample is illuminated with diffused and focused white light, and the transmitted light is collected using an objective (Nikon Plan Fluor 10x, 0.3 NA). The light source is a commercial halogen lamp whose broadband spectrum is shown in Fig.~\ref{fig:transm_exp}(b). The total power density of the light source in the wavelength range 400 – 1025~nm is approximately 200~mW/cm$^2$. A two-lens system is used to focus a magnified image of the reciprocal plane to a spectrometer slit, and an iris placed in the image plane is used to spatially filter out light beyond the nanoparticle lattice. A real space camera is used to monitor the sample via a beamsplitter, and the six different polarization states used in the QGT extraction are resolved with a polarizer. As we focus the reciprocal plane of the lattice to the spectrometer (IsoPlane SCT320), each position on the slit corresponds to a transmission angle $\theta_t$ at the sample. For each wavelength $\lambda_0$, the in-plane wavevector is given by $k_{||} = 2\pi/\lambda_0 \sin (\theta_t)$. The spectrometer resolves the light spectrum at each point on the slit on to a 2D CCD camera (PI Pixis 2048), where each pixel row and column correspond to a unique value of $\theta_t$ and $\lambda_0$, respectively. The maximum detectable angle is limited to 17.4\degree~by the numerical aperture of the objective. The setup enables capturing energy-resolved crosscuts of the sample dispersions in reciprocal space.
	
	In order to calculate the Stokes vector components required for the QGT extraction, the transmission measurement for each value of $\alpha$ is performed with four linear and two circular polarization configurations: horizontal, vertical, diagonal, antidiagonal, and right and left circular polarization. The linear polarization planes are fixed with the orientation of the spectrometer slit used in the setup.  
	This is achieved by fabricating multiple rotated copies of the array on the same substrate, enabling us to effectively rotate the structure with discrete steps as we measure the otherwise identical copies of the original design.

	The results from the transmission experiments in Fig.~\ref{fig:transm_exp}(c) reveal the polarization properties of the SLR bands, which are consistent with the findings in Section \ref{sec:2bandmodel}. As discussed above, the SLRs closely follow the empty lattice dispersions of the modes $(0,-1)$ (higher energy mode) and $(-1,0)$ (lower energy mode), in close agreement with Eq.~(1) in the main text. Along the $k_{x}-$axis ($\alpha=0\degree$ in Figs.~\ref{fig:ela}(b),(c) and \ref{fig:transm_exp}(a)), the bands have a parabolic and linear dependence on $k_{||}$ and are TM and TE polarized, respectively, as in Ref.~\cite{GuoPRB2017}. As the sample is rotated, $\alpha$ increases and the SLRs do not behave purely as TM or TE. For instance, horizontally polarized light mainly couples to the lower energy mode for $\alpha=20\degree$, but the upper energy band is also visible. Conversely, filtering of vertical polarization yields a strong higher energy band, with a minor but clear coupling to the lower energy SLR. The transmission measurements in Fig.~\ref{fig:transm_exp}(c) show that horizontal light couples increasingly to the higher energy mode as $k_{||}$ rotates in the reciprocal space towards the diagonal of the Brillouin zone, and decreasingly to the lower energy mode; an opposite behavior is found for vertical polarization. Close to the diagonal ($\alpha=40\degree$), both polarizations yield similar intensities for both modes, and the band energies become close to degenerate. 
	
	Our measurement procedure involves accessing the intensity of the two analyzed modes for each of the six relevant polarizations. We numerically tracked the intensity maximum $I_{max}$ of each band, as those shown in Fig.~\ref{fig:transm_exp}(c), as a function of $k_{||}$. The intensity of the bands evolves smoothly with the angle $\alpha$ as explained above, and, as a consequence, changes in the intensity for very close values of $\alpha$ may be smaller than the noise level of the measurements. To avoid the presence of non-physical effects due to noise in our QGT results, we utilize experimental results of the band intensity maximum $I_{max}(k_{||})$ for several values of $\alpha$ sufficiently spaced from each other, and interpolate the results to form an intensity grid with a spacing of $0.5\degree$ that constitutes the input for the QGT computation. We have verified that the experimentally extracted QGT remains robust independently from the chosen spacing for the grid parametrizing the $k$-space. On the other hand, the SLR modes are characterized by a finite width caused by the inherent losses of the system and consequently the bands cannot be resolved near the $\Gamma$-point where their energy difference becomes smaller than the bandwidth. The quantum metric and Berry curvature are ill-defined for degenerate bands, e.g. see Eq.~\eqref{qgt_alternativeform}, thus the experimental extraction takes place away from the $\Gamma$-point where the bands are well-resolved.
	
	The QGT calculation procedure from the experimentally obtained Stokes vector components ($S_{x}(\mathbf{k}),S_{y}(\mathbf{k}),S_{z}(\mathbf{k})$) is adapted from Refs.~\cite{BleuPRB2018,GianfrateNature2020}. The Bloch sphere angles $\theta(\mathbf{k})$ and $\phi(\mathbf{k})$ given by \eqref{angles} and the generic form of the system eigenstate \eqref{quantumstate} are inserted into the QGT definition in Eq.~(2) of the main text, which leads to the quantum metric and Berry curvature formulae:
	\begin{align}
		g_{ij}^{n}=\Re T_{ij}^{n} = \frac{1}{4}(\partial_{i} \theta \partial_{j}\theta + \sin^2{\theta}\partial_{i}\phi\partial_{j}\phi),\label{quantumetric_transm}\\
		\mathfrak{B}_{ij}^{n}=-2\Im T_{ij}^{n} = \frac{1}{2}\sin{\theta}(\partial_{i}\theta \partial_{j}\phi - \partial_{j}\theta \partial_{i}\phi).\label{berrycurv_transm}
	\end{align}
	These components are then evaluated numerically from the experimental data. 

	\begin{figure}[!ht]
		\includegraphics[width=1\columnwidth]{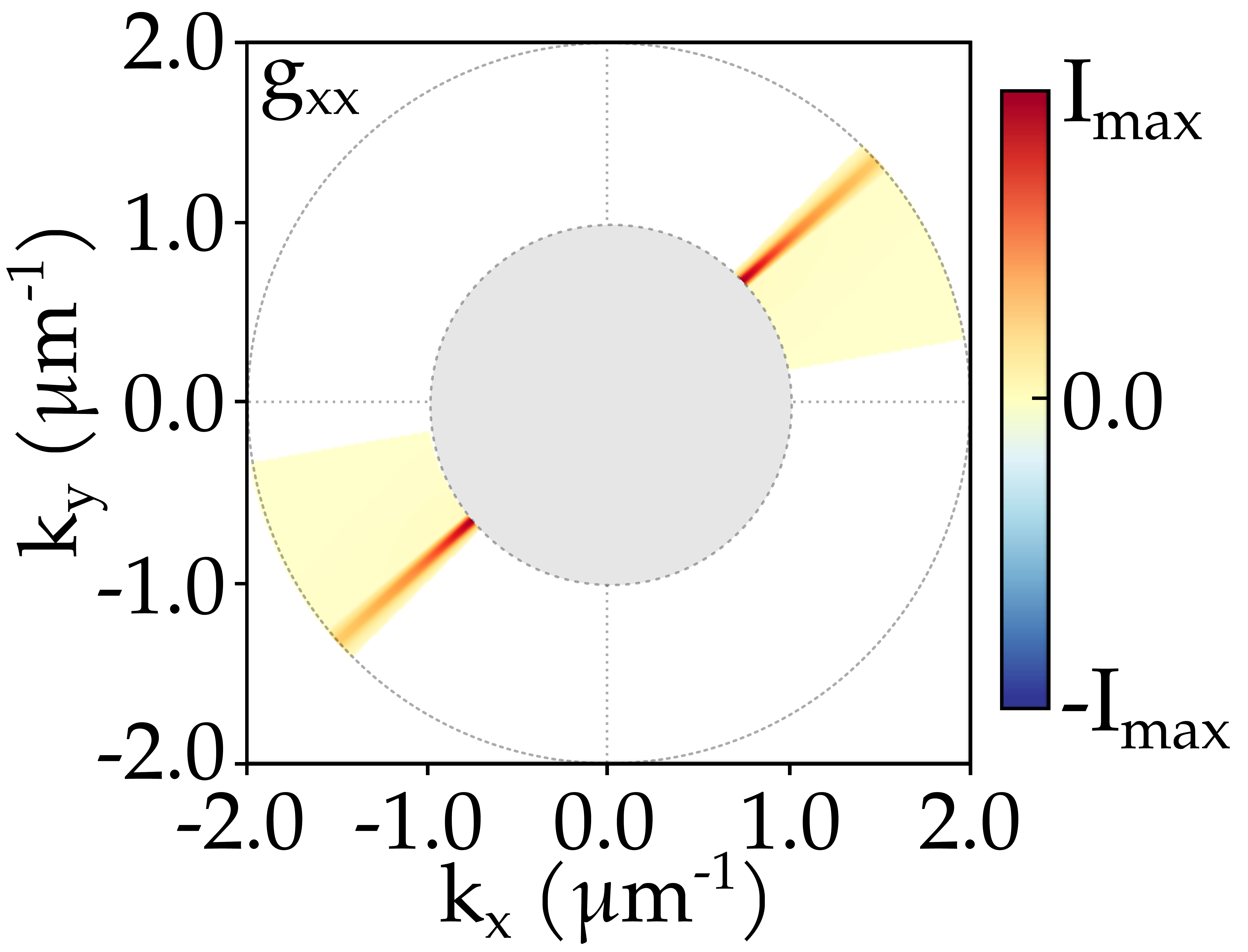}
		\caption{\label{fig:qgt_exp} General view of the reciprocal space, and component $g_{xx}$ of the quantum metric for the higher energy mode. We show results for slit angles ranging from $\alpha=10\degree$ to $45\degree$. Extension of these results to the rest of the quadrants of the Brillouin zone is immediate by the symmetry of the square lattice. The grey circle indicates points in $k$-space where band touchings are produced, preventing the calculation of the QGT. Results for $\alpha=0\degree$ are not included, since the modal dispersion relation captured by the measurements lies at a very different range of $k_{||}$ than for $\alpha\approx45\degree$.
		}
	\end{figure}
	
	\section{Derivation of the two-band model}\label{sec:2bandmodel}
	We next provide a simple two-band model to intuitively interpret the experimental QGT results. The two bands correspond to the two polarization directions of the plasmonic-photonic modes, and the dispersion of the SLR modes is encoded in the model in a simplified form. We start by considering the coupling of far-field light polarization with the two TE/TM SLR modes studied in Ref.~\cite{CuerdaPRB2023}. 
	We introduce a minimal two-band model consisting in a $2\times 2$ Hamiltonian matrix with momentum-dependent parameters \cite{KavokinPRL2005}:
	\begin{equation}\label{2bandHam}
		\hat{H}=\epsilon(\mathbf{k})I_{2\times 2}+\mathbf{\Omega}(\mathbf{k})\cdot\boldsymbol{\sigma},
	\end{equation}
	where $\boldsymbol{\sigma}\equiv(\sigma_{x},\sigma_{y},\sigma_{z})$ is a vector containing the Pauli matrices:
	\begin{align}
		\sigma_{x}=
		\begin{pmatrix}
			0 & 1 \\
			1 & 0
		\end{pmatrix}
		\textrm{,}
		\ \sigma_{y}=
		\begin{pmatrix}
			0 & -i \\
			i & 0
		\end{pmatrix}
		\textrm{,}
		\ \sigma_{z}=
		\begin{pmatrix}
			1 & 0 \\
			0 & -1
		\end{pmatrix}
		\textrm{,}\ 
	\end{align}
	and $I_{2\times 2}$ is the identity matrix. As in Eq.~\eqref{2bandHam}, any two-band system can be parametrized into a model that describes the interaction of an effective magnetic field $\mathbf{\Omega}\equiv(\Omega_{x},\Omega_{y},\Omega_{z})$ with a two-state transition \cite{KKavokinPRL2004}. A generic eigenstate of such a system is characterized by angles $\theta$ and $\varphi$ of the Bloch sphere:
	\begin{equation}\label{quantumstate}
		|u_{n,\mathbf{k}}\rangle\equiv\begin{pmatrix}
			u_{R}\\
			u_{L}\\
		\end{pmatrix}=\begin{pmatrix}
			\cos\left(\frac{\theta}{2}\right)e^{i\phi}\\
			\sin\left(\frac{\theta}{2}\right)\\
		\end{pmatrix},
	\end{equation}
	where
	\begin{equation}\label{angles}
		\theta=\arccos S_{z}\textrm{,}\quad\phi=\arctan(S_{x},S_{y})\textrm{,}
	\end{equation}
	and $S_{i}$ are the components of the Stokes vector $\mathbf{S}\equiv(S_{x},S_{y},S_{z})$, defined as the averages of the Pauli matrices:
	\begin{equation}\label{stokes_pauli}
		S_{i}=\langle\sigma_{i}\rangle=\langle u_{n,\mathbf{k}}|\sigma_{i}|u_{n,\mathbf{k}}\rangle.
	\end{equation}
	The basis $\{(1,0)^{T}, (0,1)^{T}\}$ defines the two bands which, depending on the system, may correspond to e.g.~the Bloch bands of a two-site unit cell lattice or spin up and spin down states 
	\cite{HaldanePRL1988,BernevigSuperconductors,KavokinPRL2005}. In our system, the polarization acts as a \emph{pseudospin} degree of freedom, and the basis of the Hamiltonian~\eqref{2bandHam} defines states of right- and left-circularly polarized light in the poles of the Poincar\'e sphere \cite{GianfrateNature2020}. We thus label the states as $|R\rangle=(1,0)^{T}$ and $|L\rangle=(0,1)^{T}$, respectively. In this basis, horizontal and vertical linearly polarized states read as $|H\rangle=(1/\sqrt{2})(1,1)^{T}$ and $|V\rangle=(1/\sqrt{2})(1,-1)^{T}$, respectively. Note that ``vertical'' and ``horizontal'' here denote the two in-plane directions of the 2D lattice, $x$ and $y$ (see Figs.~1(a) and 2(c) of the main text), that is, ``vertical'' is not the out-of-plane ($z$) direction. We define the diagonal and antidiagonal linear polarizations accordingly: $|D\rangle=(1/\sqrt{2})(i,1)^{T}$, $|A\rangle=(1/\sqrt{2})(1,i)^{T}$.
	
	The components of the Stokes vector defined in Eq.~\eqref{stokes_pauli} that characterize the two-level state in Eqs.~\eqref{quantumstate} and \eqref{angles} may be rewritten in terms of the amplitudes of right- ($u_{R}$) and left- ($u_{L}$) circularly polarized light as:
	\begin{equation}\label{stokes_intensities}
		S_{x}=\dfrac{I_{H}-I_{V}}{I_{H}+I_{V}},\ S_{y}=\dfrac{I_{D}-I_{A}}{I_{D}+I_{A}},\ S_{z}=\dfrac{I_{R}-I_{L}}{I_{R}+I_{L}},
	\end{equation}
	where $I_{R}=|u_{R}|^{2}$, $I_{L}=|u_{L}|^{2}$, and $I_{H}=|u_{H}|^{2}$, $I_{V}=|u_{V}|^{2}$, $I_{D}=|u_{D}|^{2}$, $I_{A}=|u_{A}|^{2}$, with:
	\begin{align}
		u_{H}&=\frac{1}{\sqrt{2}}\left(u_{R}+u_{L}\right),\\
		u_{V}&=\frac{1}{\sqrt{2}}\left(u_{R}-u_{L}\right),\\
		u_{D}&=\frac{1}{\sqrt{2}}\left(e^{i\pi/4}u_{R}+e^{-i\pi/4}u_{L}\right),\\
		u_{A}&=\frac{1}{\sqrt{2}}\left(e^{i\pi/4}u_{R}-e^{-i\pi/4}u_{L}\right).
	\end{align}
	Expressions \eqref{stokes_pauli} and \eqref{stokes_intensities} show that the averages of the Pauli matrices $\sigma_{x}$, $\sigma_{y}$ and $\sigma_{z}$ correspond to the degrees of polarization $H$/$V$, $D$/$A$ and $R$/$L$, respectively. Therefore, by judiciously choosing the effective magnetic field components $\Omega_{i}$ of Eq.~\eqref{2bandHam}, we can design a Hamiltonian that accounts for both the dispersion and the polarization-dependent properties of the two TE/TM modes studied above. From that Hamiltonian, the eigenvectors of the system in the RC/LC basis and thus the QGT (see Eq.~\eqref{qgt_alternativeform}) are readily obtained. 
	
	We next describe how the main properties of the two SLR bands related to the empty lattice modes (-1,0) and (0,-1), studied in Ref.~\cite{CuerdaPRB2023}, may be encompassed into a Hamiltonian that reads as in Eq.~\eqref{2bandHam}. For this, we demand the following conditions: \emph{(i)} At the $\Gamma-$point, TE and TM bands must be degenerate. \emph{(ii)} Away from the $\Gamma-$point, the SLR bands closely follow the empty lattice approximation, and the momentum-dependence of the energy dispersion bands is ruled by Eq.~(1) in the main text. \emph{(iii)} Our model must be able to reproduce the degeneracy lifting found in Ref.~\cite{CuerdaPRB2023} at the diagonal of the Brillouin zone.  
	
\begin{figure}[t!]
		\includegraphics[width=0.99\columnwidth]{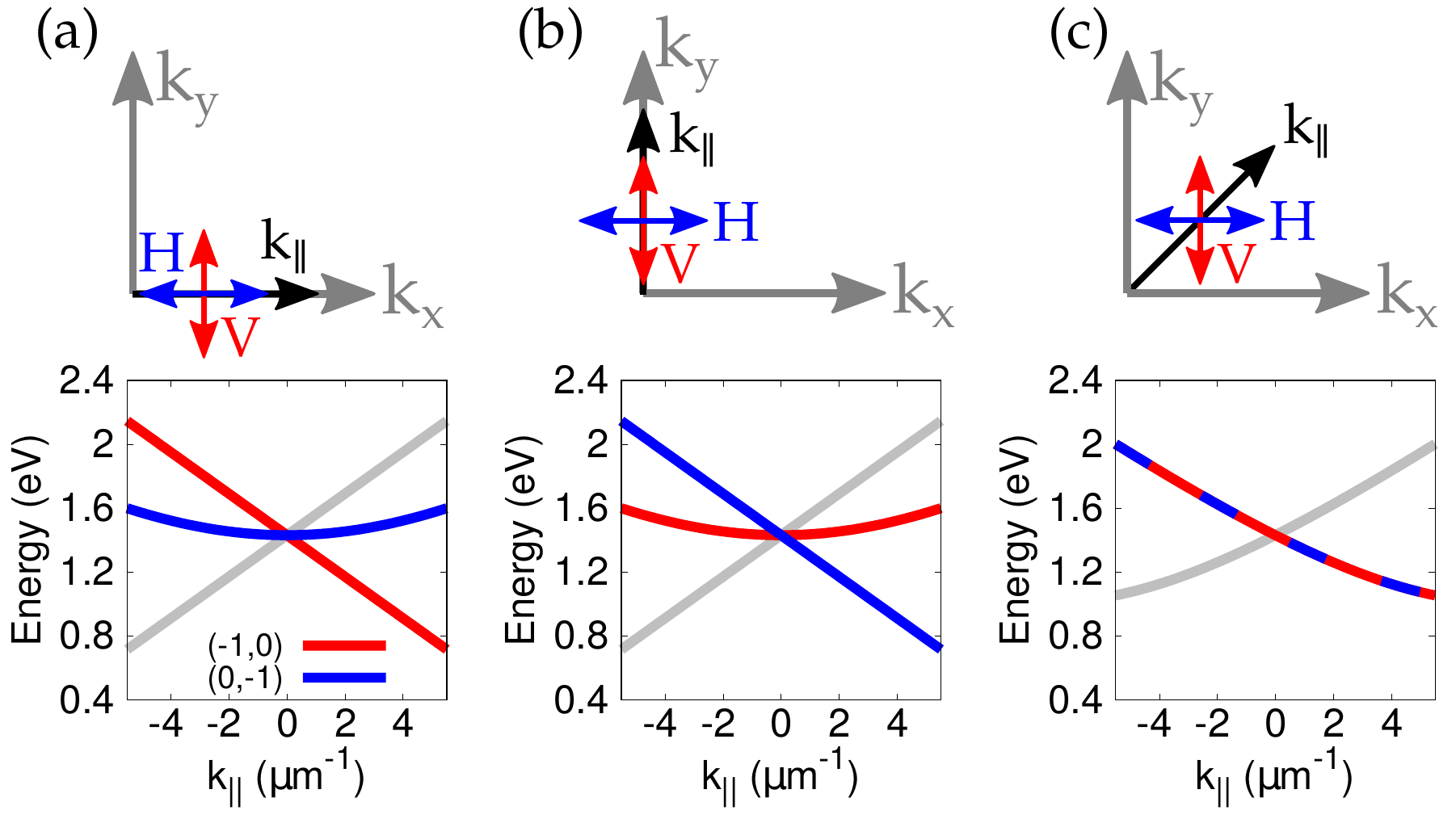}
		\caption{\label{fig:polprops_ela}Polarization properties of the empty lattice modes $(0,-1)$ and $(-1,0)$ of a square plasmonic lattice. We show how the empty lattice modes selectively couple to horizontal (blue lines) and vertical (red lines) linear polarization, for three representative cases: (a) $\mathbf{k}_{||}$ points along the $k_{x}-$axis, (b) $\mathbf{k}_{||}$ points along the $k_{y}-$axis, and (c) $\mathbf{k}_{||}$ points along the main diagonal of the Brillouin zone. Empty lattice band dispersions $(1,0)$ and $(0,1)$ are also shown (grey lines).} 
		\vspace*{-2mm}
	\end{figure}

	We first address points \emph{(i)} and \emph{(ii)} above, and perform a mapping that encapsulates the empty lattice dispersion energies of the considered two modes in a Hamiltonian with the form \eqref{2bandHam} that incorporates the polarization degree of freedom to those modes. We choose the functions $\Omega_{x}(\mathbf{k})$ and $\epsilon(\mathbf{k})$ in the first quadrant of the Brillouin zone ($k_{x,y}>0$), as follows:
	\begin{align}
		\epsilon(\mathbf{k})&=\dfrac{1}{2}(E_{-1,0}(\mathbf{k})+E_{0,-1}(\mathbf{k}))\label{epk},\\
		\Omega_{x}(\mathbf{k})&=\dfrac{1}{2k^2}\left(k_{x}^{2}-k_{y}^{2}\right)\left(E_{0,-1}(\mathbf{k})-E_{-1,0}(\mathbf{k})\right),\label{omega_x}
	\end{align}
	where
	\begin{align}
		E_{-1,0}(\mathbf{k})&=\dfrac{\hbar c}{n_{h}}\sqrt{(k_{x}-G)^{2}+k_{y}^{2}},\label{ela_energy1}\\
		E_{0,-1}(\mathbf{k})&=\dfrac{\hbar c}{n_{h}}\sqrt{k_{x}^{2}+(k_{y}-G)^{2}}.\label{ela_energy2}
	\end{align}
	We next study the energy dispersion bands of the two SLR modes with $\Omega_{y}=\Omega_{z}=0$ in Eq.~\eqref{2bandHam}, and verify that they fulfill the expected polarization properties, see Refs.~\cite{GuoPRB2017,QuantumPlasmonics}. In Ref.~\cite{CuerdaPRB2023}, we introduced these SLRs as TE and TM modes with respect to the $k_{x}$-axis, along which horizontal ($x$-) polarized light is TM to the in-plane momentum, and vertical ($y$-) polarized light is TE, see Fig.~\ref{fig:polprops_ela}(a). Therefore, the Hamiltonian \eqref{2bandHam} with parameters \eqref{epk} and \eqref{omega_x} fulfills the conditions $\langle H|\hat{H}| H\rangle=E_{-1,0}(\mathbf{k})$ and $\langle V|\hat{H}| V\rangle=E_{0,-1}(\mathbf{k})$, as expected. Due to the symmetry of the square lattice, we find the same TE and TM properties of SLRs for in-plane momentum parallel to the $k_{y}$-axis, but now horizontally polarized light is TE to in-plane mode propagation, and vertically polarized light is TM (see Fig.~\ref{fig:polprops_ela}(b)), resulting in $\langle H|\hat{H}| H\rangle=E_{0,-1}(\mathbf{k})$ and $\langle V|\hat{H}| V\rangle=E_{-1,0}(\mathbf{k})$. We also note that, along the diagonal of the Brillouin zone (Fig.~\ref{fig:polprops_ela}(c)), the two-band Hamiltonian with the coupling parameters \eqref{epk} and \eqref{omega_x} produces degenerate dispersion bands for both vertical and horizontal polarizations (in agreement with the empty lattice bands in Fig.~\ref{fig:ela}(c) for $\alpha=45\degree$): $\langle H|\hat{H}| H\rangle=\langle V|\hat{H}| V\rangle=E_{d}(k_{d})$, where $E_{d}(k_{d})\equiv E_{-1,0}(k_{d})=E_{0,-1}(k_{d})$ with $k_{x}=k_{y}\equiv k_{d}$ in Eqs.~\eqref{ela_energy1} and \eqref{ela_energy2} ($k_{||}=\sqrt{2}k_{d}$ in Fig.~\ref{fig:polprops_ela}(c)). The two considered modes are not TE nor TM polarized along this high-symmetry axis, and both couple to horizontal and vertical linearly polarized light. Band degeneracy along the diagonal indicates that our model is not yet complete, hence we address condition \emph{(iii)} next.
	
	As explained in Section \ref{sec:ela}, the empty lattice band structure of a square lattice given by the terms \eqref{epk} and \eqref{omega_x} does not feature a band opening at the diagonal of the first Brillouin zone, and for diagonal and antidiagonal linear polarizations we find $\langle D|\hat{H}|D\rangle=\langle A|\hat{H}|A\rangle=E_{d}(k_{d})$, as with horizontal and vertical polarizations above. However, our T-matrix simulations in Ref.~\cite{CuerdaPRB2023} include the influence of the metallic nanoparticles and reveal the band structure of the complete system, with a lifting of the degeneracy along the diagonal, except for the band touching at $k_{x}=k_{y}=0$. Such a behavior can be introduced in our simple two-band model with an additional $k$-dependent term: 
	\begin{equation}\label{omega_y}
		\Omega_y(\mathbf{k})=\dfrac{g}{2}\sqrt{k_x^2 + k_y^2}\textrm{,}
	\end{equation}
	where $g$ is a constant. Using the terms \eqref{epk}, \eqref{omega_x} and \eqref{omega_y} altogether in the Hamiltonian \eqref{2bandHam} yields two separate bands at the diagonal: $\langle D|\hat{H}|D\rangle=E_{d}(k_{d})-(g/2)|\mathbf{k}|$ and $\langle A|\hat{H}|A\rangle=E_{d}(k_{d})+(g/2)|\mathbf{k}|$. Therefore, a band splitting of size $g|\mathbf{k}|$ occurs at every point $\mathbf{k}$ on the main diagonal of the Brillouin zone. Interestingly, including the term \eqref{omega_y} in the Hamiltonian \eqref{2bandHam} does not alter the band structure under horizontally or vertically polarized light, and we find that $\langle H|\hat{H}| H\rangle=\epsilon(\mathbf{k})+\Omega_{x}(\mathbf{k})$ and $\langle V|\hat{H}| V\rangle=\epsilon(\mathbf{k})-\Omega_{x}(\mathbf{k})$, for every $\mathbf{k}$ in the first quadrant of the Brillouin zone. Hence from this simple model, we conclude that the band splitting yields two modes with well-defined linear polarizations; namely, the higher energy mode is TE polarized with respect to the diagonal of the Brillouin zone, and the lower energy mode is TM polarized. We confirm this behavior in Fig.~2(b) of the main text.
	
	The size of the band splitting in our simplified two-band model is controlled by the parameter $g$ in the term \eqref{omega_y}. We estimate its realistic value from the band energy difference found in numerical simulations of Ref.~\cite{CuerdaPRB2023} to qualitatively model the effect of the energy splitting induced by the nanoparticles composing the lattice. Note, however, that the degeneracy lifting was found for both the real and imaginary parts of the eigenfrequency, see Ref.~\cite{CuerdaPRB2023}. Therefore we may assume that the constant $g$ is in general complex-valued:
	\begin{equation}
		g=g'+ig'',
	\end{equation}
	where $g'$ controls the splitting of the energy bands at the diagonal of the Brillouin zone, and $g''$ is utilized to qualitatively introduce the difference in modal losses that is observed in the numerical results in Ref.~\cite{CuerdaPRB2023}. A finite $g''$ is a necessary condition for having a difference in the modal losses, but its value does not equal the loss difference. Thus manually varying the constant $g''$ allows examining the consequences of non-hermiticity in the system.
	The two-band model is a heuristic approach aimed at modeling the effects of band structure and relevant removal of degeneracies in the system. The terms $\Omega_{x,y}$ are defined to provide a minimal model that captures the relevant effects that produce quantum metric and Berry curvature in our system. Both in our experiments and in the simulations of Ref.~\cite{CuerdaPRB2023}, no assumption is made over the band structure or losses of the system, yet the relevant effects are found around the diagonal of the first Brillouin zone. This has motivated us to introduce the loss effects only in the term $\Omega_{y}$ that controls the degeneracy removal at the diagonals of the Brillouin zone. Interestingly, the dependence of both energy and modal losses is similar along the diagonal of the Brillouin zone (i.e. monotonously increasing with $k$ as we move away from the $\Gamma$-point along the diagonal - see Ref.~\cite{CuerdaPRB2023}). Therefore, as a first approximation, it is reasonable to assume that the momentum-dependence along that high-symmetry axis is linear ($\Omega_{y}\sim k$) and the same for both quantities, and that the quantitative difference can be modeled with multiplicative constants $g'$ (energy) and $g''$ (modal losses).
	
	Using a form of the Hamiltonian as in Eq.~\eqref{2bandHam}, with nonzero terms $\Omega_{x}$ and $\Omega_{y}$, suffices to provide a minimal model to study quantum metric and Berry curvature in a square plasmonic lattice. Berry curvature may be complementarily introduced by for example breaking the sub-lattice symmetry in a honeycomb lattice, and time-reversal symmetry could be broken by using magnetic materials \cite{BahariScience2017,KatajaNatComm2015} or chiral particle shapes \cite{GoerlitzerAdvMat2020}. Opening a gap between the right- and left-circular polarization enters in Eq.~(\ref{2bandHam}) as a nonzero, Zeeman-like $\Omega_{z}$ term \cite{BleuPRB2018}. Time-reversal symmetry breaking may lift degeneracies in the band structure in such a way that topological bands are created instead of just local Berry curvature, leading to topologically protected chiral modes \cite{HaldanePRL2008}. We leave such possibilities as directions for further work, and consider $\Omega_{z}=0$ here.
	
	The matrix form of the general two-band Hamiltonian (\ref{2bandHam}) in the RC/LC basis reads as follows:
	\begin{align}\label{2HamMatrix}
		H=
		\begin{pmatrix}
			\epsilon(\mathbf{k}) & \Omega_{x}(\mathbf{k})-i\Omega_{y}(\mathbf{k}) \\
			\Omega_{x}(\mathbf{k})+i\Omega_{y}(\mathbf{k}) & \epsilon(\mathbf{k})
		\end{pmatrix}
		\textrm{.}
	\end{align}
	If $\Omega_{x,y}$ are real-valued for every $\mathbf{k}$, then the matrix (\ref{2HamMatrix}) is Hermitian. This gives real eigenvalues for the energies of the two bands\\
	\begin{equation}\label{energies_hermitian}
		E_{\pm}^{H}=\epsilon\pm\sqrt{\Omega_{x}^{2}+\Omega_{y}^{2}}\textrm{,}
	\end{equation}
	together with the eigenvectors
	\begin{align}\label{eigvecs_hermitian}
		|u_{\pm}\rangle=\dfrac{1}{\sqrt{2}}\begin{pmatrix}
			\pm e^{i\phi_{\mathbf{k}}^{H}}\\
			1
		\end{pmatrix},
	\end{align}
	where
	\begin{equation}\label{Hermitian_phase}
		\phi_{\mathbf{k}}^{H}=\arctan(-\Omega_{y}(\mathbf{k})/\Omega_{x}(\mathbf{k}))\textrm{.}
	\end{equation}
	However, as discussed above, energies of the two-band system are in general complex, and the lifting of degeneracy along the diagonals of the first Brillouin zone is also produced for the imaginary part. Therefore we assume a complex gap opening at the diagonals: $\Omega_{y}=\Omega'_{y}+i\Omega''_{y}$, with real-valued $\Omega'_{y}$ and $\Omega''_{y}$ for every $\mathbf{k}$. In this case, the Hamiltonian matrix (\ref{2HamMatrix}) becomes non-Hermitian. For a diagonalizable $N\times N$ non-Hermitian Hamiltonian, a family of $N$ \emph{right} eigenvectors $|R_{n}\rangle$ and an equal number of \emph{left} eigenvectors $|L_{n}\rangle$ correspond to the same complex eigenvalue $E_{n}$, $n=\{+,-\}$ such that \cite{SinghalArxiv2022}
	\begin{equation}\label{def:right_left_eigs}
		H|R_{n}\rangle=E_{n}|R_{n}\rangle,\qquad\langle L_{n}|H=E_{n}\langle L_{n}|, 
	\end{equation}
	where the orthogonality condition is preserved: $\langle L_{n}|R_{n'}\rangle=0$ for non-degenerate eigenvalues $E_{n}\neq E_{n'}$ and $n\neq n'$. In our two-band model, eigenvalues and eigenvectors of the non-Hermitian system read as
	\begin{equation}\label{energies_nonherm}
		E_{\pm}^{NH}=\epsilon\pm\sqrt{\Omega_{x}^2-\left(\Omega''_{y}-i\Omega'_{y}\right)^2}
	\end{equation}
	and
	\begin{align}
		|R_{\pm}\rangle=\sqrt{\dfrac{|A_{\mathbf{k}}|}{|A_{\mathbf{k}}|+|B_{\mathbf{k}}|}}\begin{pmatrix}
			\pm\sqrt{\dfrac{|B_{\mathbf{k}}|}{|A_{\mathbf{k}}|}}e^{-i\Delta\phi/2}\\
			1
		\end{pmatrix}
		\textrm{,}\label{righteig}\\
		|L_{\pm}\rangle=\sqrt{\dfrac{|B_{\mathbf{k}}|}{|A_{\mathbf{k}}|+|B_{\mathbf{k}}|}}
		\begin{pmatrix}
			\pm\sqrt{\dfrac{|A_{\mathbf{k}}|}{|B_{\mathbf{k}}|}}e^{-i\Delta\phi/2}\label{lefteig}\\
			1
		\end{pmatrix}
		\textrm{,}
	\end{align}
	respectively. Here
	\begin{align}
		|A_{\mathbf{k}}|&=\sqrt{(\Omega_{x}-\Omega''_{y})^2+(\Omega'_{y})^2}\label{sup:A_const},\\
		|B_{\mathbf{k}}|&=\sqrt{(\Omega_{x}+\Omega''_{y})^2+(\Omega'_{y})^2}\label{sup:B_const},
	\end{align}
	and
	\begin{equation}\label{phasedifference}
		\Delta\phi=\phi_{1}^{NH}(\mathbf{k})-\phi_{2}^{NH}(\mathbf{k}),
	\end{equation}
	where
	\begin{align}
		\phi_{1}^{NH}&=\arctan\left(\dfrac{|\Omega'_{y}|}{|\Omega_{x}-\Omega''_{y}|}\right),\\  
		\phi_{2}^{NH}&=\arctan\left(\dfrac{-|\Omega'_{y}|}{|\Omega_{x}+\Omega''_{y}|}\right).
	\end{align}
	When the imaginary part of $\Omega_y$ is set to zero for all points in $k-$space, eigenvalues and eigenvectors converge to the Hermitian case: $E_{\pm}^{NH}=E_{\pm}^{H}$ and $|L_{\pm}\rangle=|R_{\pm}\rangle=|u_{\pm}\rangle$. 
	
	We next calculate the quantum geometric tensor using the two-band model: 
	\begin{equation}\label{qgt_alternativeform}
		T_{ij}^{n}=\sum_{m\neq n}\dfrac{\langle u_{m,\mathbf{k}}|\partial_{k_{i}}\hat{H}|u_{n,\mathbf{k}}\rangle\langle u_{n,\mathbf{k}}|\partial_{k_{j}}\hat{H}|u_{m,\mathbf{k}}\rangle}{{(E_{m}-E_{n})}^2}.
	\end{equation}
	In our simplified model, four components of the QGT can be calculated for each band: $T_{xx}^{\pm}$, $T_{yy}^{\pm}$,  $T_{xy}^{\pm}$, and $T_{yx}^{\pm}$. We first examine the Hermitian case for which the eigenenergies $E_{\pm}^{H}$ and eigenvectors $|u_{\pm}\rangle$ are utilized. The components of the quantum metric are given by $g_{ij}^{\pm}=\Re T_{ij}^{\pm}$; further, one easily finds from Eq.~(\ref{qgt_alternativeform}) that $g_{ij}\equiv g_{ij}^{+}=g_{ij}^{-}$ for this model.
	
	In the Hermitian regime, the non-zero quantum metric is caused by the energy band splitting. We find
	\begin{align}
		&\langle u_{-}|\partial_{k_{i}}\hat{H}|u_{+}\rangle=i[(\partial_{k_{i}}\Omega_{x})\sin\phi_{\mathbf{k}}^{H}\nonumber\\
		&\qquad\qquad\qquad\qquad\qquad\qquad+(\partial_{k_{i}}\Omega_{y})\cos\phi_{\mathbf{k}}^{H}]\label{bracket}
	\end{align}
	and $\langle u_{+}|\partial_{k_{i}}\hat{H}|u_{-}\rangle=-\langle u_{-}|\partial_{k_{i}}\hat{H}|u_{+}\rangle$. Hence, for a band splitting set to zero ($\Omega_{y}=0$), the term \eqref{bracket} equals zero and the QGT components yield a trivial value, even with a non-zero $\Omega_{x}$ term in the Hamiltonian. The Berry curvature $\mathfrak{B}_{ij}^{\pm}=-2\Im T_{ij}^{\pm}$ is zero in the Hermitian limit at every point of $k-$space. This is expected in the absence of a non-zero Zeeman term $\Omega_{z}$ when the Hamiltonian matrix \eqref{2HamMatrix} is Hermitian \cite{BleuPRB2018}. However, we next show that the Berry curvature can be different from zero in our lossy system, where the matrix \eqref{2HamMatrix} is actually non-Hermitian with $\Omega_{y}\in\mathbb{C}$.  
	
	In the non-Hermitian regime, two sets of right and left eigenvectors are solutions of the Hamiltonian~\eqref{2bandHam}, hence four possible expressions of the QGT arise according to Eq.~\eqref{qgt_alternativeform}. As in a recent study concerned with the non-Hermitian Berry phase \cite{SinghalArxiv2022}, here we assume that the adiabatic theorem holds even for our non-Hermitian system, and that the system remains in either the right or the left eigenstate up to an overall phase factor. From this perspective, we consider only two expressions of the QGT:
	\begin{align}
		(T_{L})_{ij}^{n}&=\sum_{m\neq n}\dfrac{\langle L_{m}|\partial_{k_{i}}\hat{H}|L_{n}\rangle\langle L_{n}|\partial_{k_{j}}\hat{H}|L_{m}\rangle}{{(E_{m}^{NH}-E_{n}^{NH})}^2}\label{qgt_left},\\
		(T_{R})_{ij}^{n}&=\sum_{m\neq n}\dfrac{\langle R_{m}|\partial_{k_{i}}\hat{H}|R_{n}\rangle\langle R_{n}|\partial_{k_{j}}\hat{H}|R_{m}\rangle}{{(E_{m}^{NH}-E_{n}^{NH})}^2}\label{qgt_right},
	\end{align}
	with energies $E_{\pm}^{NH}$ given by \eqref{energies_nonherm}. Our two-band model provides analytical expressions of the quantum metric and non-Hermitian Berry curvature:
	\begin{align}
		(g_{R})_{ij}^{+}&=\dfrac{\alpha_{R,i}^{-}\alpha_{R,j}^{+}+\beta_{i}\beta_{j}}{4{\left(|A_{\mathbf{k}}|+|B_{\mathbf{k}}|\right)}^{2}\left(\Omega_{x}^{2}-{\left(\Omega''_{y}-i\Omega'_{y}\right)}^{2}\right)}\label{quantummetric_right},\\
		(g_{L})_{ij}^{+}&=\dfrac{\alpha_{L,i}^{-}\alpha_{L,j}^{+}+\beta_{i}\beta_{j}}{4{\left(|A_{\mathbf{k}}|+|B_{\mathbf{k}}|\right)}^{2}\left(\Omega_{x}^{2}-{\left(\Omega''_{y}-i\Omega'_{y}\right)}^{2}\right)}\label{quantummetric_left},\\
		(\mathfrak{B}_{R})_{ij}^{+}&=\dfrac{\alpha_{R,i}^{-}\beta_{j}-\alpha_{R,j}^{+}\beta_{i}}{2{\left(|A_{\mathbf{k}}|+|B_{\mathbf{k}}|\right)}^{2}\left(\Omega_{x}^{2}-{\left(\Omega''_{y}-i\Omega'_{y}\right)}^{2}\right)}\label{berrycurv_right},\\
		(\mathfrak{B}_{L})_{ij}^{+}&=\dfrac{\alpha_{L,i}^{-}\beta_{j}-\alpha_{L,j}^{+}\beta_{i}}{2{\left(|A_{\mathbf{k}}|+|B_{\mathbf{k}}|\right)}^{2}\left(\Omega_{x}^{2}-{\left(\Omega''_{y}-i\Omega'_{y}\right)}^{2}\right)}\label{berrycurv_left},
	\end{align}
	with
	\begin{align}
		&(\alpha_R)_{i}^{\pm}=\left(|A_{\mathbf{k}}|-|B_{\mathbf{k}}|\right)(\partial_{k_{i}}\epsilon)\nonumber\\
		&\qquad\qquad\qquad\pm 2\sqrt{|A_{\mathbf{k}}||B_{\mathbf{k}}|}(\partial_{k_{i}}\Omega''_{y})\cos\left(\tfrac{\Delta\phi}{2}\right)\label{alpha_nonhermqgtR},\\
		&(\alpha_L)_{i}^{\pm}=-\left(|A_{\mathbf{k}}|-|B_{\mathbf{k}}|\right)(\partial_{k_{i}}\epsilon)\nonumber\\
		&\qquad\qquad\qquad\pm 2\sqrt{|A_{\mathbf{k}}||B_{\mathbf{k}}|}(\partial_{k_{i}}\Omega''_{y})\cos\left(\tfrac{\Delta\phi}{2}\right)\label{alpha_nonhermqgtL},\\
		&\beta_{i}=2\sqrt{|A_{\mathbf{k}}||B_{\mathbf{k}}|}\Big[\cos\left(\tfrac{\Delta\phi}{2}\right)(\partial_{k_{i}}\Omega'_{y})\nonumber\\
		&\qquad\qquad\qquad\qquad\qquad-\sin\left(\tfrac{\Delta\phi}{2}\right)(\partial_{k_{i}}\Omega_{x})\Big],\label{beta_nonhermqgt}
	\end{align}
	where $i=x,y$. The left and right forms of the quantum metric and the Berry curvature are obtained from the left and right QGT in Eqs.~\eqref{qgt_left} and \eqref{qgt_right} above, respectively: $(g_{L,R})_{ij}^{n}=\Re (T_{L,R})_{ij}^{n}$, $(\mathfrak{B}_{L,R})_{ij}^{n}=-2\Im (T_{L,R})_{ij}^{n}$. The lower energy band components are related to Eqs.~\eqref{quantummetric_right}-\eqref{berrycurv_left} as follows: $(g_{L,R})_{ij}^{-}=(g_{L,R})_{ij}^{+}$ and $(\mathfrak{B}_{L,R})_{ij}^{-}=-(\mathfrak{B}_{L,R})_{ij}^{+}$.
	
	Examining Eqs.~\eqref{quantummetric_right}-\eqref{beta_nonhermqgt} brings insight into the physical origin of the quantum metric and the Berry curvature. In the Hermitian limit, $\Omega''_{y}=0$, $\Delta\phi=-2\phi_{\mathbf{k}}^{H}$ and $|A_{\mathbf{k}}|=|B_{\mathbf{k}}|$, thus we find that $(\alpha_{R,L})_{i}^{\pm}\equiv\alpha^{H}_{i}=0$, and
	\begin{align}
		&\beta_{i}\equiv\beta_{i}^{H}=2\sqrt{\Omega_{x}^{2}+\Omega_{y}^{'2}}[(\partial_{k_{i}}\Omega'_{y})\cos\phi_{\mathbf{k}}^{H}\nonumber\\
		&\qquad\qquad\qquad\qquad\qquad\qquad+(\partial_{k_{i}}\Omega_{x})\sin\phi_{\mathbf{k}}^{H}].\label{beta_hermitian}
	\end{align}
	Inserting \eqref{beta_hermitian} into Eqs.~\eqref{quantummetric_right} and \eqref{quantummetric_left} shows that the right and left quantum metric converge to the same value in the Hermitian limit: $(g_{L,R})_{ij}^{n}\equiv (g^{H})_{ij}^{n}$. In this limit, the quantum metric depends exclusively on the energy band splitting, as setting $\Omega'_{y}=0$ in Eq.~\eqref{beta_hermitian} leads to $(g^{H})_{ij}^{n}=0$. However, the most general non-Hermitian expressions \eqref{quantummetric_right} and \eqref{quantummetric_left} of the quantum metric also contain terms related to losses, and do not become zero even with $\Omega'_{y}=0$. The results of both the right and left quantum metric, shown in Fig.~3(a) of the main text, highly resemble those found for the Hermitian limit ($g''=0$)  provided that $g''\approx 10^{-2}g'$ as in realistic T-matrix simulations~\cite{CuerdaPRB2023}.~Therefore, we conclude that expressions \eqref{quantummetric_right} and \eqref{quantummetric_left} add only small non-Hermitian corrections to our findings above, and the emergence of quantum metric in a plasmonic lattice is dominantly ruled by the energy band splitting in the diagonals of the Brillouin zone.
	
	Importantly, Fig.~3(b) of the main text shows that the left eigenmodes possess non-trivial Berry curvature that relates to the right Berry curvature by a minus sign: $(\mathfrak{B}_{L})_{ij}^{+}\approx-(\mathfrak{B}_{R})_{ij}^{+}$. The relation between the right and left Berry curvature is in general more complex, see Eqs.~\eqref{berrycurv_right} and \eqref{berrycurv_left}. However, with our selection of parameters we find that $|\partial_{k_{i}}\epsilon|\gg|\partial_{k_{i}}\Omega''_{y}|$ in Eqs.~\eqref{alpha_nonhermqgtR} and \eqref{alpha_nonhermqgtL}; therefore, $(\alpha_{R})_{i}^{\pm}\approx-(\alpha_{L})_{i}^{\pm}$, leading to an opposite sign between the panels in Fig.~3(b). 
	
	\section{Origin of the non-zero Berry curvature and Stokes vector analysis}
	In this section, we discuss the physical origin of the experimentally observed and theoretically obtained Berry curvature. In particular, we show that there are two potential sources of Berry curvature in our system. First, there is a net circularly polarized component coming from the imbalance in the right/left circularly polarized light ratio, due to losses. A second contribution originates from the spin-orbit coupling in combination with non-Hermiticity that creates the phase difference $\Delta\phi$ in Eq.~\eqref{phasedifference}. 
	
	We first discuss on the emergence of a circular component by losses. We begin our discussion by making some important remarks:\\
	\begin{itemize}
		\item The term $\Omega_{x}$ is in general non-zero, but becomes zero \emph{exactly} at the diagonal of the Brillouin zone.
		\item The term $\Omega_{y}''$ is zero in the Hermitian limit, and non-zero in the non-Hermitian regime.
		\item If one of the two terms above ($\Omega_{x}$ or $\Omega_{y}''$) is zero, then $|A|$ and $|B|$ in Eqs.~\eqref{sup:A_const} and \eqref{sup:B_const} are equal: $|A|=|B|$. Conversely, if $\Omega_{x}\neq0$ and $\Omega_{y}''\neq0$, then $|A|\neq|B|$.
		\item The inequality $|A|\neq|B|$ ensures the existence of an LCP/RCP imbalance and, consequently, a net circular component and non-zeroness of Berry curvature. However, $|A|=|B|$ does not immediately prohibit a non-zero Berry curvature.
	\end{itemize}
	We explore different regimes listed in four subsections below: \hyperlink{sup:subsec_Herm_offdiag}{A.} Hermitian regime ($\Omega_{y}''=0$) (away from the diagonal of the Brillouin zone, $\Omega_{x}\neq0$, but close to it). \hyperlink{sup:subsec_Herm_diag}{B.} Hermitian regime at the diagonal ($\Omega_{x}=0$). \hyperlink{sup:subsec_nonHerm_diag}{C.} Non-Hermitian regime $\Omega_{y}''\neq0$ at the diagonal. \hyperlink{sup:subsec_nonHerm_offdiag}{D.} Non-Hermitian regime away from the diagonal, but close to it.
	\subsection{Hermitian regime, close to but not at diagonal ($\Omega_{y}''=0$, $\Omega_{x}\neq0$)}\label{sup:subsec_Herm_offdiag}
	In this case, we find $|A|=|B|$ and both right and left eigenvectors in Eqs.~\eqref{righteig} and \eqref{lefteig} read as follows:
	\begin{equation}\label{eigenvec_case1}
		|R_{\pm}\rangle=|L_{\pm}\rangle=\dfrac{1}{\sqrt{2}}
		\begin{pmatrix}
			\pm e^{-i\Delta\phi(\mathbf{k})/2}\\
			1
		\end{pmatrix}    
	\end{equation}
	The eigenvectors are written in the right/left circular basis, thus a generic eigenvector reads as $|u\rangle\equiv(u^{R},u^{L})^{T}$, where $u^{R/L}$ are right/left circular (RC/LC) polarized amplitudes. Therefore, it becomes clear from Eq.~\eqref{eigenvec_case1} that there is no imbalance between RC and LC polarizations, i.e. there is no net circular component in this case, because $|u^{R}|/|u^{L}|=1$. 
	However, expression \eqref{eigenvec_case1} contains a phase difference $\Delta\phi(\mathbf{k})$ that is $k$-dependent, and this is the origin of the non-zero quantum metric both in the theory and in the experiments.
	\subsection{Hermitian regime, diagonal ($\Omega_{y}''=0$, $\Omega_{x}=0$)}\label{sup:subsec_Herm_diag}
	At the diagonal, we find that the term: 
	\begin{equation}
		\Omega_{x}(\mathbf{k})=\dfrac{1}{2k^2}\left(k_{x}^{2}-k_{y}^{2}\right)\left(E_{0,-1}(\mathbf{k})-E_{-1,0}(\mathbf{k})\right)
	\end{equation}
	becomes zero because $k_{x}=k_{y}$ and the empty lattice energies $E_{0,-1}$ and $E_{-1,0}$ are equal along the high symmetry axis. Therefore $\Omega_{x}=0$ at the diagonal, and the phase difference $\Delta\phi$, as defined in Eq.~\eqref{phasedifference}, is no longer momentum-dependent: $\Delta\phi=\pi$. Thus $\pm e^{i\pi}=\pm i$, and
	\begin{equation}\label{diag_anti_eigenvecs}
		|u_{+}\rangle=\dfrac{1}{\sqrt{2}}\begin{pmatrix}
			i\\
			1
		\end{pmatrix}
		=|D\rangle
		;\ 
		|u_{-}\rangle=\dfrac{1}{\sqrt{2}}\begin{pmatrix}
			-i\\
			1
		\end{pmatrix}
		=|A\rangle .
	\end{equation}
	We conclude that in the Hermitian limit, the TE and TM modes at the diagonal are the D and A eigenstates of the Poincar\'e sphere. In this case, the quantum metric becomes zero (since the phase is not $k$-dependent). This together with the results of subsection \hyperlink{sup:subsec_Herm_offdiag}{A} explain why the quantum metric components are non-zero only at angles slightly deviating from the diagonal, but not on top of it. 
	
	Naturally, the states $D$ and $A$ do not yield any net circular component (or LCP/RCP imbalance) and $|u^{R}|/|u^{L}|=1$ is of course found as in subsection \hyperlink{sup:subsec_Herm_offdiag}{A}. This together with the fact that $\Omega_{y}''=0$ in the Hermitian case make the Berry curvature zero, as can be seen from Equations \eqref{berrycurv_right}-\eqref{beta_nonhermqgt}.
	\subsection{Non-Hermitian regime, diagonal ($\Omega_{y}''\neq0$, $\Omega_{x}=0$)}\label{sup:subsec_nonHerm_diag}
	Even in the non-Hermitian regime, we found a zero value for both the quantum metric and for the Berry curvature at the diagonal of the Brillouin zone. This is explained using similar arguments as in subsections \hyperlink{sup:subsec_Herm_offdiag}{A} and \hyperlink{sup:subsec_Herm_diag}{B} above: At the diagonal $\Omega_{x}=0$ and thus $|A|=|B|$, so a circular component is again not present.
	
	The quantum metric is also zero as the phase difference in Eq.~\eqref{phasedifference} is not momentum-dependent:
	\begin{equation}\label{phasedif_notTETM}
		\Delta\phi=2\arctan\left(\dfrac{g'}{g''}\right)\textrm{.}
	\end{equation}
	However, the result in Eq.~\eqref{phasedif_notTETM} has an important implication: the modes in the non-Hermitian regime do not behave as TE/TM modes. This is obvious when introducing Eq.~\eqref{phasedif_notTETM} into \eqref{eigenvec_case1} and then comparing with the $|D\rangle$ and $|A\rangle$ eigenvectors of Eq.~\eqref{diag_anti_eigenvecs}.
	\vspace{0.35cm}
	\subsection{Non-Hermitian regime, close to but not on the diagonal ($\Omega_{y}''\neq0$, $\Omega_{x}\neq0$)}\label{sup:subsec_nonHerm_offdiag}
	In this case, we find that $|A|\neq|B|$ and this creates an RCP/LCP imbalance in Eq.~\eqref{eigenvec_case1} since $|u^{R}|/|u^{L}|\neq1$. Therefore, in this regime, a net circular component is predicted by the two-band model from time-reversal symmetry breaking by losses.

Further, in this regime, the phase difference $\Delta\phi$ defined in Eq.~\eqref{phasedifference} has a momentum-dependence and thus the quantum metric components are non-zero away from the diagonal of the Brillouin zone. This agrees completely with the observed experimental phenomena shown in Fig.~2(d) of the main text and with the microscopic T-matrix simulations.
\begin{figure*}[htp!]
	\centering
	\includegraphics[width=0.98\textwidth]{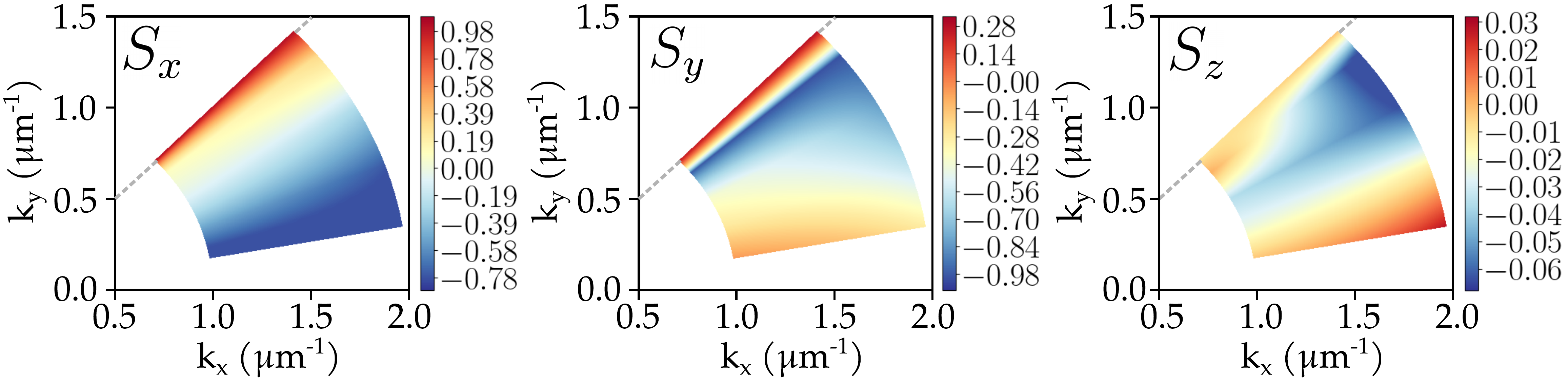}
	\caption{\label{fig:stokes}{Components of the Stokes vector, as extracted from the experimentally measured data. While $S_{x}$ and $S_{y}$ present an arguably non-zero distribution close to the diagonal of the Brillouin zone (grey dashed line), we find that the $S_{z}$ component is approximately zero (within the uncertainty imposed by the noise level in the experiments) at every point of the analyzed portion of $k$-space.}}
\end{figure*}
\\

Our experimental results correspond to subsections \hyperlink{sup:subsec_nonHerm_diag}{C} and \hyperlink{sup:subsec_nonHerm_offdiag}{D} of the discussion above, and accordingly one would expect that the circular component (given by the imbalance between the left and right polarization in areas slightly off from the diagonal of the Brillouin zone, see subsection \hyperlink{sup:subsec_nonHerm_offdiag}{D} in the above discussion) shows up as a non-zero $S_{z}$ component of the Stokes vector. Fig.~\ref{fig:stokes} shows that while the components $S_{x}$ and $S_{y}$ are clearly non-zero, $S_{z}\approx0$ to a reasonable uncertainty estimation $\Delta S_{z}\approx 0.07$.

The response to right and left circular polarization in our experiment is indeed very similar (see the rightmost panels of Fig.~2(b)). Given the striking similarity of our results with those of the two-band model, we believe in the existence of a circular component in the non-Hermitian regime in accordance with our two-band model. The fact that this circular component cannot be resolved from our experimental data (i.e., it is very small) leads us to conclude that the Berry curvature that we observe in the experiment is not caused by this circular component, although it is not incompatible with its existence.

In what follows, we provide an explanation of the origin of non-zero Berry curvature in our system, which is not given by the imbalance of right/left circular response but by the momentum-dependent phase difference $\Delta\phi(\mathbf{k})$ combined with non-Hermitian effects. From expressions \eqref{berrycurv_right}-\eqref{beta_nonhermqgt}, it is clear that the Berry curvature can remain non-zero and significant in magnitude even for the case $|A|=|B|$ (or $|A|\simeq|B|$), i.e.~when there is no (or negligible) imbalance between the left and right circular polarization and thus no (observable) net circular component. However, as shown by these equations, both $\Omega''_y$ and $\Omega'_y$ need to be non-zero and momentum-dependent for this to be possible. Thus, the combination of the spin-orbit coupling ($\Omega'_y(\mathbf{k})$) with non-Hermitian character (losses, $\Omega''_y(\mathbf{k})$) is sufficient to produce a sizable Berry curvature related to in-plane phase change $\Delta \phi$, even if the net circular component ($|A|-|B|$) remains negligibly small. 
\section{Comparison of transmission and eigenmode calculations}
\begin{figure*}[htp!]
	\centering
	\includegraphics[width=0.98\textwidth]{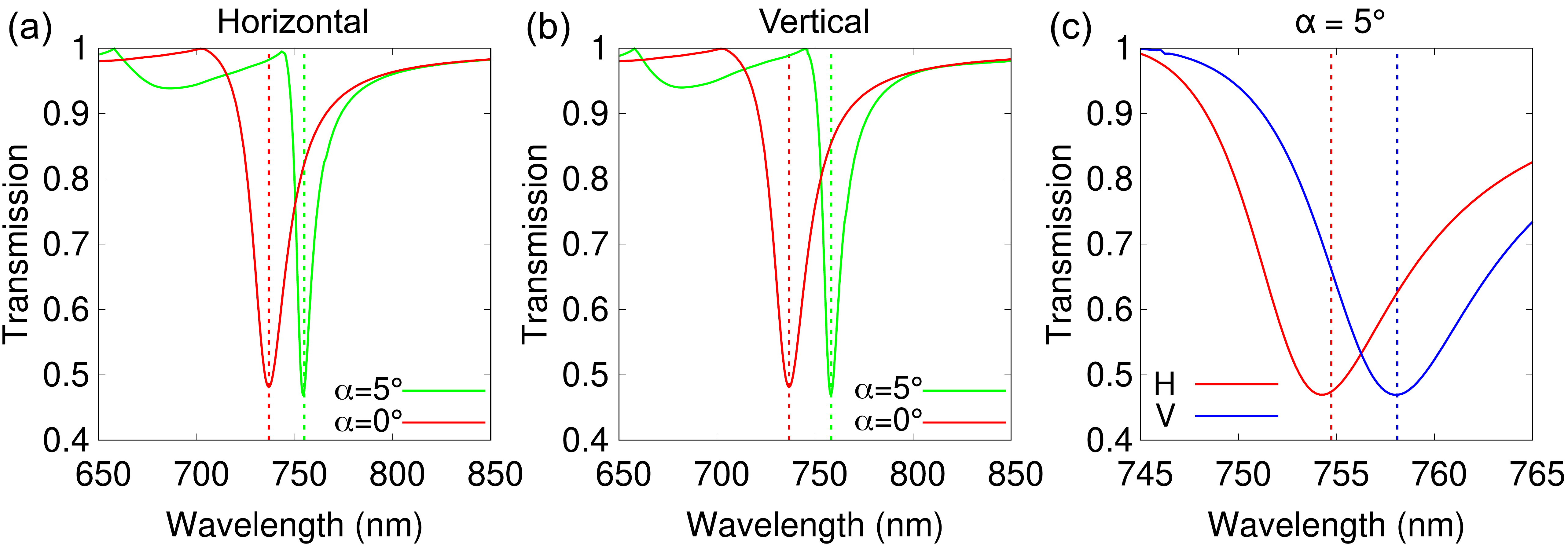}
	\caption{\label{fig:eig_vs_transm}Comparison of the zeroth-order transmission (solid lines) with the eigenfrequency from separate eigenmode calculations (dashed lines), both obtained with the finite element method. Transmission spectra are calculated for both normal incident linearly polarized light (angle of incidence $\alpha=0$) and an angle $\alpha=5\degree$ with respect to the normal direction. (a) Transmission obtained for horizontal (H) linear polarization. (b) Same for vertical (V) linear polarization. (c) Close-up on the transmission minima obtained with $\alpha=5\degree$ for both H and V polarizations. The polarizations H and V are defined in the same way as in Figs.~2(a),(b) of the main text.}
\end{figure*}
The transmission resulting from our measurement procedure accurately characterizes both the band structure and the polarization-dependent properties of the eigenmodes in our non-Hermitian system. While traditional methods of interpreting measurement data $-$ in this case, using extinction maxima as an input for the QGT calculation (see eqs.~\eqref{quantumetric_transm}, \eqref{berrycurv_transm}, \eqref{angles} and \eqref{stokes_intensities}) $-$ may not be suitable for certain non-Hermitian systems~\cite{ZhenNature2015}, this is not the case for our system. 

We next prove this with microscopical (finite element) numerical simulations. In particular, we simulate the band splitting along the diagonal of the Brillouin zone (see Fig.~\ref{fig:eig_vs_transm}) and show that the polarization properties in Fig.~2(b) are reproduced. Figs.~\ref{fig:eig_vs_transm}(a),(b) show that the transmission minima (which correspond to the maxima in Fig.~2(b)) agree very well with separate eigenmode simulations, both at the $\Gamma$ point (red lines, with normal incident radiation $k_{||}=0$) and at a point on the diagonal of the Brillouin zone (green lines, with $(k_{x},k_{y})=(k_{0}\sin{\alpha}/\sqrt{2})(1,1)$ and $\alpha=5\degree$). Moreover, while the obtained transmission spectrum is identical at normal incidence for both horizontal (H) and vertical (V) incoming light polarization (defined as in Figs.~2(a),(b)), Fig.~\ref{fig:eig_vs_transm}(c) shows that the transmission minima for H and V polarizations lie at different wavelengths, revealing the band splitting. We note that the ordering of the H/V minima agrees with that of the maxima found in Fig.~2(b). The polarization of the eigenmodes (dashed lines) agrees with that of the transmission minima and is identified from the near fields shown in Section \ref{supp:FEM} of this Supplementary (see also the discussion therein).

\vspace{0.5cm}
\section{Connection of the two-band model to related models in the literature}
Here we discuss our two-band model in the context of the available literature. The two-band Hamiltonian in Eq.~\eqref{2bandHam} is of the generic form that is sufficient for describing basic topological phenomena~\cite{BernevigSuperconductors}. A similar Hamiltonian has been successfully used to calculate the QGT for example in exciton-polariton systems both in the Hermitian \cite{GianfrateNature2020} and in the non-Hermitian \cite{LiaoPRL2021} regimes. The momentum-dependent couplings $\Omega_{i}(\mathbf{k})$ have well established interpretations in such systems as the Hamiltonian matrix, in the Hermitian limit, reads as follows \cite{BleuPRB2018}:
\begin{align}\label{HamMatrixLikeMalpuech}
	\hat{H}=
	\begin{pmatrix}
		\epsilon(\mathbf{k})+\Omega_{z} & \beta_{0}k^{2}e^{2i\theta} \\
		\beta_{0}k^{2}e^{-2i\theta} & \epsilon(\mathbf{k})-\Omega_{z}
	\end{pmatrix}
	\textrm{.}
\end{align}
In Ref.~\cite{BleuPRB2018}, the term $\epsilon(\mathbf{k})$ represents the unperturbed photonic bands sustained by a planar microcavity, of which our empty lattice bands constitute an analogy. The Zeeman term $\Omega_{z}$ creates chiral modes and, as discussed above, is zero in our system. Terms such as a constant $\Omega_{x}$ in Eq.~\eqref{2bandHam}, that accounts for birefrigence~\cite{BleuPRB2018}, are absent here as well. The off-diagonal, $k$-dependent terms in Eq.~\eqref{HamMatrixLikeMalpuech} have corresponding terms in our system. In microcavity polariton systems, these spin-orbit coupling terms originate from TE-TM band splitting at Dirac points with linear crossings~\cite{GianfrateNature2020}. The underlying physical reason of the TE-TM splitting is, typically, the different penetration depth of TE and TM modes in the Bragg mirrors of the microcavity or a similar difference in the electromagnetic environment of the TE and TM modes \cite{PanzariniPRB1999,RibeiroChemSci2018}. In our case the non-zero spin-orbit coupling terms $\Omega_x$ and $\Omega_y$ (Eqs.~\eqref{omega_x} and \eqref{omega_y}) correspond to a TE-TM splitting as well. However, the physical origin of the splitting is different: it arises from the highly directional and polarization-dependent nature of the radiation emerging from the nanoparticles. In particular, the radiative coupling of the nanoparticles through the diffraction orders of the lattice creates a collective electromagnetic mode and imposes a phase to the individual LSPR (or to the dipole moment, in the simplest case) of each particle. This is explained in Figs. 1(a) and 2(c) in the main text.

Unlike in e.g.~Ref.~\cite{GianfrateNature2020}, our two-band model is non-Hermitian. In our case, the non-hermiticity enters as an imaginary term multiplying $\sigma_{y}$ in Eq.~\eqref{2bandHam}. A similar term, accounting for an imaginary effective field, also makes the Hamiltonian non-Hermitian in studies of the quantum metric near exceptional points~\cite{SolnyshkovPRB2021}. While the non-Hermitian term therein is constant, Eq.~\eqref{omega_y} is momentum-dependent in our Hamiltonian. Overall, both the studies in Refs.~\cite{SolnyshkovPRB2021,LiaoPRL2021} and ours deal with the non-Hermitian implications in the QGT; whereas the former report on non-Hermitian quantum metric, here we present both Hermitian quantum metric (with non-Hermitian corrections) and Berry curvature of non-Hermitian nature.


%